\documentclass[pre, twocolumn, amsmath, amssymb, superscriptaddress,floatfix,aps]{revtex4-1}
\usepackage{graphicx}
\usepackage[caption=false]{subfig}
\usepackage{color}
\usepackage{xcolor}
\usepackage{textcomp}
\definecolor{magenta}{HTML}{FF5477}
\definecolor{blue}{HTML}{498BB6}
\definecolor{purple}{HTML}{523E6F}
\definecolor{brown}{HTML}{AD8A69}

\begin{document}


\title{Predictability of Critical Transitions}



\author{Xiaozhu Zhang}
\email[]{xzhang@nld.ds.mpg.de}
\affiliation{Network Dynamics, Max Planck Institute for Dynamics and Self-Organization}
\author{Christian Kuehn}
\email[]{ck274@cornell.edu}
\affiliation{Institute for Analysis and Scientific Computing, Vienna University of Technology}
\author{Sarah Hallerberg}
\email[]{shallerberg@nld.ds.mpg.de}
\affiliation{Network Dynamics, Max Planck Institute for Dynamics and Self-Organization}

\date{\today}

\begin{abstract}
Critical transitions in multi-stable systems have been discussed as models for a variety of phenomena ranging from the extinctions of species to socio-economic changes and climate transitions between ice-ages and warm-ages.
From bifurcation theory we can expect certain critical transitions to be preceded by a decreased recovery from external perturbations.
The consequences of this critical slowing down have been observed as an increase in variance and autocorrelation prior to the transition.
However especially in the presence of noise it is not clear, whether these changes in observation variables are statistically relevant such that they could be used as indicators for critical transitions. 
In this contribution we investigate the predictability of critical transitions in conceptual models. 
We study the quadratic integrate-and-fire model and the van der Pol model, under the influence of external noise.
We focus especially on the statistical analysis of the success of predictions and the overall predictability of the system. 
The performance of different indicator variables turns out to be dependent on the specific model under study and the conditions of accessing it.
Furthermore, we study the influence of the magnitude of transitions on the predictive performance.
\end{abstract}


\maketitle


\section{\label{sec:intro}Introduction}
Evidences of abrupt drastic shifts, so-called \textit{critical transitions} or \textit{tipping points}, have been reported in very different complex systems including, but not limited to, ecosystems \cite{Schefferetal1,SchefferCarpenter}, medical conditions \cite{Venegasetal,McSharrySmithTarassenko}, financial markets \cite{MayLevinSugihara} and climate 
\cite{Lentonetal,Alleyetal}.
When a critical transition is reached, the system undergoes a sudden, relatively rapid and often irreversible change.
The idea that a wide variety of critical transitions (CTs) could be preceded by \textit{early-warning signs} \cite{Wiesenfeld1,wissel1984universal} is summarized in the short review \cite{scheffer2009critical}.
Early-warning signs are theoretically justified through understanding a system that is likely to undergo a critical transition as a dynamical system close to a bifurcation point.
In this case, perturbations from stable states decay more slowly in comparison to the situation far away from the bifurcation.
This effect is well known as \textit{critical slowing down} (CSD) \cite{Wiesenfeld1,wissel1984universal}.
The consequences of a CSD can be monitored in several variables, that can serve to predict CTs.
Examples for these predictors in stochastic systems are an \textit{increase of the variance} \cite{biggs2009turning,carpenter2006rising,scheffer2009early} and an \textit{increase in autocorrelation} \cite{ives1995measuring,scheffer2009early}.
This approach has been applied to models \cite{kuehn2013mathematical,DonangeloFortDakosSchefferNes,GuttalJayaprakash1} but also to experimental data \cite{DrakeGriffen,MeiselKlausKuehnPlenz} and field data \cite{Wangetal}.
However, the above mentioned examples of critical transitions that have been predicted and observed only once \cite{DrakeGriffen, scheffer2009early} do not allow us to access the quality of the predictors in any statistically relevant way.
It is in fact common for indicators of extreme events \cite{Physa, hallerberg2007precursors, Hallerberg2008a, BogachevBundeEPL2009, BogachevBundePRE2009, HallerbergDeWijn2014, Miotto2014} to be tested with standard measures for the quality of classifiers, such as skill scores \cite{Brier,roulston}, contingency tables \cite{egan1975signal} or receiver operator characteristic curves (ROC curves) \cite{egan1975signal}.
Additionally it is common to test for the dependence of the prediction success on parameters related to the estimation of predictors, the prediction procedure and the events under study.
Such parameters could be, for example, the length of the data record used to estimate the predictor, the lead time (time between issuing the forecast and the observation of the event) or the magnitude and relative frequency of the event under study \cite{Hallerberg2008a}.
Similar tests for indicators associated with CSD are\textemdash apart from one study \cite{BoettingerHastings}\textemdash still missing.

Although theoretical aspects of a relation between CSD and CTs can be formalized mathematically using the theory of fast-slow stochastic dynamical systems \cite{kuehn2011mathematical,kuehn2013mathematical}, relatively little is known about the practical relevance of the indicators associated with CSD.
There are several mathematical hypotheses in the theoretical results, which may not be satisfied in practical applications.
In fact, real-life data sets are extremely difficult to analyze with respect to critical slowing down \cite{KuehnMartensRomero,MeiselKuehn,DitlevsenJohnsen}.
In the vicinity of many bifurcations, even relatively small perturbations can cause a drastic shift from a marginally stable state of the system to a contrasting state \cite{scheffer2009early}, leading to noise-induced transitions \cite{AshwinWieczorekVitoloCox, berglund2006noise}, which are more difficult to predict \cite{BoettingerHastings1} as they are intermediate between a large deviation \cite{FreidlinWentzell} and a bifurcation regime.

In this contribution we evaluate the statistical relevance of predictor variables associated with CSD.
In contrast to single event predictions made for critical transitions in experiments or large scale models, we study the success of predictors using data sets that contain $10^{6}$ CTs.
To generate these data sets, we use two low-dimensional conceptual models, introduced in Sec.~\ref{sec:models}.
In Sec.~\ref{sec:detection}, we explain how we verify the occurrence of CTs in the generated time series.
Details concerning the estimation of predictors associated with CSD are presented in Sec.~\ref{sec:predicting}.
The predictive power of these predictors is evaluated in Sec.~\ref{sec:predpower} with a focus on varying the length of the observation used for estimating the predictors, and the lead time, i.e. the time between issuing the forecast and the occurrence of the event.
%
%
Additionally, we study in Secs.~\ref{sec:mag_setup} and \ref{sec:detailedmag} whether the magnitude of the transitions has an effect of their predictability. 
%
%
%
%
We summarize our results in Sec.~\ref{sec:conclusions}.
%
%
\section{Conceptual Models for Critical Transitions}
\label{sec:models}
One basic characterization of CTs is the separation of system into two distinct phases: (a) a slow drift of system variables without any large variations and (b) a fast transition phase with a very drastic sudden change of at least one of the system variables. Hence, it is natural to propose fast-slow dynamical systems \cite{KuehnBook} as a basic building block to model these effects \cite{kuehn2011mathematical}. 
A deterministic planar fast-slow system of ordinary differential equations (ODEs) is given by
\begin{eqnarray}
\epsilon \frac{dx}{d\tau} &=& \dot{x}=f(x,y),  \label{eq:fast-slow_general_det1}\\ 
         \frac{dy}{d\tau} &=& \dot{y}=g(x,y),  \label{eq:fast-slow_general_det2}
\end{eqnarray}
where $(x,y)\in \mathbb{R}^2$, $0<\epsilon \ll1$ is a small parameter so that $x$ is a \textit{fast variable} in comparison to the \textit{slow variable} $y$. Although one may also study more general fast-slow systems with $(x,y)\in\mathbb{R}^m\times \mathbb{R}^n$, we shall restrict to the planar case as it already illustrates the effects we are interested in. 
The set 
\begin{equation}
C_0:=\{(x,y)\in \mathbb{R}^2:f(x,y)=0\}
\end{equation}
is called the \textit{critical manifold}; in fact, we are going to assume that $C_0$ is a smooth manifold for the examples we consider.
%
%
%
Considering the limit $\epsilon\rightarrow 0$ in \eqref{eq:fast-slow_general_det1} then leads to the slow subsystem on $C_0$ given by
\begin{equation}
\frac{dy}{d\tau} = \dot{y}=g(h(y),y).  \label{eq:slow_flow}
\end{equation} 
For $\epsilon>0$, it is possible to rescale time in \eqref{eq:fast-slow_general_det1}-\eqref{eq:fast-slow_general_det2} via $t:=\tau/\epsilon$ and then consider $\epsilon\rightarrow 0$ to obtain the fast subsystem
\begin{eqnarray}
\frac{dx}{dt} &=& x'=f(x,y),  \label{eq:fast_det1}\\ 
\frac{dy}{dt} &=& y'=0,  \label{eq:fast_det2}
\end{eqnarray}
which is a parametrized system with parameter $y$ (or frozen slow variable).
%
%
%
%

To incorporate the effect of noise, which may arise, {e.g.}, via external forcing, finite-system size or internal small-scale fluctuations, one usually extends the system \eqref{eq:fast-slow_general_det1} and \eqref{eq:fast-slow_general_det2} by noise terms \cite{kuehn2013mathematical} and studies fast-slow stochastic ordinary differential equations (SODEs) given by 
\begin{eqnarray}
\dot{x}  &=& \frac{1}{\epsilon}f(x,y) + \frac{\sigma_1}{\sqrt{\epsilon}}\,\eta_1(\tau), 
\label{eq:fast-slow_general_phys1}\\ 
\dot{y} &=& g(x,y) + \sigma_2 \, \eta_2(\tau),  
\label{eq:fast-slow_general_phys2}
\end{eqnarray}
where $\eta_1(\tau)$ and $\eta_2(\tau)$ are independent Gaussian white noises.
%
%

There are different mechanisms that explain how large sudden jumps can be induced in fast-slow SODEs \cite{AshwinWieczorekVitoloCox}. 
The two main classes are bifurcation-induced (\textit{B-tipping}) or noise-induced (\textit{N-tipping}) transitions. 
Regarding early-warning signs, it is well known that before B-tipping we expect a growth invariance and autocorrelation \cite{carpenter2006rising,Wiesenfeld1}. 
For example, the variance of the fast variable near the attracting part of the slow manifold approaching a fold point at $(x_c,y_c)$ satisfies
\begin{equation}
\textnormal{Var}(x(t))= \mathcal{O}\left(\frac{1}{\sqrt{y-y_c}}\right),\quad 
\text{as $y\searrow y_c$}\label{eq:fold_var_scaling}
\end{equation} 
as long as $(x(t),y(t))$ is outside a small $\epsilon$-dependent neighborhood near the fold point \cite{kuehn2013mathematical,berglund2002metastability} and suitable limits in $\epsilon$ relative to the noise level are considered.
A general variance increase up to a threshold and/or scaling laws such as \eqref{eq:fold_var_scaling} can be used for predicting a jump at a fold point under suitable conditions on a sufficiently long data set and under the condition that the slow drift speed governed by $\epsilon$ is sufficiently large in comparison to the noise level $\sigma:=\sqrt{(\sigma_1)^2+(\sigma_2)^2}$; for a fold, this condition is more precisely expressed by assuming that $\sigma=\sigma(\epsilon)$ depends upon $\epsilon$,
\begin{equation}
\label{eq:drift_cond}
0< \sigma(\epsilon)=\sigma\ll \sqrt\epsilon\ll 1,\quad \text{as $\epsilon \rightarrow 0$}
\end{equation}
However, if the condition fails and the opposite relation $\sigma\gg \sqrt\epsilon$ holds, we are in the N-tipping regime and the noise drives large jumps with high probability before the fold point governed by large deviation theory \cite{FreidlinWentzell}. 
Purely noise-induced jumps are very difficult, if not impossible, to predict.
%
%
\subsection{The Quadratic Integrate-and-Fire (QIF) Model}

\begin{figure}[t!]
\centerline{
\includegraphics[width=1.0\columnwidth]{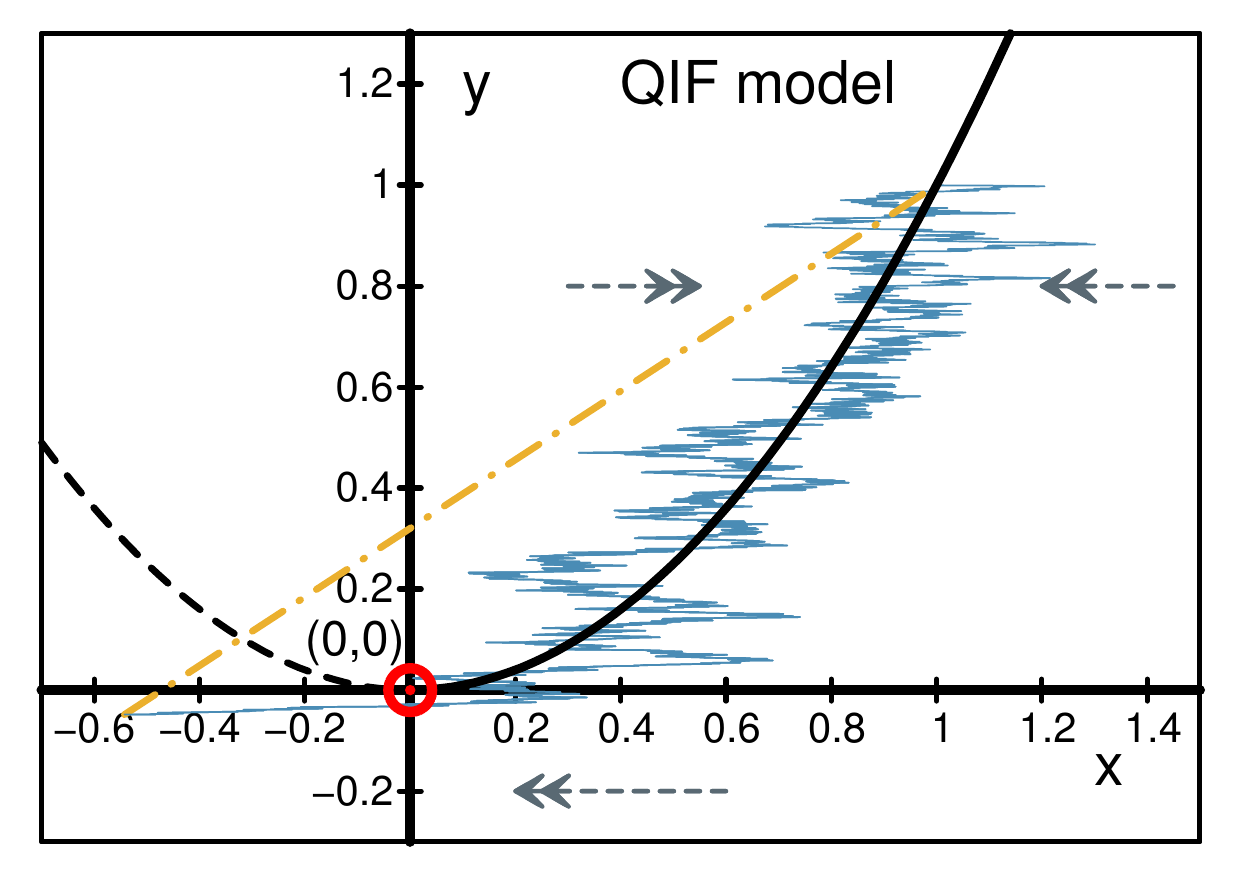}
}
\caption{\label{fig:phase_0502}The dynamics of the QIF model illustrated by the phase portrait. The parameters $(\epsilon,\delta,\sigma_1) = (0.02,0.5,0.2)$.
The black curves represent the critical manifold $C^{\text{QIF}}_0=\left\lbrace(x,y):y=x^2\right\rbrace$, a solid line for the attracting part and dashed line for the repelling part. 
The fold point $(0,0)$ is marked as a red (gray) circle. 
The gray dashed double arrows indicate the fast subsystem flow, {i.e.}, when the system is not near $C_0^{\text{QIF}}$.
A numerical solution is presented as a blue (gray) solid line, and its path of reset as a yellow (light gray) dot-dashed line.}
\end{figure}     

One natural model to start with is to restrict to a fold normal form with a reset after the jump at the fold; see Fig.~\ref{fig:phase_0502}.
In particular, if we let $f(x,y)=y-x^2$, $F(x,y)\equiv1$, $g(x,y)=-1$, and $G(x,y)\equiv0$ in \eqref{eq:fast-slow_general_phys1} and \eqref{eq:fast-slow_general_phys2}, then we obtain a local fold normal form with noise acting on the slow variable 
\begin{eqnarray}
\dot{x} &=& \frac{1}{\epsilon}(y-x^2)  +  \frac{\sigma_1}{\sqrt{\epsilon}}\eta_1(t), \label{eq:QIFmodel1}\\ 
\dot{y} &=& -1. \label{eq:QIFmodel2} 
\end{eqnarray}
We are also going to use a deterministic reset in a certain phase-space region described below and refer to the resulting model as the quadratic integrate-and-fire (QIF) model. 
It resembles the classical integrate-and-fire model \cite{SacerdoteGiraudo} except that we use a quadratic function for the integration phase before the firing phase. 
Near its bifurcation point the QIF also resembles the theta model for excitable neurons \cite{latham2000intrinsic, ermentrout1996type}. 
In the noiseless case, the value of $y$ determines the number of equilibria of the fast flow of $x$. 
For a positive $y$, we have two equilibrium branches $C_0^{a\pm}=\{x=\pm\sqrt{y},y>0\}$, whose union together with the fold point at $(0,0)$ is the critical manifold $C_0$. 
Note that $C_0^{a+}$ is attracting, while $C_0^{a-}$ is repelling. When $y$ is negative, the fast subsystem has no equilibria at all. 
In particular, a saddle-node (or fold) bifurcation occurs at $y=0$. 
Therefore, the critical manifold of the QIF model is attracting in quadrant I ($x>0,y>0$) and repelling in quadrant II ($x<0,y>0$) as is illustrated in Fig.~\ref{fig:phase_0502}. 

When $0<\epsilon\ll 1$ and starting from the point $(1,1)$ and uniformly decreasing $y$, the trajectory of the solution travels near the attracting critical manifold $C^{a+}_0$ towards the fold point $(0,0)$.
Shortly before reaching the fold point, depending upon the noise level, the system may perform a noise-induced jump across the flattening potential barrier between the stable and the unstable equilibria and arrive in quadrant II.
In quadrant II, the repelling critical manifold drives the system further and further towards negative infinity in $x$. 
In our model, however, the system is considered to be in an excited state, when the fast variable $x$ is below a threshold $-\delta$ and reset to the initial condition $(1,1)$.

Numerical simulations of equations \eqref{eq:QIFmodel1} and \eqref{eq:QIFmodel2} using the Euler-Maruyama method \cite{Higham} generate time series of $N$ observations $\{x_n\}$ and $\{y_n\}$ at discrete time steps $t_n=t_0 + n \Delta t$. 
Here, $t_0$ denotes the initial time, $n=0,1,\ldots,N-1$ is the index of each time step, and $\Delta$ is a constant time interval of numerical integration.

\begin{figure}[t!]
\centerline{
\includegraphics[width=1.0\columnwidth]{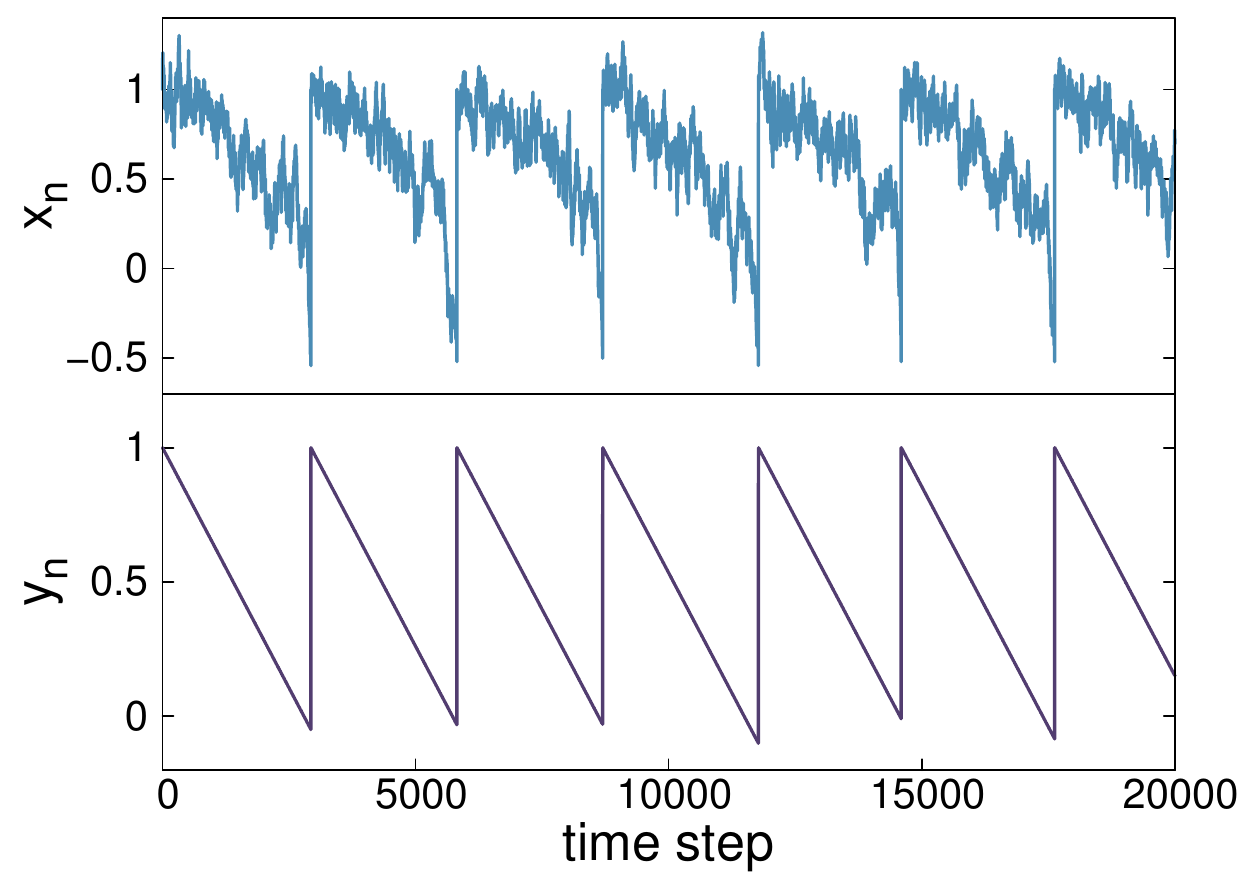}
}
\vspace{-0.2cm}
\caption{\label{fig:ts_xyt}Time series of the fast variable $x$ and the slow variable 
$y$ of the QIF model, simulated using the Euler-Maruyama method with parameter values 
$(\epsilon,\delta,\sigma_1) = (0.02,0.5,0.2)$.
}
\end{figure}
As illustrated in Fig.~\ref{fig:ts_xyt}, CTs can be observed in the time series of the fast variable $x$ while the slow variable $y$ acts as the slowly changing bifurcation parameter.
Since the system is reset to the initial state $(1,1)$ after $x$ exceeds a certain threshold $-\delta$, we can use the QIF model to generate an arbitrary amount of CTs, which we are going to investigate below from a statistical perspective. 

The stochastic QIF model may not only have direct relevance for many transitions in neuroscience \cite{ErmentroutTerman,SacerdoteGiraudo} as a local normal form to model the subthreshold dynamics before spiking or bursting but the QIF model could also be viewed as useful for any applications with local fold dynamics and global resets.

\subsection{The van der Pol Model}
\label{sec:vdPmodel}
In addition to the purely local QIF model with resets, it is also natural to compare it to a model, where the resets are via a smooth global nonlinearity. 
The classical example to study are van der Pol \cite{vanderPol1} (or FitzHugh-Nagumo \cite{FitzHugh}) relaxation oscillators \cite{Grasman}. 
In particular, we consider $f(x,y) = y-\frac{27}{4\delta^3}x^2(x+\delta)$, $F(x,y)\equiv1$, $g(x,y) = -\frac{\delta}{2}-x$, $G(x,y)\equiv0$ and obtain a version of the van der Pol (vdP) system
\begin{eqnarray}
\dot{x} &=& \frac{1}{\epsilon}\left(y-\frac{27}{4\delta^3}x^2(x+\delta)\right) 
+ \frac{\sigma_1}{\sqrt{\epsilon}}\eta_1(t), \label{eq:vdPmodel1} \\
\dot{y} &=& -\frac{\delta}{2}-x. \label{eq:vdPmodel2}
\end{eqnarray}
The precise choice of the form of the model will be motivated in more detail below, particularly with respect to the parameter $\delta$. 
When the external stimulus exceeds a certain threshold, the behavior of the system changes qualitatively from a stable fixed point to a limit cycle undergoing a Hopf bifurcation. 

\begin{figure}[t!]
\centerline{
  \includegraphics[width=1.0\columnwidth]{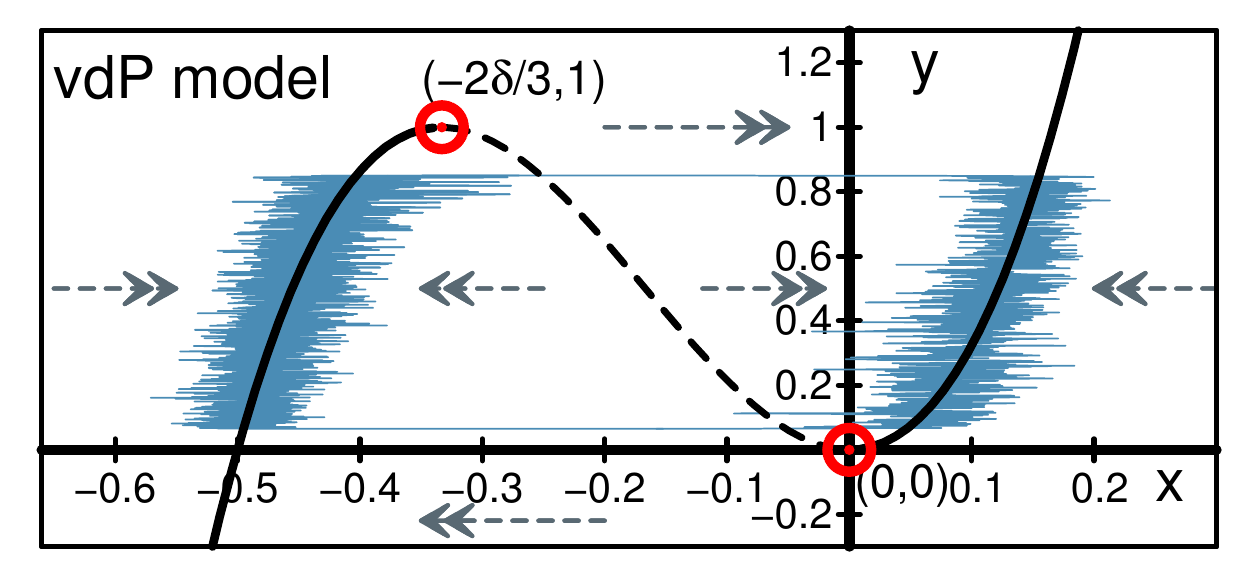}}
\vspace{-0.2cm}
\caption{\label{fig:phase_0601}The dynamics of the vdP model. The parameters $(\epsilon,\delta,\sigma_1) = (0.02,0.5,0.1)$.
Analogous to Fig.~\ref{fig:phase_0502}, the critical manifold (black lines, solid for the attracting part and dashed for the repelling part), the fold points at $(-\frac{2}{3}\delta,1)$ and $(0,0)$ (red circles) and the numerical solution trajectory [blue (gray) solid line] are plotted in state space. 
The dashed double arrows indicate the orientation of the relaxations in the noisy case and the noise-induced transitions for the fast varia
}
\end{figure}

The deterministic version of the model, {i.e.}, $\sigma_1=0$, has one fixed point 
$(x_{\text{FP}},y_{\text{FP}}) = (-\frac{\delta}{2},\frac{27}{32})$, which is unstable 
under the assumptions $\delta\in(-\sqrt{3},0)$ and $0<\epsilon\ll1$. 
A trajectory of the stochastic vdP ($\sigma_1 \neq 0$) forms a noisy relaxation-oscillation-type periodic orbit involving two rapid transitions and two slow drifts as is illustrated in Fig.~\ref{fig:phase_0601}. 
Since the critical manifold of the vdP model has two fold points at $(-\frac{2}{3}\delta,1)$ and $(0,0)$, the manifold is naturally split into three parts (left, middle and right)
\begin{eqnarray}
 C_{\text{vdP}}^l & = & C_{\text{vdP}}\cap\left\lbrace (x,y):x<-\frac{2}{3}\delta\right\rbrace, \\
 C_{\text{vdP}}^m &= & C_{\text{vdP}}\cap\left\lbrace (x,y):-\frac{2}{3}\delta<x<0\right\rbrace, \\
 C_{\text{vdP}}^r &=& C_{\text{vdP}}\cap\left\lbrace (x,y):x>0\right\rbrace.
\end{eqnarray}
By investigating the stability of the equilibria of of the fast variable $x$ for a fixed $y$ (in the $\epsilon\rightarrow0$ limit), we see that $C_{\text{vdP}}^l$, $C_{\text{vdP}}^r$ are normally hyperbolic attracting parts of the critical manifold and $C_{\text{vdP}}^m$ is a normally hyperbolic repelling part of the critical manifold. 
Therefore, in the vicinity of $C_{\text{vdP}}^l$ and $C_{\text{vdP}}^r$, the system is resilient to perturbations for long times, {i.e.}~it is able to relax quickly back to the critical manifold \cite{berglund2006noise}. 
Furthermore, the system travels slowly along $C_{\text{vdP}}^l$ and $C_{\text{vdP}}^r$, driven by the slow variable $y$. 

In quadrant I of the $x$-$y$ plane, $\dot{y}$ is negative so the system moves downwards along $C_{\text{vdP}}^r$. 
Under conditions of the form \eqref{eq:drift_cond} the system recovers from small random disturbances with high probability until it is close to the fold point $(0,0)$. 
When near $(0,0)$, the potential barrier between the stable equilibrium on $C_{\text{vdP}}^r$ and on $C_{\text{vdP}}^l$ flattens, so that it becomes more likely that the system jumps away from $C_{\text{vdP}}^r$ to a region to the left of $C_{\text{vdP}}^m$ and then it gets repelled to a region near the other attracting part $C_{\text{vdP}}^l$. 
Note very carefully that if we consider the fast subsystem precisely at $y=0$ and for $\sigma_1=0$, then we have 
\begin{equation}
x' = \left(-\frac{27}{4\delta^3}x^2(x+\delta)\right) 
\end{equation}
so that the second fast subsystem equilibrium besides $x=0$ is given by $x=-\delta$. So we may define the magnitude of a CT as the distance $\delta$ for the jump at, respectively near, the lower fold. 
Analogous analysis applies for the second fold point as we have a fast subsystem at $y=1$
given by
\begin{equation}
x' = \left(1-\frac{27}{4\delta^3}x^2(x+\delta)\right) 
\end{equation}
with fast subsystem steady states at $x=-\frac23\delta$ and $x=\frac13\delta$, so the distance of the jump at the second fold is again of size $\delta$. 
\begin{figure}[!t]
\begin{center}
\centerline{\includegraphics[width=1\columnwidth]{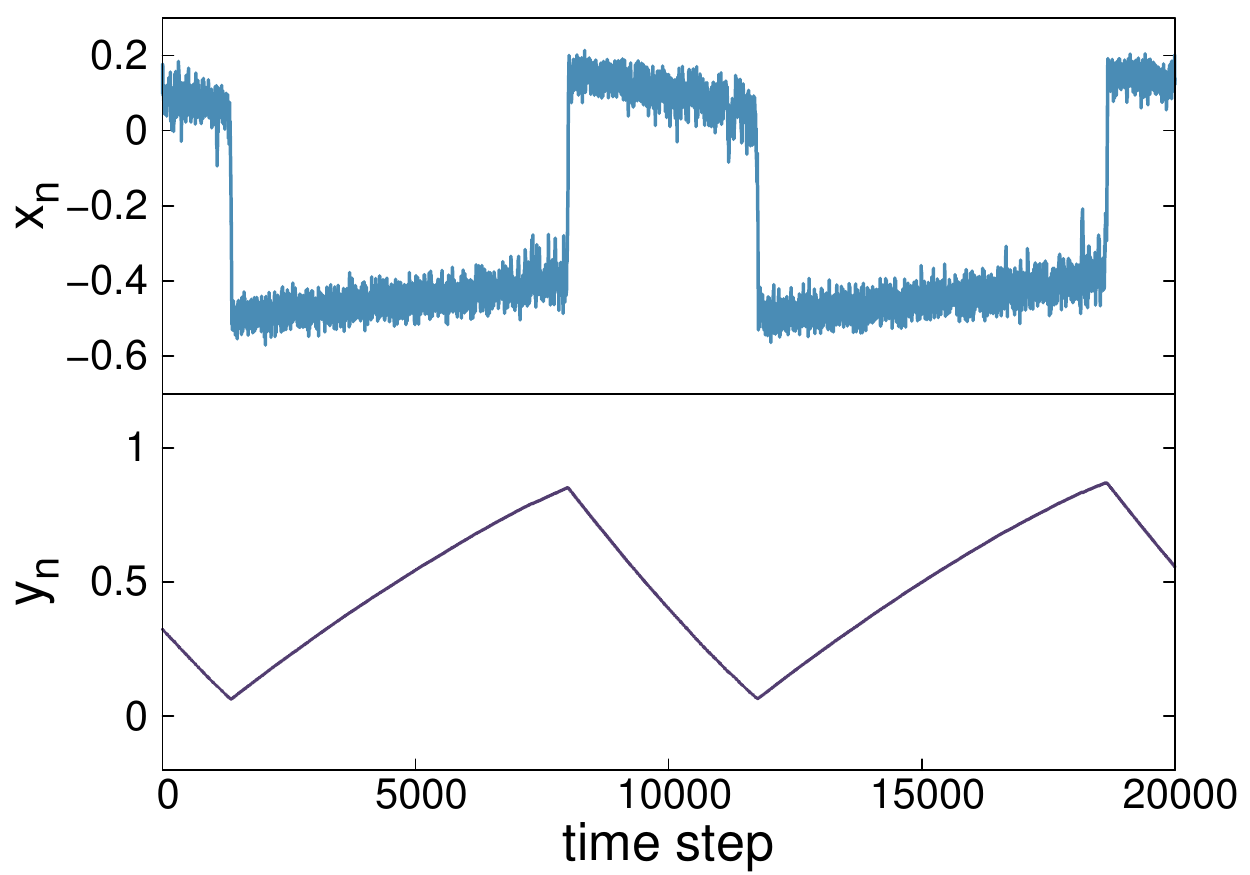}}
\end{center}
\vspace{-1cm}
\caption{\label{fig:ts_xytvdP}Time series of the fast variable $x$ and the slow variable $y$ of the vdP model, simulated using the Euler-Maruyama method with parameter values $(\epsilon,\delta,\sigma_1) = (0.02,0.5,0.1)$.}
\end{figure}
This allows us to change $\delta$ and associate with it the size of both CTs during a relaxation oscillation. 
The only difference between the branches is that the speed of upward drift in the left branch is slower than the downward drift in the right branch, leading to a longer residence time for the left branch.

Simulating Eqs.~\eqref{eq:vdPmodel1} and \eqref{eq:vdPmodel2} can again be used to generates time series of $N$ observations $\{x_n\}$ and $\{y_n\}$ at discrete time steps $t_n$; see Fig.~\ref{fig:ts_xytvdP}. 
In contrast to the QIF model no (potentially artificial) resets are necessary for the vdP model.

\section{Detecting Critical Transitions as Extreme Events in Time Series}
\label{sec:detection}
 
\begin{figure}[!t]
\centerline{
  \includegraphics[width=1\columnwidth]{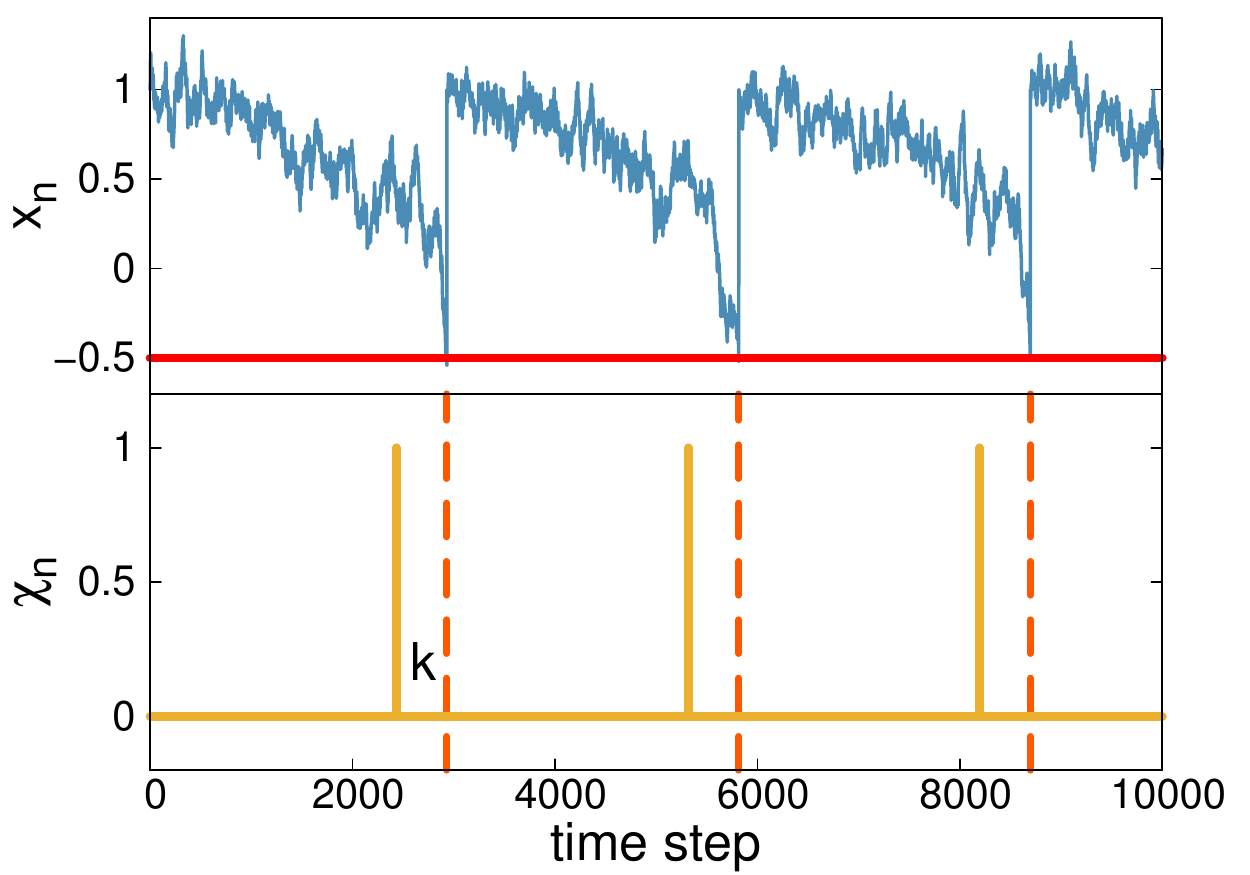}}
\vspace{-0.2cm}
\caption{Time series $\{x_n\}$ and tracer time series $\{\chi_n\}$ of the QIF model with $(\epsilon,\delta,\sigma_1) = (0.02,0.5,0.2)$.
In the upper panel, the threshold for reset $x=-\delta$ is shown as a red solid line.
Both time series are aligned such that the time points $t_n$ when the CTs occur (dashed lines) and the ones $t_m$ when the predictions are made (orange (light gray) impulses of the tracer time series) can be compared. 
The forecast lead time between $t_n$ and $t_m$ is set to $k=500$ for demonstration.}\label{fig:ts_tracer_QIF}
\end{figure}

\begin{figure}[!t]
\centerline{
\includegraphics[width=1\columnwidth]{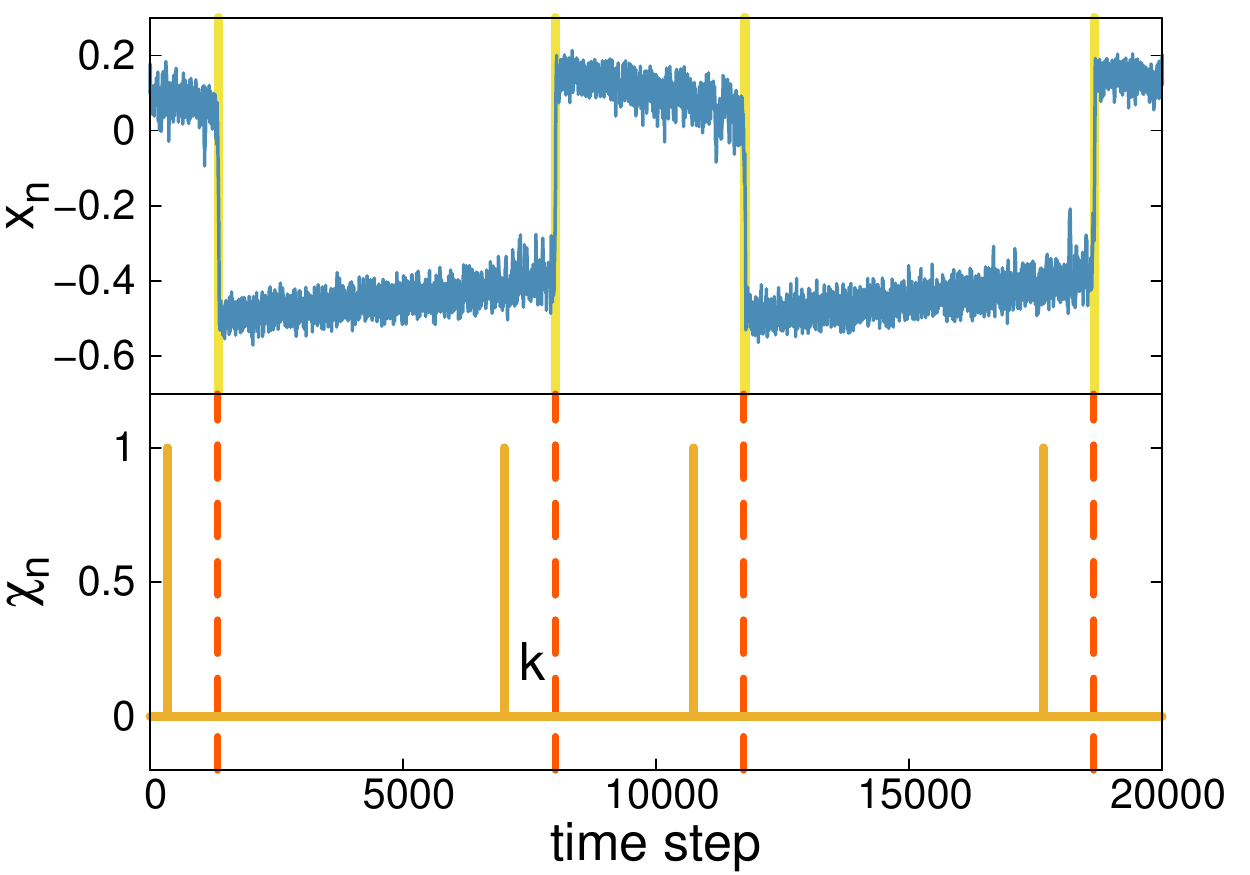}
}
\vspace{-0.3cm}
\caption{Time series $\{x_n\}$ and tracer time series $\{\chi_n\}$ of the vdP model with $(\epsilon,\delta,\sigma_1) = (0.02,0.5,0.1)$. 
In the upper panel, the transition phases are shaded with yellow (very light gray).
Both time series are aligned such that the time points $t_n$ when the CTs occur (dashed lines) and the ones $t_n$ when the predictions are made (orange (light gray) impulses of the tracer time series) can be compared. 
The forecast lead time $k$ is set to  $k=1000$ for demonstration.}
\label{fig:ts_tracer_vdP}
\end{figure}

The two types of CTs observed in the QIF and the vdP model, correspond to two types of extreme events: \textit{threshold-crossings} and \textit{increments}. 
Critical transitions in the QIF model can be categorized as threshold-crossings with the threshold $-\delta$ (Fig.~\ref{fig:ts_tracer_QIF}). 
Relevant examples for extreme events that consist in threshold crossings are river levels outreaching a flood warning threshold after a heavy rain, or temperatures dropping below the freezing point. 
Increments, however are defined as the sudden change of the observed variable larger than a certain value. 
Prominent examples for increments are surges in electric loads and drastic decreases of stock market indices. 
In the time series of the fast variable generated by the vdP model CTs can be observed as increments with magnitude $\delta$ (see Fig.~\ref{fig:ts_tracer_vdP}). 

In order to monitor the occurrence of threshold crossings and increments we introduce a secondary time series called the \textit{tracer time series} $\{\chi_n\}$ with the time index $n=0,1,\ldots,N$ and the same sampling interval $\Delta$. 
The binary \textit{tracer variable} $\chi_n$ stores information about whether an extreme event, {i.e.}, here a CT happens at a future time point $t_n+k\Delta$. 
It is not used as an indicator variable for predicting future CTs, but only for marking the start time of CTs, as will be explained in detail in the following.
The time $k\Delta$, {i.e.}, the time between a forecast made at $t_n$ and occurrence of an event at $t_n + k\Delta$, is sometimes called \textit{lead time} of a prediction.

Identifying threshold crossings in the QIF model and keeping track of their occurrence in a tracer time series is straightforward: Once the value of $x$ drops below the threshold $-\delta$ at $t_n$, the occurrence of a threshold crossing is identified and the tracer variable $ \chi_n(\delta, k)$ at $t_n-k\Delta$ is set to $1$:
\begin{eqnarray}
 \chi_n(\delta, k) &=& \left\lbrace
\begin{array}{ccc}
1& : & \mathrm{if}\hspace{0.1cm}x_{n+k}\leq-\delta \\
0& : & \mathrm{if}\hspace{0.1cm}x_{n+k}\geq-\delta.
\end{array}\right.
\label{eq:tracer_QIF}
\end{eqnarray}

\begin{figure}[!t]
\centerline{
\includegraphics[width=1.1\linewidth]{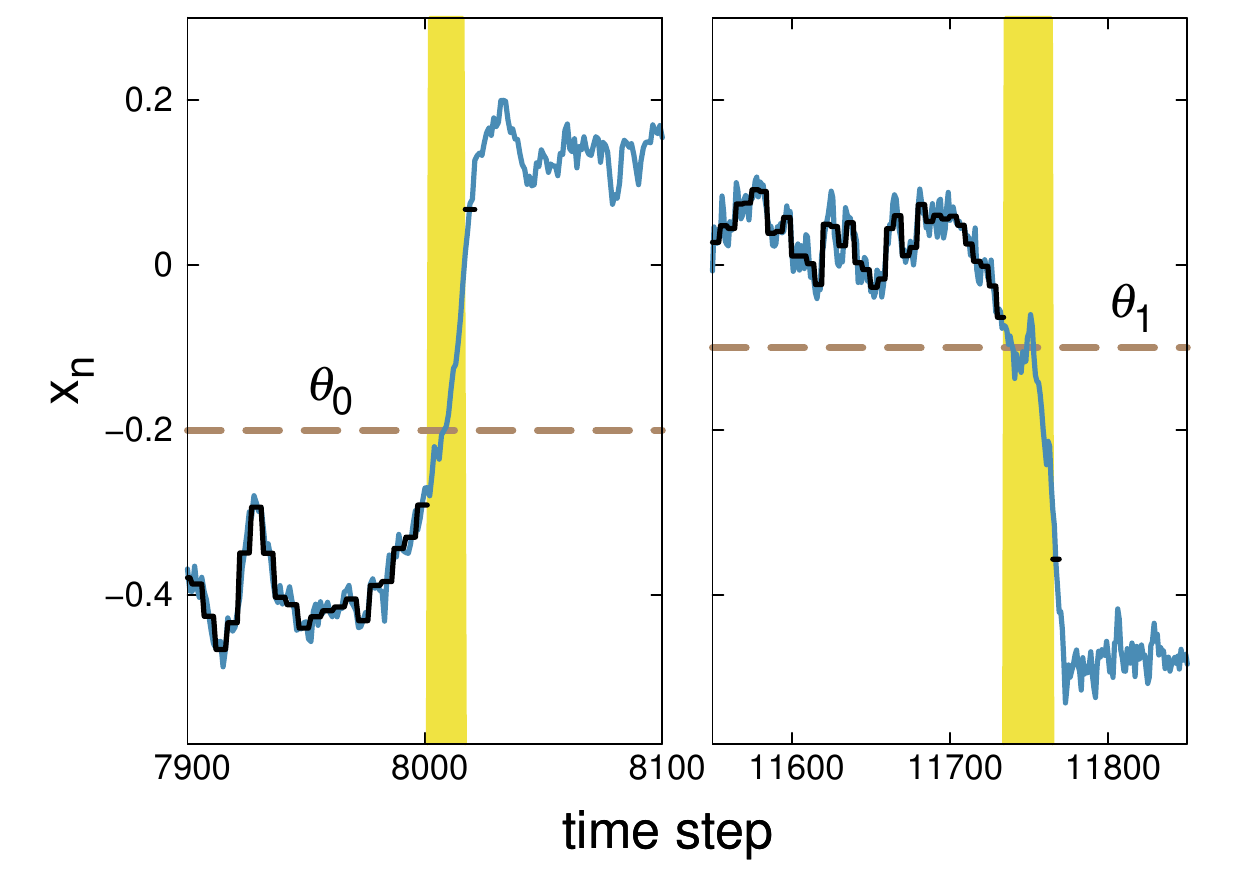} 
}
\vspace{-0.4cm}
\caption{The detection method for positive and negative increments in the vdP model. 
The time series $\{x_n\}$ is shown as blue (gray) lines and the finite time average prior to CTs as black steps.
The finite-time average is used to locate the onset of the transitions by comparing it to the thresholds (dashed lines, $\theta_0$ for the lower state and $\theta_1$ for the upper state).
The narrowed down transition phases (upwards-jumping in the left panel and downwards-jumping in the right panel) are colored yellow (light gray).
The values of the parameters $(M,w_{id},w_f,\theta_0,\theta_1)$ are chosen empirically as $(300,5,20,-0.2,-0.1)$.}
\label{fig:ts_id_vdP}                                                 
\end{figure}

Identifying the onsets of the CTs in the vdP model requires us to distinguish among large fluctuations, meta stable states, and true CTs.
We therefore introduce two auxiliary thresholds: $\theta_0$, the upper limit of the lower state, and $\theta_1$, the lower limit of the upper state.
In addition to the occurrence of a large change $\left|x_{n+k+M} - x_n \right| \geq \delta$ with respect to the current value $x_n$, we also require that the finite-time average after the change $\overline{x}^{n+k+M}_f$ remains above $\theta_1$ or below $\theta_0$ (see Fig.~\ref{fig:ts_id_vdP}).
In more detail, the tracer variable detecting the occurrence of a CT is defined as follows:

\begin{eqnarray}
  \chi_{n+(p-1)w_{id}}(\delta,k) &= &\left\lbrace
\begin{array}{crl}
1& :\mathrm{if} & x_{n+k+M} - x_n \geq\delta\\
& & \wedge\; \overline{x}^{n+k+M}_f\geq\theta_1\\
& & \wedge\; \overline{x}^{n+k}_{0},...,\overline{x}^{n+k}_{p-1}<\theta_0\\
& & \wedge\; \overline{x}^{n+k}_{p} \geq\theta_0 \\
& &\mathrm{(transition\ upwards)}, \\
\\
&\mathrm{or}& x_{n+k+M} - x_n\leq-\delta\\
& & \wedge\; \overline{x}^{n+k+M}_f\leq\theta_0\\
& & \wedge\; \overline{x}^{n+k}_{0},...,\overline{x}^{n+k}_{p-1}>\theta_1\\
& & \wedge\; \overline{x}^{n+k}_{p} \leq\theta_1 \\
& &\mathrm{(transition\ downwards)}; \\
\\
0& : & \mathrm{otherwise}.
\end{array}\right. \label{eq:tracer_vdP} \\
&\mbox{with} & \quad \overline{x}^{n+k+M}_f = \frac{1}{w_{f}} \sum_{i=n+k+M}^{n+k+M+w_f} x_i \;, \nonumber\\
&\mbox{and} & \quad \overline{x}^{n+k}_{p} = \frac{1}{w_{id}} \sum_{i=n+k+({p}-1)w_{id}}^{n+k+{p}w_{id}} x_i \;. \nonumber 
\end{eqnarray}

Here, we use non-overlapping windows of $w_{id}$ time steps and the values of the parameters, $M$, $w_{id}$, $w_f$, $\theta_0$ and $\theta_1$ are set such that they capture the CTs in each specific data set.  
%
%
More specifically, choosing $M$ to be slightly larger than the typical duration of the fast transition phase for a CT and $\delta$ to be larger than the size of typical fluctuations around each state allows us to capture true CTs while excluding large fluctuations.
Additionally, testing for the future behavior after large increments as summarized by $\overline{x}^{n+k+M}_f$ allows us to distinguish true CTs from transitions to meta-stable states.%

%
The exact onset of a transition is determined as follows: We start from the point where a large change is detected, $t_{n+k}$, which should be somewhere before the transition, and calculate the finite-time average in non-overlapping windows. If in the $p$-th window starting from time $t_{n+k}$ the average $\overline{x}^{n+k}_{p}$ crosses the threshold, the onset of the transition is determined to be $t_{n+k}+(p-1)w_{id}\Delta$ (Eq.~\ref{eq:tracer_vdP}).
The end of the transitions is determined analogously, i.e., starting from $t_{n+k+M}$ and calculating the finite time average backwards until it crosses the other threshold.
%
\section{Predicting Critical Transitions}
\label{sec:predicting}
Traditionally, one would think that the prediction of the future implies a precise modeling and understanding of the system under study, such as, e.g., in weather forecasting and climate modeling.
However, whenever precise information about the system under study is unknown, or if the system is too complex to allow for a detailed numerically expensive modeling, then data-based predictions can be a suitable alternative \cite{Physa, hallerberg2007precursors}.  
As more and more large data sets, machine learning methods, and computation power become available \cite{laney01controlling3v}, data-based predictions can also be understand as a generic first choice that does not require human resources, but only the availability of sufficiently many data records of events.
Apart from approaches that predict values of the future trajectories of systems \cite{Hirata2014,Runge2015}, discrete events can be successfully predicted by observing relevant early-warning signals (precursors, predictors).
Methods to identify relevant predictors and prediction margins consist in statistical inference 
\cite{Rish2001,Physa,hallerberg2007precursors,BogachevBundeEPL2009,HallerbergDeWijn2014,Miotto2014}, artificial neural networks \cite{LeCun1989, HinSal06} or other pattern recognition and machine learning approaches \cite{Cortes1995}. 
Here we apply statistical inference (naive Bayesian classifiers \cite{Rish2001,naivebayes}) to predict CTs based on time series generated by the models introduced in Sec.~\ref{sec:models}.
\subsection{Estimating Indicators}
\label{sec:precursors}
Knowing that CTs are preceded by a CSD, we can expect all quantities that are affected by the slowing down to possess some predictive power.
From a practical point of view, finite-time estimates of the standard deviation and the autocorrelation can be easily estimated from a time series.
In more detail, we estimate the following indicators:
\begin{itemize}
\item[P1] a sliding window estimate of the variance ($v_n$),
\item[P2] a sliding window estimate of the standard deviation~($s_n$),
\item[P3] a sliding average of $s$ denoted as $\langle \mbox{$s$} \rangle_n$ and
\item[P4] a sliding window estimate of the lag-$1$~autocorrelation~($a_n$) \cite{priestley2001spectral}.
\end{itemize}
To mimic a realistic prediction scenario, we choose sliding windows of lengths $w\Delta$ such that the resulting indicator variable contains only information from the present data $x_n$ and from the past $x_{n-w+1}, x_{n-w+2}, ..., x_{n-1}$, but not from the future.
In other words, indicators at time $t_n$, estimated from a sliding window between $x_{n-w+1}$ and $x_{n-1}$, are given by
\begin{eqnarray}
\mbox{P1:} &\it{}& \quad v_n = v(x)_{t_n} = \frac{1}{w}\sum_{i=n-w+1}^{n}(x_i-\overline{x}_{t_n})^2, \label{eq:var}\\
&\it{}& \quad \mbox{with}\; \; \; \overline{x}_{t_n} = \dfrac{1}{w}\sum_{i=n-w+1}^{n}x_i; \nonumber \\
\mbox{P2:}&\it{}& \quad s_n = s(x)_{t_n}  = \sqrt{v(x)_{t_n}}; \label{eq:sd}\\
\mbox{P3:}&\it{}&\quad\langle s \rangle_n =  \frac{1}{w_{\text{avg}}} \sum_{i=n-w_{\text{avg}}+1}^{n} s(x)_{t_n}; \label{eq:avsd}\\
\mbox{P4:}&\it{}& \quad a_n(1)  = \dfrac{\displaystyle\sum_{i=n-w+1}^{n-1}\left(x_i-\overline{x}_{t_n}\right)\left(x_{i+1}-\overline{x}_{t_n}\right)}{\displaystyle\sum_{i=n-w+1}^{n}\left(x_i-\overline{x}_{t_n}\right)^2}.\label{eq:corr1}
\end{eqnarray} 
\begin{figure}[htp!]
\begin{center}
\includegraphics[width=\linewidth]{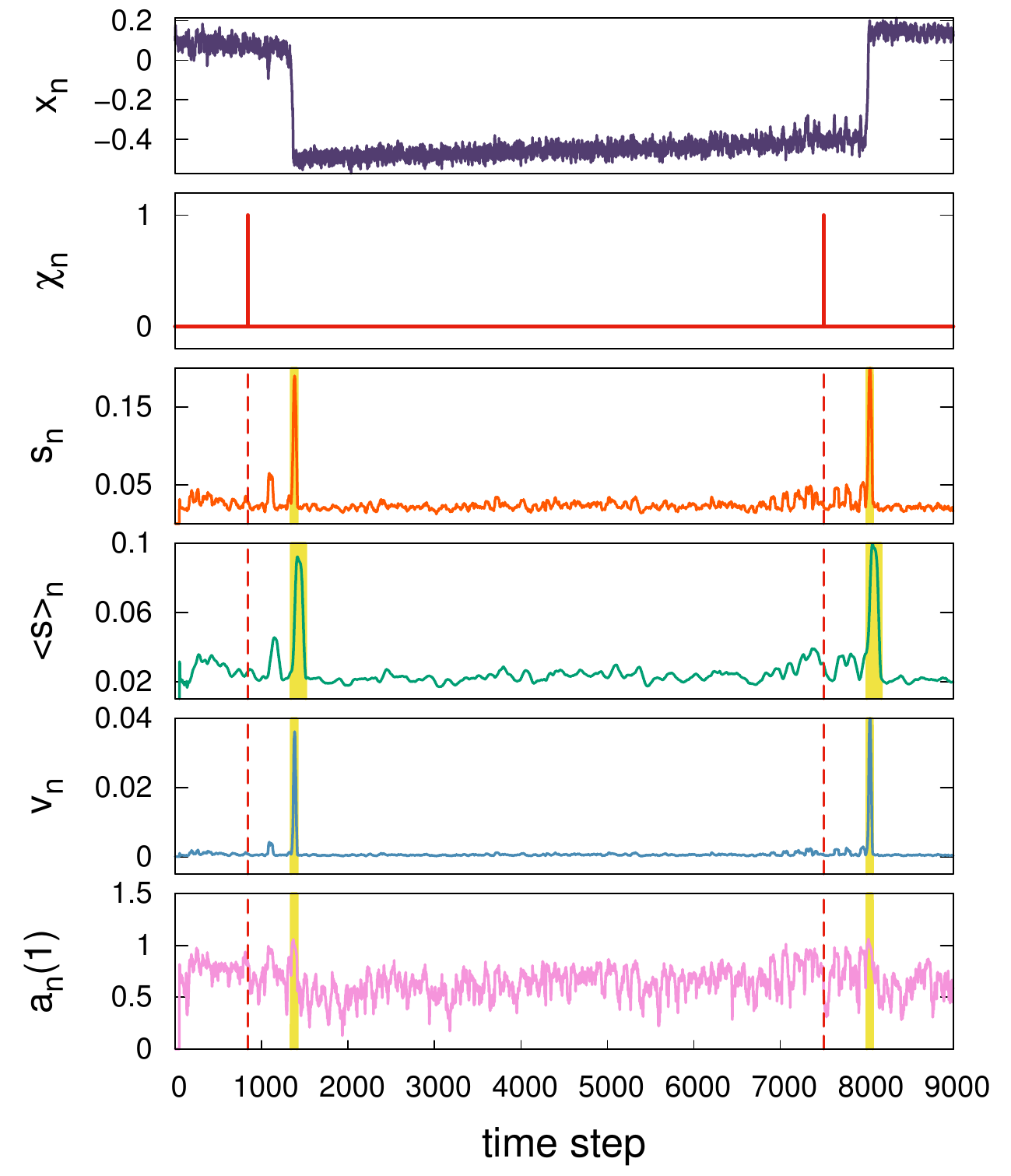}  
\end{center}
\vspace{-0.6cm}
\caption{Time series of tested indicator variables. 
From top to bottom, shown in the panels are the time series of $x$ (vdP model with $\delta=0.5$), of tracer variable $\chi$, and of the four proposed indicators. 
The standard deviations and the autocorrelations are calculated in a sliding window with length $w=50$ time steps. 
The window width for averaging standard deviations $w_{\text{avg}}=100$ time steps. 
The tracer time series in the second panel is shown with a forecast lead time $k=500$ time steps. 
The time when predictions are made is marked with a dashed line in the four panels of indicators.
The indicators within the intervals marked yellow (light gray) are excluded from the statistics for prediction.}
\label{fig:prec_comp}
\end{figure} 
The indicators estimated from a sliding window establish an \textit{indicator time series} (Fig.~\ref{fig:prec_comp}), which provides information for prediction at each time step.
The indicator values estimated from the time series including the critical transitions (shaded yellow in Fig.~\ref{fig:prec_comp}) can be exceptional, especially for the standard deviations. These values have nothing to do with alarming the critical transitions, therefore they are excluded from further analysis.
%
%
\subsection{Predictions}
\label{sec:prediction}
In order to distinguish between relevant and non-relevant values of indicator variables, we use a naive Bayesian classifier~\cite{Rish2001} {i.e.}, a conditional probability distribution function (CPDF),
\begin{equation}
\mathbb{P}(\chi_n(\delta)=1|\psi_n),
\end{equation}
where $\chi_n(\delta)$ is the tracer variable as defined in Eqs.~\eqref{eq:tracer_QIF} and \eqref{eq:tracer_vdP} and $\psi_n$ denotes any indicator variable, such as, e.g., the ones defined in Eqs.~\eqref{eq:var}-\eqref{eq:corr1}.
\begin{figure}[!t]
\vspace{-0.2cm}
\centerline{
\includegraphics[width=0.8\linewidth]{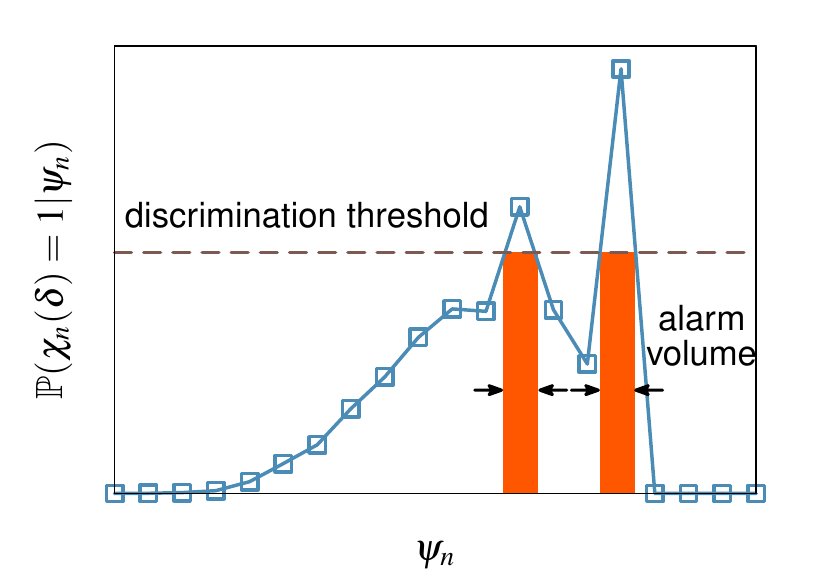} 
}
\vspace{-0.5cm}
\caption{An example for decision making via thresholds of CPDFs, as specified in Eq.~(\ref{eq:decision variable_roc}).
By gradually lowering the discrimination threshold and correspondingly extending the alarm volume, we can generate an ROC curve from a CPDF. The solid line with hollow squares represents a numerically determined CPDF.
The discrimination threshold choosing high-performance indicators is shown as a dashed line and the corresponding (disconnected) alarm volumes (filled rectangles) are indicated by black arrows.}
\label{fig:pdf_discr}
\end{figure}
The CPDFs are obtained numerically from time series of indicator variables $\psi_n$ by estimating the marginal $\mathbb{P}(\psi_n)$
and the joint probability distribution function $\mathbb{P}(\chi_n(\delta)=1,\psi_n)$ and then using Bayes' theorem \cite{bayes}.
%
%
More precisely, the number of bins is fixed to $20$ for both marginal and joint PDFs, which prevents over-fitting without discarding too much information.
The most meaningful value of a indicator variable, the value most likely to be followed by an event, is the one that maximizes the CPDF.
Relevant indicators should lead to non-flat CPDFs with one or several maxima.
Apart from being sensitive to the occurrence of the event, a meaningful indicator should also be specific, i.e., not occur by chance without being related to an event.
One indication of specificity is that the maximum of the CPDF does not coincide with a maximum of the marginal PDF. 
An alarm $A_n$ for CTs is raised at time $t_n$ if the value of the CPDF is above a discrimination threshold $d$, {i.e.}, let $V(d)=\{\psi_n:\mathbb{P}(\chi_n(\delta)=1|\psi_n)\geq d\}$ and consider
\begin{equation}
  A_n[\psi_n,V(d)]= \left\lbrace
\begin{array}{ccc}
1,&  & \text{if}\hspace{0.1cm}\psi_n\in V(d); \\
0,&  & \text{otherwise.}
\end{array}\right. 
\label{eq:decision variable_roc}
\end{equation}
Gradually lowering the discrimination threshold $d$ and determining the corresponding alarm volume $V$ is illustrated in Fig.~\ref{fig:pdf_discr}.
\begin{figure}[!t]
  \vspace{-0.3cm}
\centerline{
\includegraphics[width=0.6\linewidth]{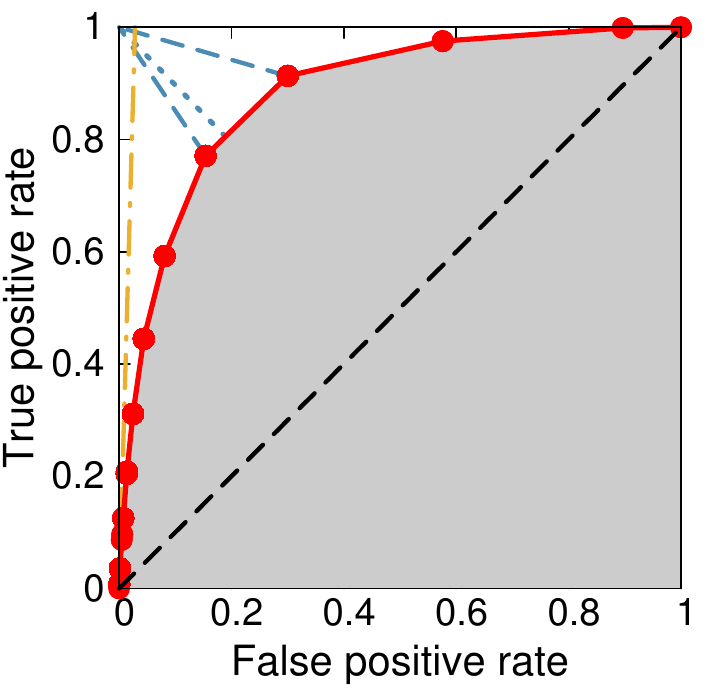} 
}
\vspace{-0.4cm}
\caption{\label{fig:roc_example} An example for an ROC curve (red dots connected by solid lines) and its summary indices.
The AUC (area under the curve, shaded gray), and the KSD (Kolmogorov Smirnov distance, blue dotted line representing the distance between the upper left corner $(0,1)$ and the ROC-curve) summarize the overall behavior of the ROC-curves.
The two blue (gray) dashed lines connect the closest and the second closest points on the curve to the $(0,1)$ point, which are needed in the calculation of KSD.
The corresponding likelihood ratio $\Lambda(\psi_n^{\ast})$, {i.e.}, the slope of the ROC curve in the vicinity of $(0,0)$ is shown as a yellow dot-dashed line.
The black dashed diagonal represents an ROC-curve generated by random predictions.}             
\end{figure}

\subsection{Quantifying Prediction Success}
\label{sec:rocs}
A common method to evaluate the performance of a binary classifier is the \textit{receiver operating characteristic curve} (ROC curve) \cite{egan1975signal,fawcett2006introduction}.
The idea of the ROC curve is to plot the rate of true positives $r_c$ (or \textit{hit rate}) versus the rate of false positives $r_f$ (or \textit{false alarm rate}) as the discrimination threshold of the classifier is varied.
Each value of the threshold $d$ corresponds to a single point in the ROC curve $(r_f, r_c)$ (see Fig.~\ref{fig:roc_example}).
The ideal classifier is expected to predict all events and issue no false alarms.
Consequently, it should generate ROC coordinates close to the upper left corner, {i.e.}, the point $(0,1)$.
The point $(0,0)$ at the lower left corner of an ROC graph represents a value of $d$ that is higher than any entry of the CPDF. 
A classifier whose predictive performance is better than random guesses corresponds to a point in the upper triangle with $r_c>r_f$. 
The shape of the ROC curve characterizes the predictability of the positives and the quality of the applied classification strategy. One can show \cite{metz1978basic} that the slope of a {\sl proper} \cite{egan1975signal} ROC curve must decrease as one moves up and to the right on the curve if decisions are made using a naive Bayesian classifier.

To compare several ROC curves, scalar \textit{summary indices} of ROC curves have been introduced.
Common summary indices, such as the minimal distance to $(0,1)$ (Kolmogorov 
Smirnov distance, KSD), the area under curve (AUC), and the slope of the ROC at $(0,0)$ (\textit{likelihood ratio}) are illustrated in Fig.~\ref{fig:roc_example}.
In this contribution, the KSD is calculated as the distance between $(0,1)$ and the line connecting the two closest points to $(0,1)$ [see the dashed blue (gray) lines in Fig.~\ref{fig:roc_example}].
%
%

\section{Comparing indicators and the ways they are estimated}
\label{sec:predpower}
Although theoretically postulated from the phenomenon of CSD \cite{carpenter2006rising,kuehn2011mathematical}, it is not \textit{apriori} known how useful and specific indicators are connected to an increase of fluctuations in practical prediction scenarios.
%
%
Additionally, one can also ask which indicator is most suitable for practical applications and how its predictive power depends on the way it is calculated and used and also on the events it aims to predict.
Another particular aspect to test is the forecast horizon, i.e., how large the lead time $k$ can be chosen without lowering the overall quality of predictions.
%
Last but not least, we also test how the quality of indicators depend on the width $w$ of the time window that was used to estimate them from a given time series.
In a realistic forecast scenario, i.e., when predicting only {\sl one} CT from a relatively short data record, the width $w$ might be limited by the total amount of available data.
Working with models that create repetitive CTs, and arbitrarily long time series, we can explore all useful variations of $w$.

We therefore generate time series of $2^{22}$ time steps from the QIF model and the vdP model, introduced in 
Sec.~\ref{sec:models} and then estimate the respective indicator time series $\{\psi_n\}$.
In more detail, we choose the additional parameter of the vdP model in Eq.~\eqref{eq:vdPmodel2} 
to be $\delta=0.5$.
The noise strength is chosen to be $\sigma_1=0.2$ for the QIF model and 
$\sigma_1=0.1$ for the vdP model. Choosing a smaller noise strength for the vdP model prevents 
the system from jumping back and forth frequently (flickering).
For both models, the time separation parameter $\epsilon$ is set to be $0.02$, which means that we are roughly in the intermediate regime $\sigma=\mathcal{O}(\sqrt{\epsilon})$. 
\begin{figure}[t!]
 \centerline{
   \includegraphics[width=0.5\columnwidth]{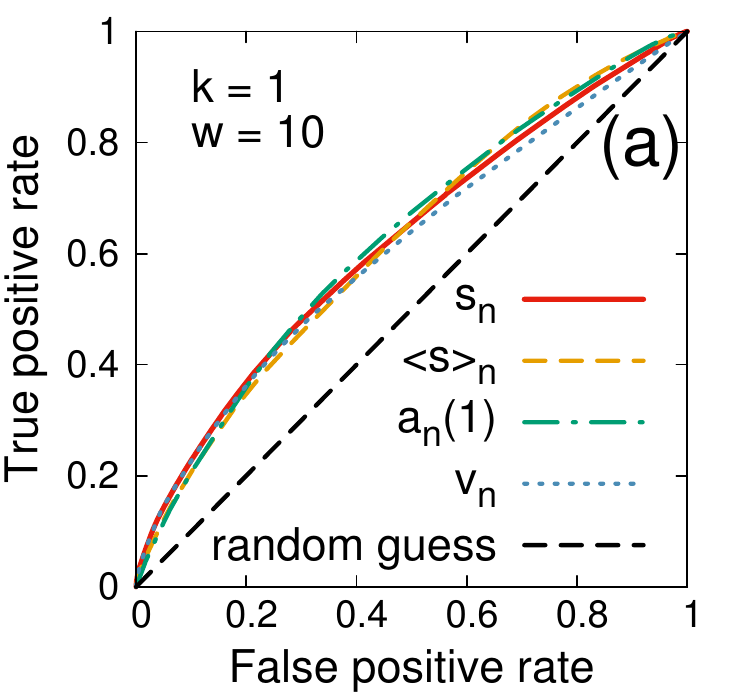}
   \includegraphics[width=0.5\columnwidth]{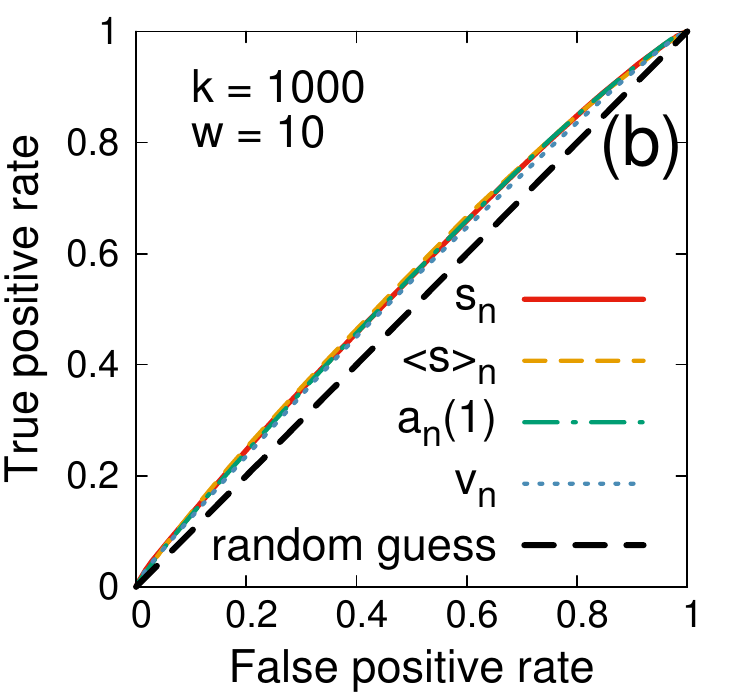}
 }
 \vspace{0.2cm}
 \centerline{
   \includegraphics[width=0.5\columnwidth]{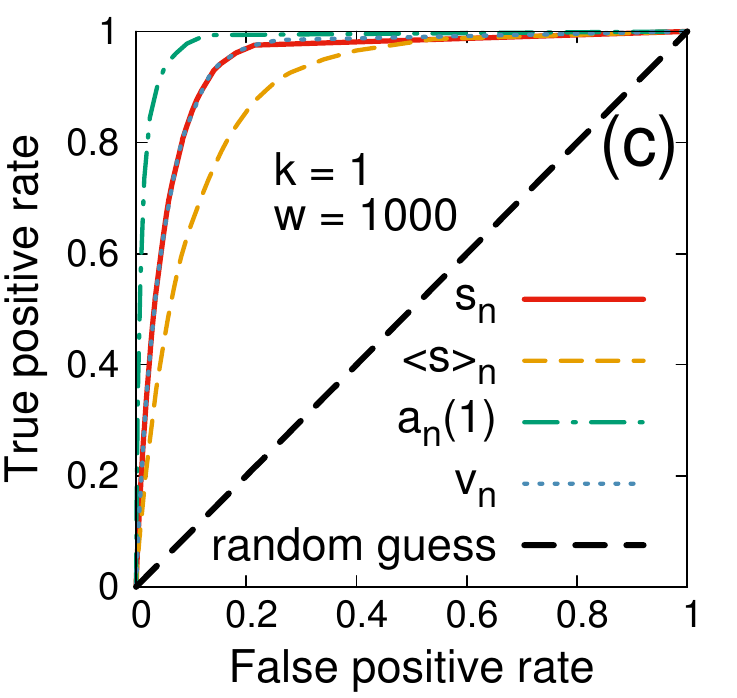}
   \includegraphics[width=0.5\columnwidth]{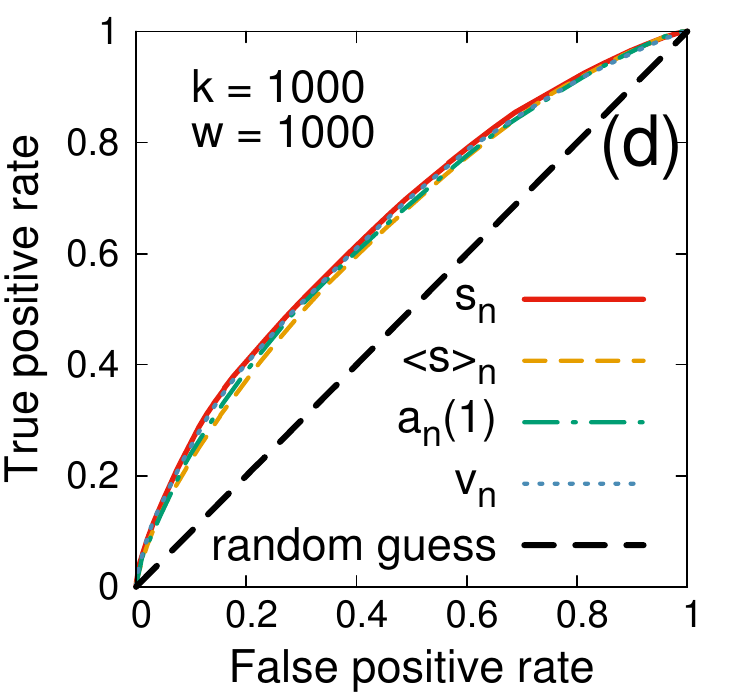}
 }
 \vspace{-0.2cm}
\caption{ROC curves generated by the indicators introduced in Sec.~\ref{sec:precursors} used to predict CTs in the QIF model. 
In general, we observe an increase in prediction quality with decreasing lead time $k$ and increasing window width $w$.
%
}
\label{fig:roc_0502_kwt}
\end{figure}
To investigate the dependence of the predictability on the window width of the indicators, we estimate multiple indicator time series $\{\psi_n\}$ for different window widths ranging from $10$ to 
$1000$.
For the sliding average of $s_n$, the window width for averaging $w_{\text{avg}}$ is set to be 100 time steps.
In order to examine the influence of the lead time on the predictability, we generate ROC curves for different $k$ ranging from $1$ to $1000$ time steps.
When exploring the $k$-$w$ space, the sum of the maximal values of $k$ and $w$ is restricted to the minimal interval between CTs.  
\begin{figure}[t!]
 \centerline{
 \includegraphics[width=0.5\columnwidth]{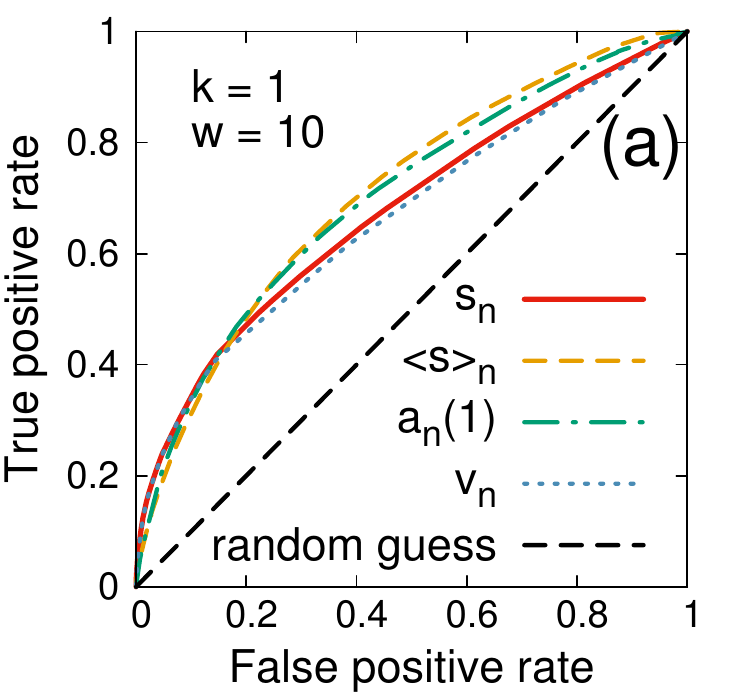}
 \includegraphics[width=0.5\columnwidth]{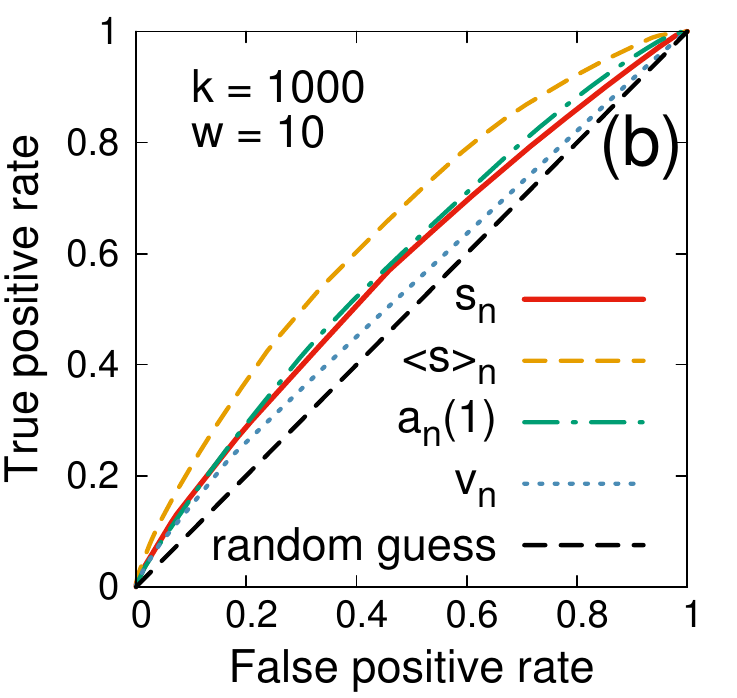}
 }
 \vspace{0.2cm}
 \centerline{
 \includegraphics[width=0.5\columnwidth]{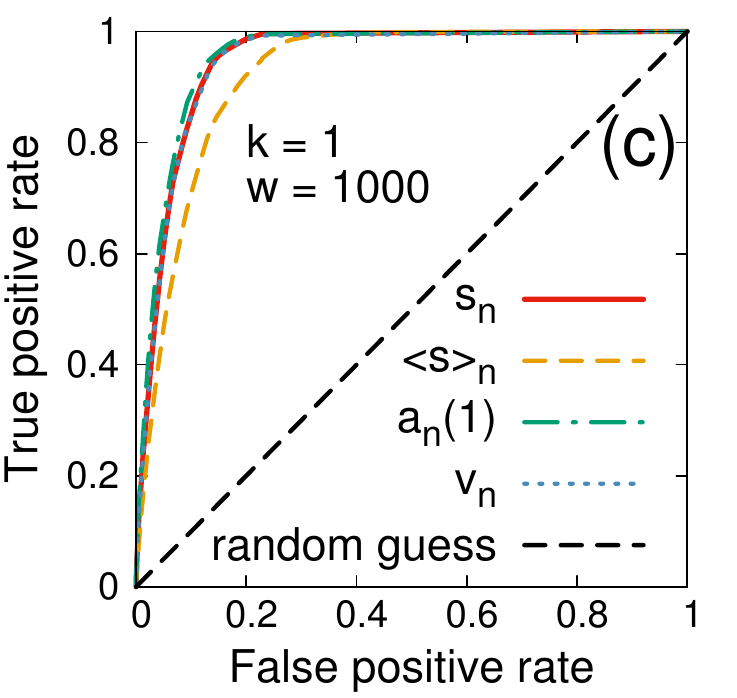}
 \includegraphics[width=0.5\columnwidth]{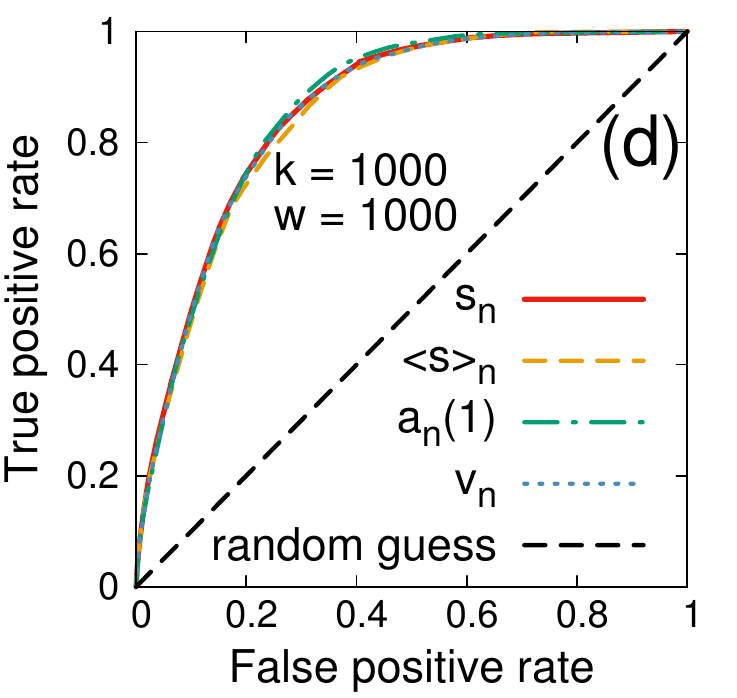}
 }
 \vspace{-0.2cm}
\caption{ROC curves generated by the indicators introduced in Sec.~\ref{sec:precursors} used to predict CTs in the VdP model. In general we observe an increase in prediction quality with decreasing lead time $k$ and increasing window width $w$. 
}
\label{fig:roc_0601_kwt}
\end{figure}
The ROC curves in Figs.~\ref{fig:roc_0502_kwt} and \ref{fig:roc_0601_kwt} illustrate the dependence of the prediction success on both variables, the forecast lead time $k$ and the window width $w$ used to estimate indicators.
We observe that the prediction performance of all tested indicators and in both models decreases with increased forecast horizon, i.e., increased lead time.
As observed for other prediction tasks, the smaller the difference between prediction and occurrence of the event, the greater the prediction success.
The closer the system is to CTs, the stronger the early warning signals and the stronger is apparently the correlation of indicator and event.
Indeed, a prediction extremely close to a CT (small $k$) may be very good; however, larger lead times would be highly desirable for applications.

Additionally, we observe a dependence of the prediction quality on the way that indicator variables are estimated.
Using very small window widths leads to predictions that are not much better than random.
Increasing the window width $w$, and thus smoothing the indicator time series, all indicators show an increasingly strong ability to predict CTs.
This increase is particularly dramatic for small lead times.
In other words, close to the transitions, meaningful indicatory signals are more likely to be hidden within stochastic fluctuations.
Consequently, increasing the window width can be an effective way to attenuate the influence of the fluctuations and to improve predictions.
\begin{figure}[t!]
\centerline{
\includegraphics[width=0.5\columnwidth]{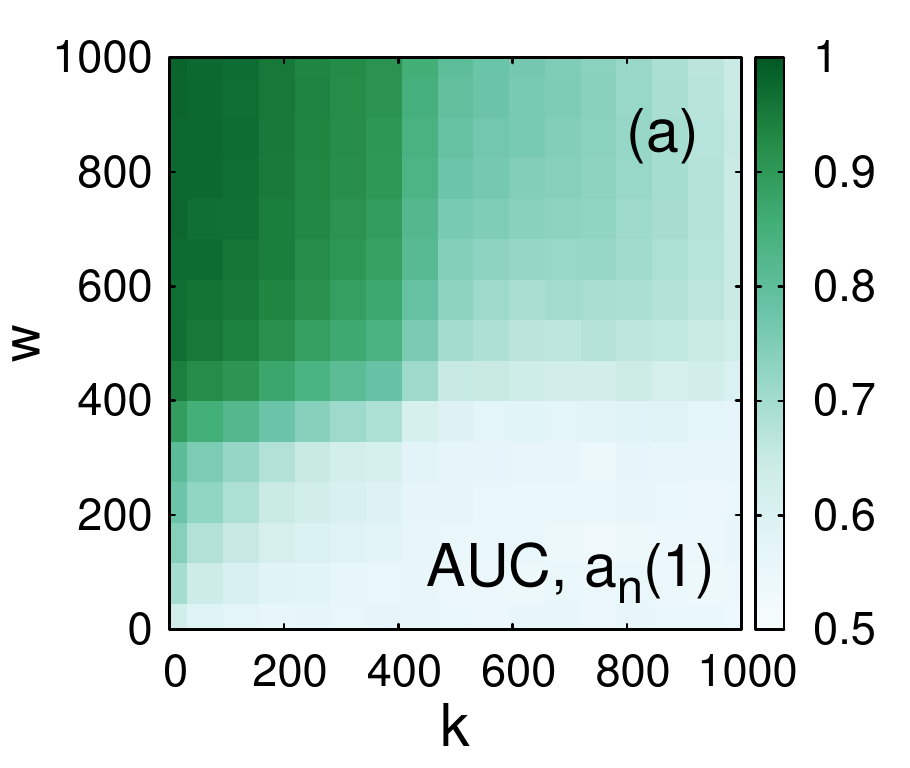}
\includegraphics[width=0.5\columnwidth]{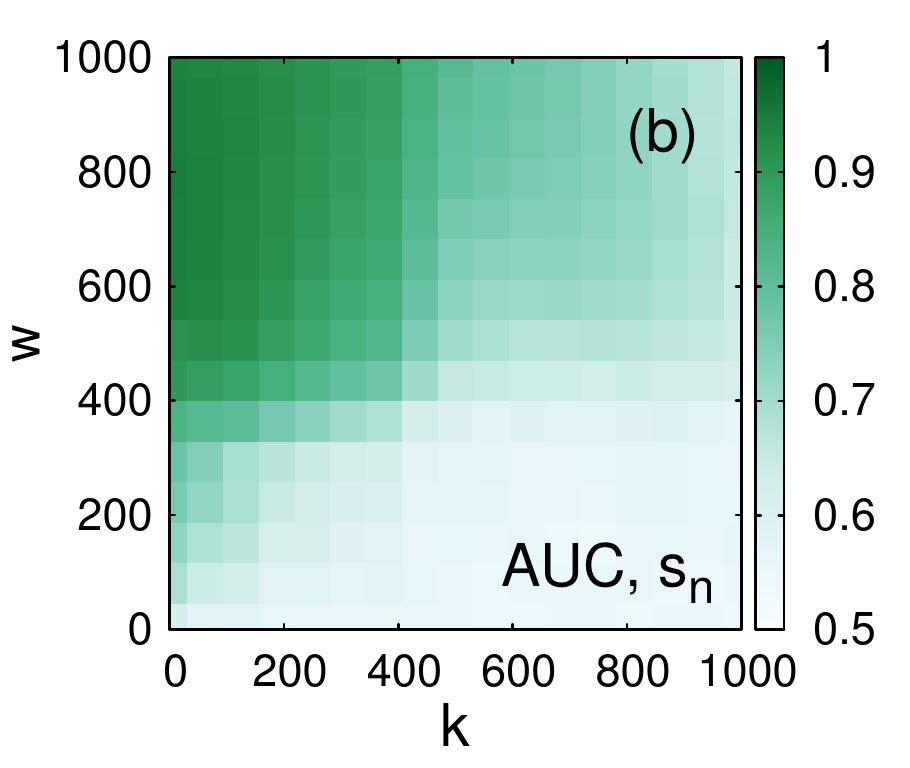}
}
\centerline{
\includegraphics[width=0.5\columnwidth]{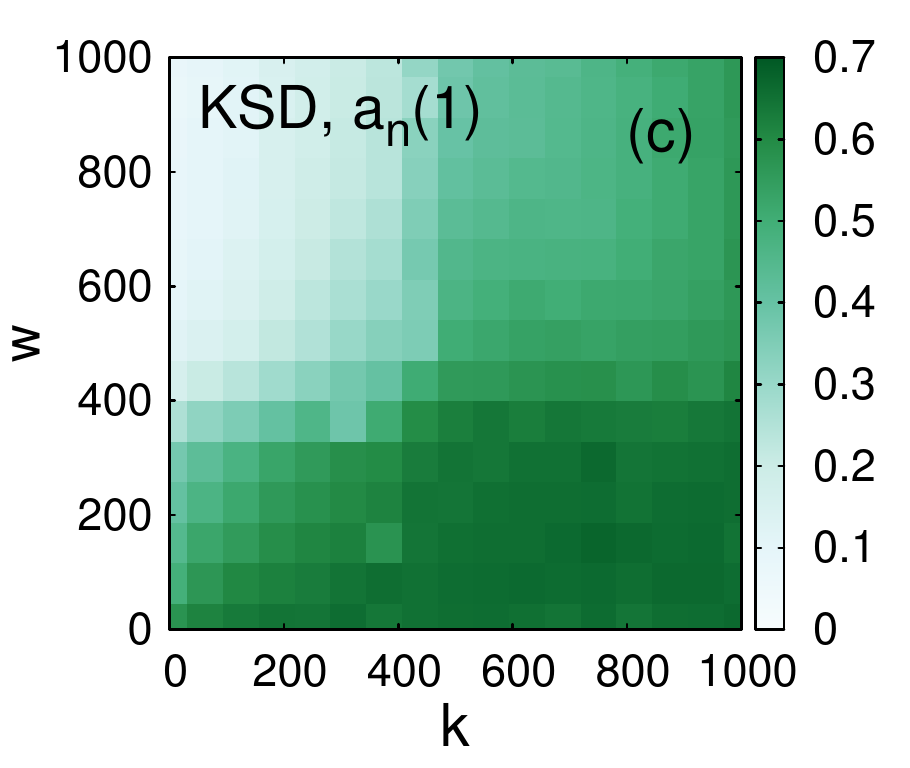}
\includegraphics[width=0.5\columnwidth]{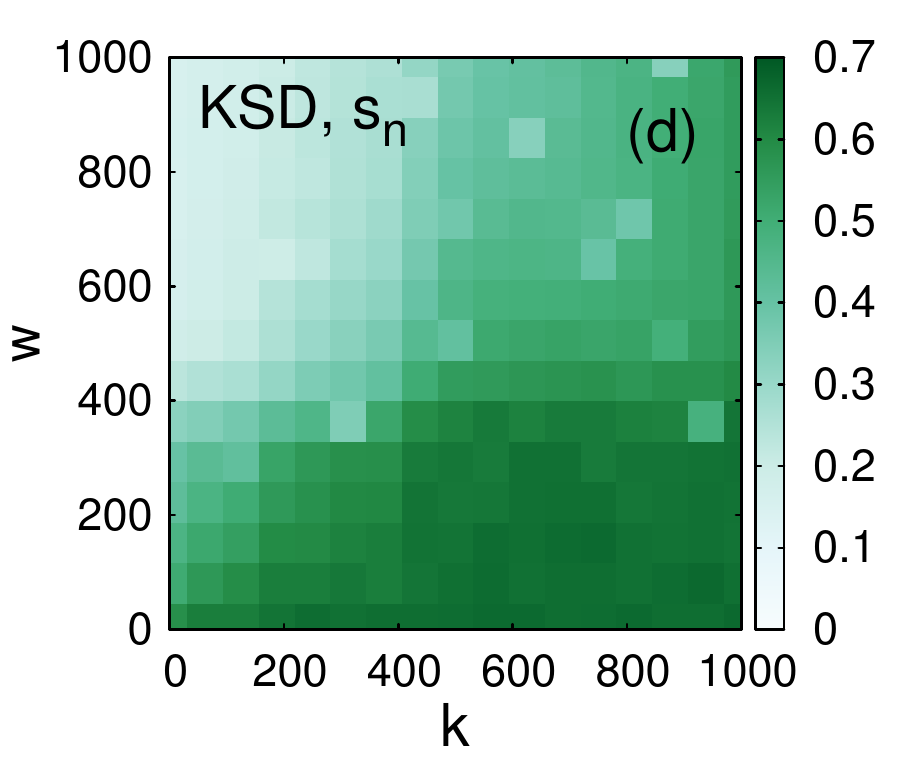}
}
\vspace{-0.4cm}
\caption{Heat maps of the ROC indices of the indicator $a_n(1)$ and $s_n$ in the $k-w$ space for the QIF model. AUCs (Area under curves) are shown in the upper row and KSDs (Kolmogorov-Smirnov distances) in the lower row.}
\label{fig:index_0502_kw}
\end{figure}
\begin{figure}[t!]
\centerline{
\includegraphics[width=0.5\columnwidth]{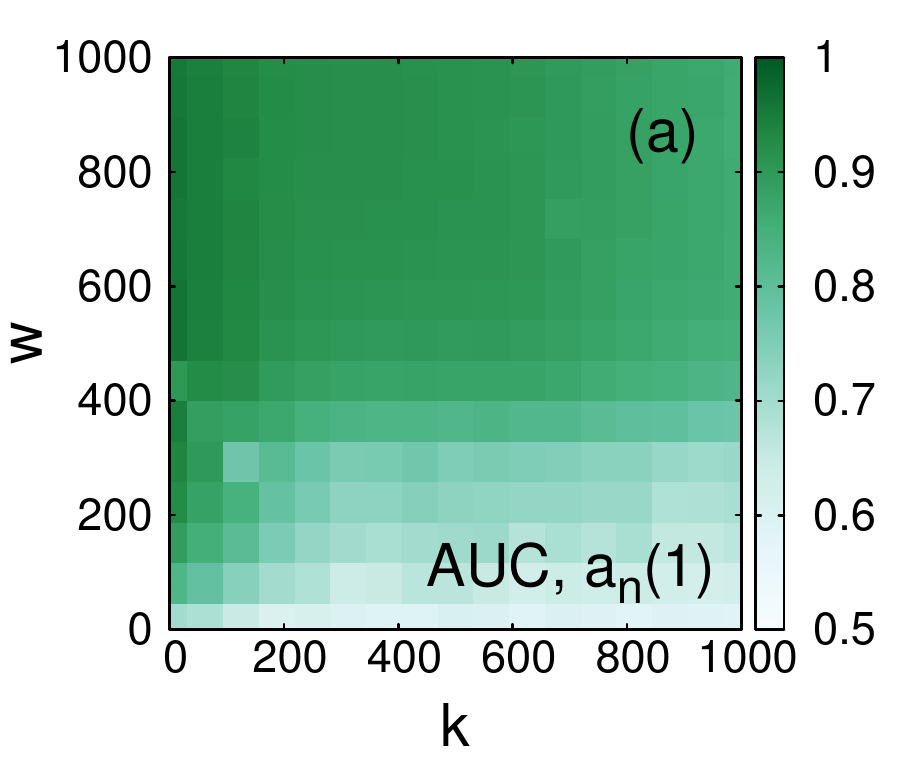}
\includegraphics[width=0.5\columnwidth]{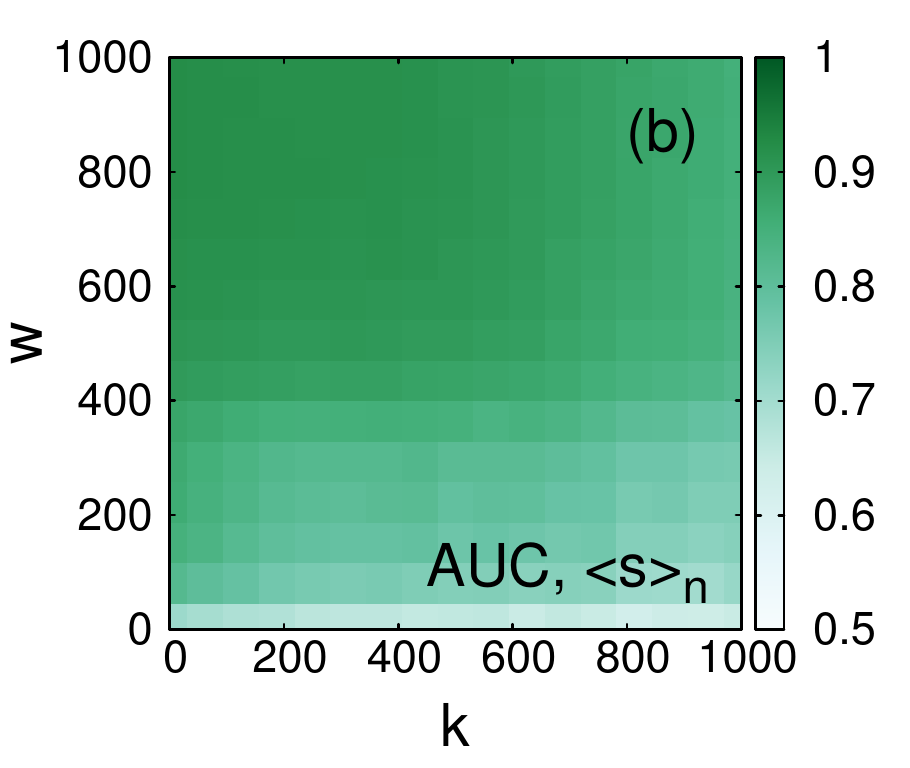}
}
\centerline{
\includegraphics[width=0.5\columnwidth]{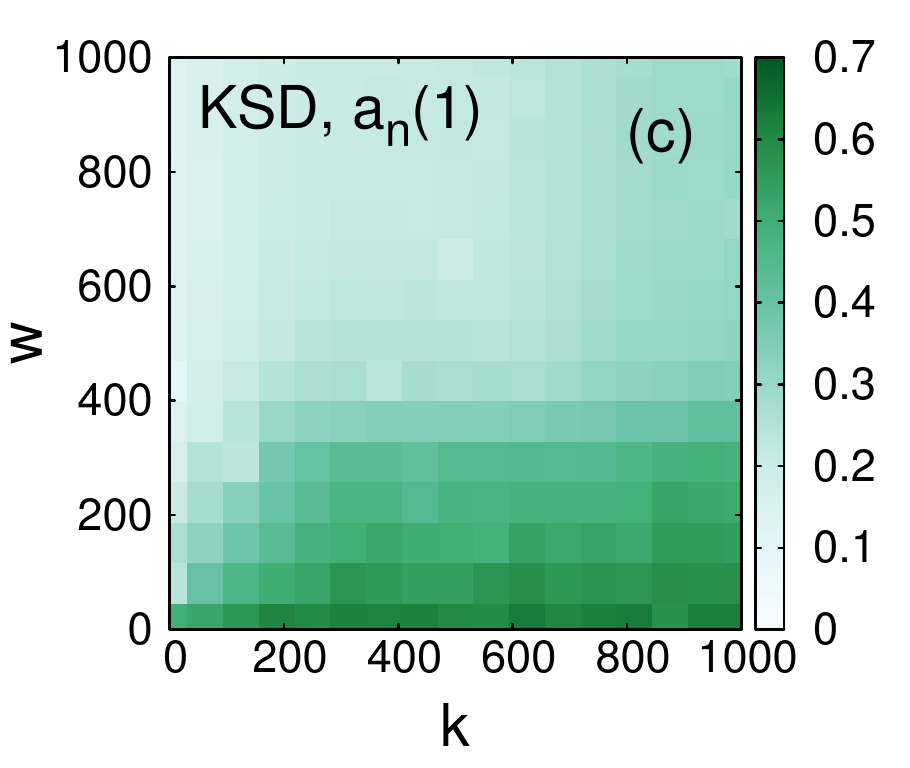}
\includegraphics[width=0.5\columnwidth]{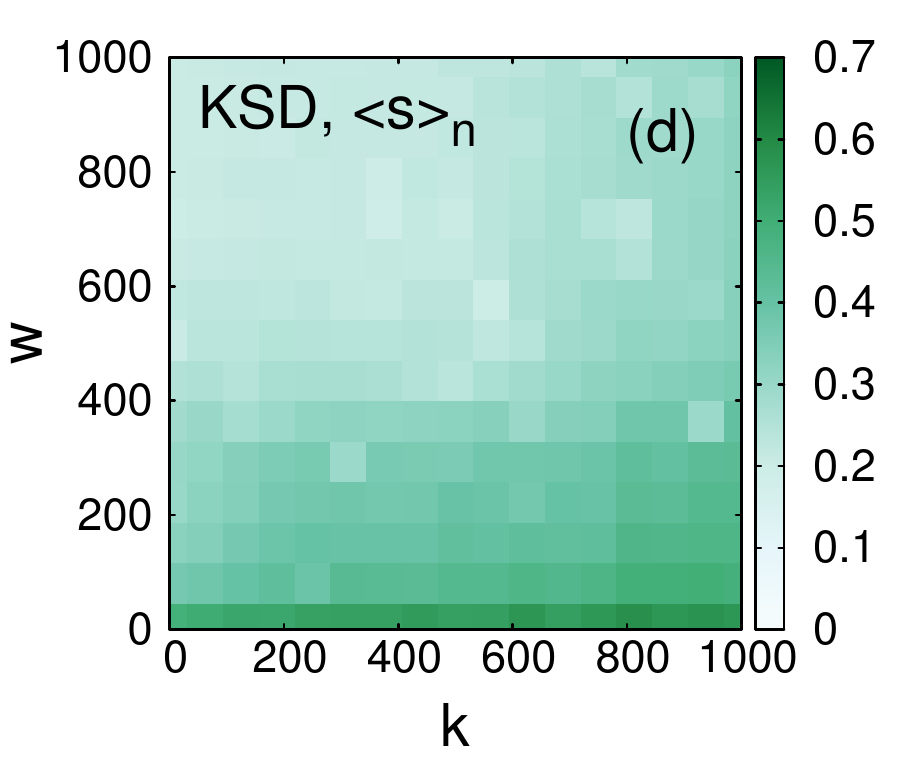}
}
\vspace{-0.4cm}
\caption{Heat maps of the ROC indices of the indicator $a_n(1)$ and $\langle s \rangle_n$ in the $k-w$ space for the vdP model. AUCs are shown in the upper row and KSDs in the lower row.}
\label{fig:index_0601_kw}
\end{figure}

The dependence of the forecast performance on the leading time $k$ and the window width $w$ is clear in the heat maps of the ROC indices in the $k-w$ parameter space (Figs.~\ref{fig:index_0502_kw} and \ref{fig:index_0601_kw}).
For the QIF model (Fig.~\ref{fig:index_0502_kw}), the lag-1 autocorrelation $a_n(1)$ (left column) is the best indicator among the four proposed indicators: at the upper-left corner of the heat maps, where we make predictions close to the CTs but use a large window size, the AUC of $a_n(1)$ is the largest and the KSD is the smallest. 
The standard deviation $s_n$ performs slightly better than its sliding average $\langle s \rangle_n$ and the variance $v_n$. However, the performances of the three are qualitatively similar.

For the vdP model, good predictions can be made even in a early stage, i.e. far from the CTs, if we use a large enough window to estimate the indicators (Fig.~\ref{fig:index_0601_kw}).
In particular, the sliding average of the standard deviation $\langle s \rangle_n$ (right column of Fig.~\ref{fig:index_0601_kw}) performs significantly better than the other indicators with smaller window widths for estimation (around $200$ time steps): Even when the predictions are made far before the transitions they can be quite successful (AUC around 0.8 for $k=800$).
The performances of $s_n$, $v_n$ are qualitatively similar to $a_n(1)$.
\section{The Dependence on the Transition Magnitude}
\label{sec:mag_setup}
\begin{figure}[t!]
  \centerline{
	\includegraphics[width=0.8\columnwidth]{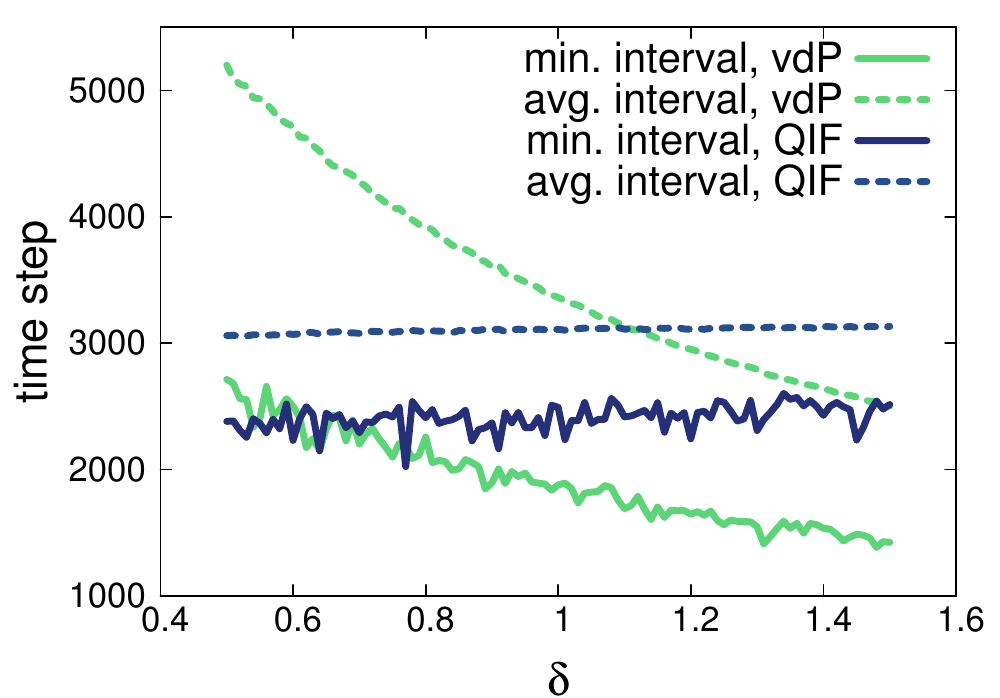}
  }
\vspace{-0.2cm}  
\caption{Magnitude dependence of the time interval between CTs. Dashed lines represent the average interval and solid lines the minimal interval between CTs. Green (gray) lines are for the vdP model and dark blue (black) ones for the QIF model.}
\label{fig:interval_mag}
\end{figure}
The magnitude of CTs generated by our models does depend on the model parameter  $\delta$ (see Fig.~\ref{fig:interval_mag}).
In the QIF model, as introduced in Sec.~\ref{sec:models}, this parameter coincides with the threshold on $x$ for the reset of the system.
In the vdP model, $\delta$ determines the distance between the fold points and the corresponding attracting branches of the critical manifold to which the system jumps from a fold. 
In this section we address the question, whether larger transitions are proceeded by a more distinguished indicatory behavior and are therefore easier to predict.
Previous studies on extreme events in stochastic processes have shown that prediction quality can sensitively depend on event magnitude and the decay of CPDFs in the limit of large events \cite{hallerberg2008influence}.
We now test whether this idea can be linked to dynamical systems for CTs and their classical early-warning signs or not. 
As test data sets for each model, we use time series generated for $100$ different values of $\delta$, {i.e.}, starting from $\delta=0.5$ to $\delta=1.5$ with increments of $0.01$. 
To detect CTs in these time series (see Sec.~\ref{sec:precursors}) the detection methods have to be adapted according to each value of $\delta$. 
For the QIF model, the threshold for identifying CTs is simply set to be $-\delta$. 
For the vdP model, the parameters $\theta_0$ and $\theta_1$ for locating the onset and the end of the transitions were (after empirically testing) set to 
\begin{eqnarray}
 \theta_0 &=& \theta_0^{\ast}-0.6\cdot(\delta-0.5)\\
 \theta_1 &=& \theta_1^{\ast}-0.1\cdot(\delta-0.5),
\label{eq:theta_mag}
\end{eqnarray}
with $\theta_0^{\ast}=-0.2$ and $\theta_1^{\ast}=-0.1$ being the thresholds for $\delta=0.5$ [see also Eq.~\eqref{eq:tracer_vdP} and Fig.~\ref{fig:ts_id_vdP}]. 

We focus on the magnitude dependence of the standard deviation and evaluate the quality of predictions using ROC-curves for four representative combinations of small and large lead time and window widths.

In total, we explore the three-dimensional parameter space of $k$, $w$, and $\delta$, which is a key practical contribution and has been missing in previous work on CTs. 
The maximal values of $k$ and $w$ are set to be $500$, so that the sum of $k$ and $w$ does not exceed the minimal interval between transitions. 
We also observed that the minimal interval between transitions decreases by about a half with increasing $\delta$ for the vdP model (Fig.~\ref{fig:interval_mag}). 
This effect can be understood as a consequence of an increase in the velocity of the slow variable $y$ at the left branch $\dot{y}\in\left[\frac{1}{6}\delta,\frac{1}{2}\delta\right]$, which drives the system towards the fold point $(-\frac{2}{3}\delta,1)$ [see also Eqs.~\eqref{eq:vdPmodel1} and \eqref{eq:vdPmodel2}]. 

\subsection{Results for the QIF Model}
\begin{figure}[t!]
	\centerline{
	\includegraphics[width=0.5\columnwidth]{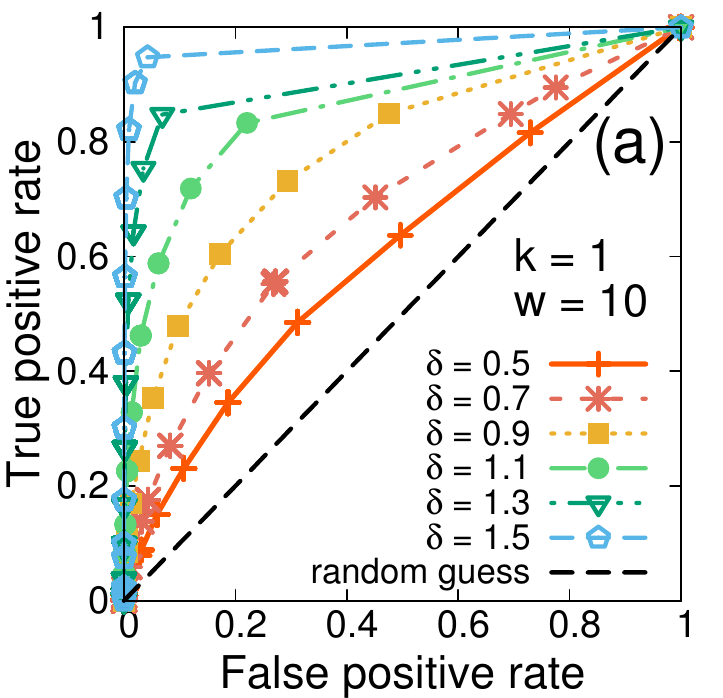}
	\includegraphics[width=0.5\columnwidth]{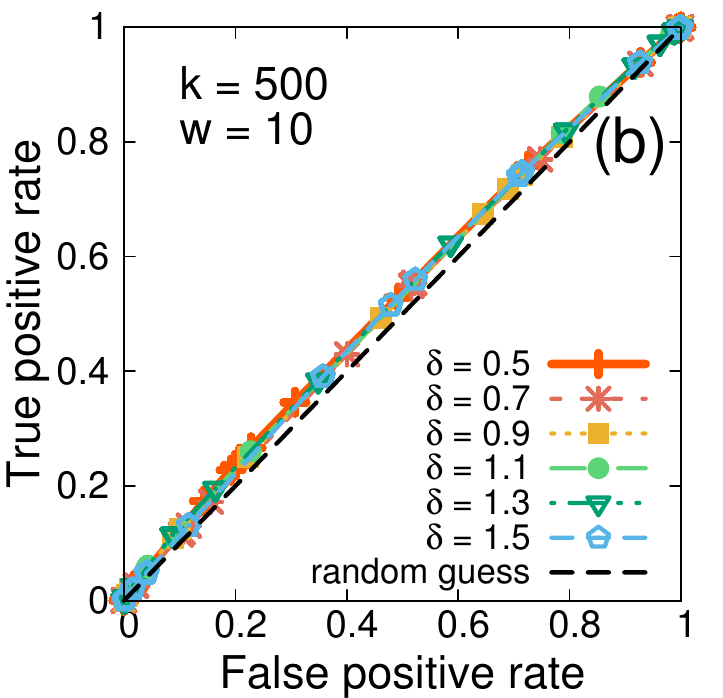}
        }
        \vspace{0.2cm}
        \centerline{
	\includegraphics[width=0.5\columnwidth]{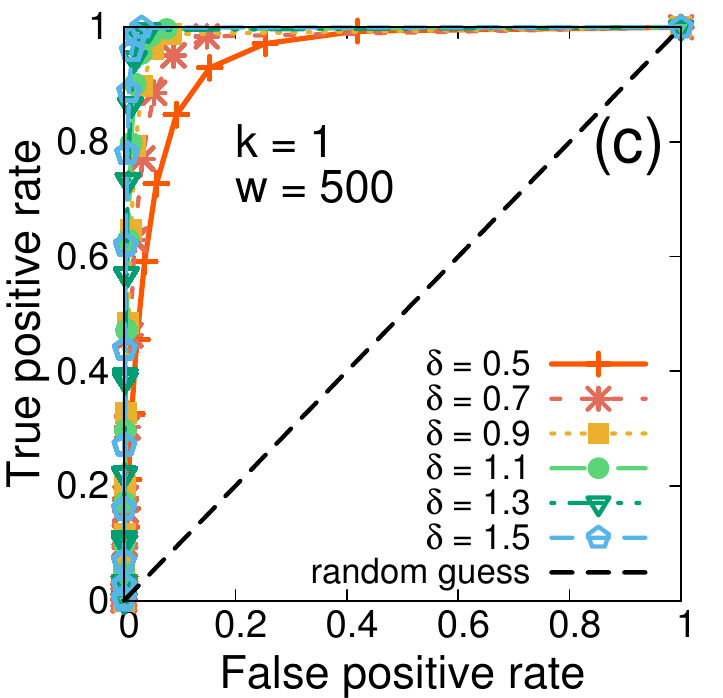}
	     \includegraphics[width=0.5\columnwidth]{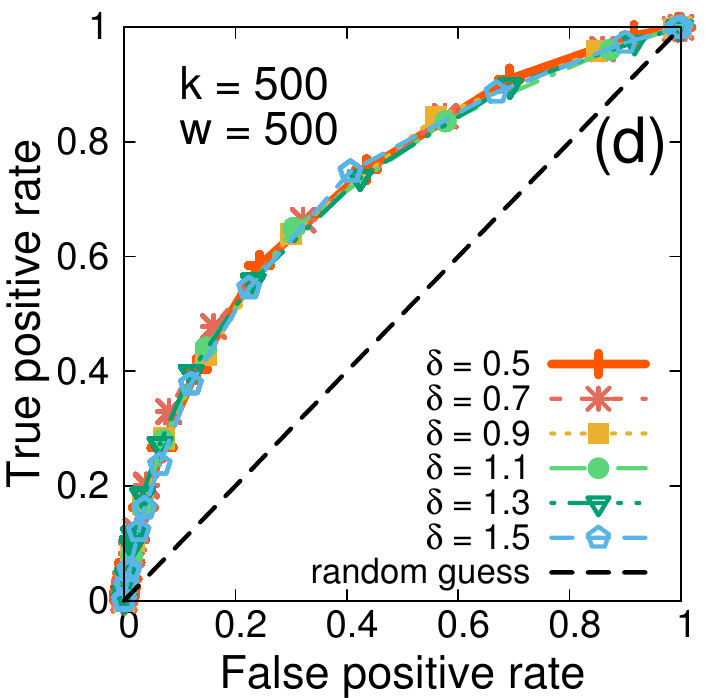}
        }
        \caption{Magnitude dependence of the ROC curves (a)-(d) of $s_n$ in the QIF model for different combinations of lead times $k$ and window widths $w$. 
        For small lead times we observe an increase in prediction quality with an increasing CT magnitude $\delta$.}\label{fig:roc_mag_QIF}
\end{figure}
\begin{figure}[t!]
  \centerline{
    \includegraphics[width=0.5\columnwidth]{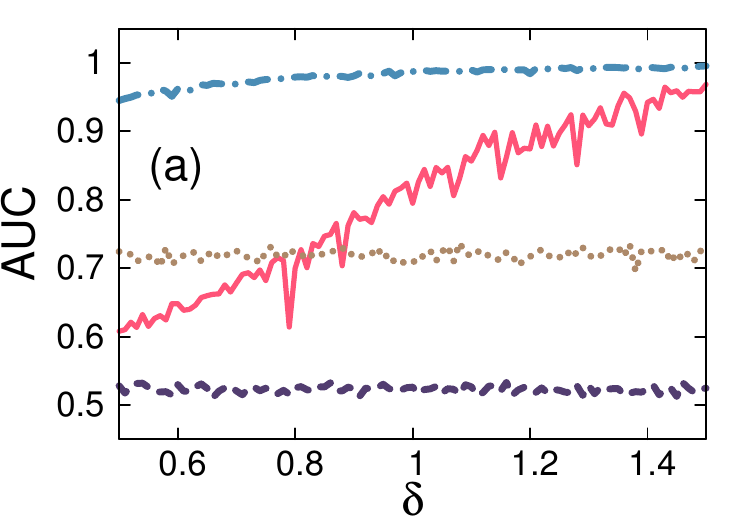}
    \includegraphics[width=0.5\columnwidth]{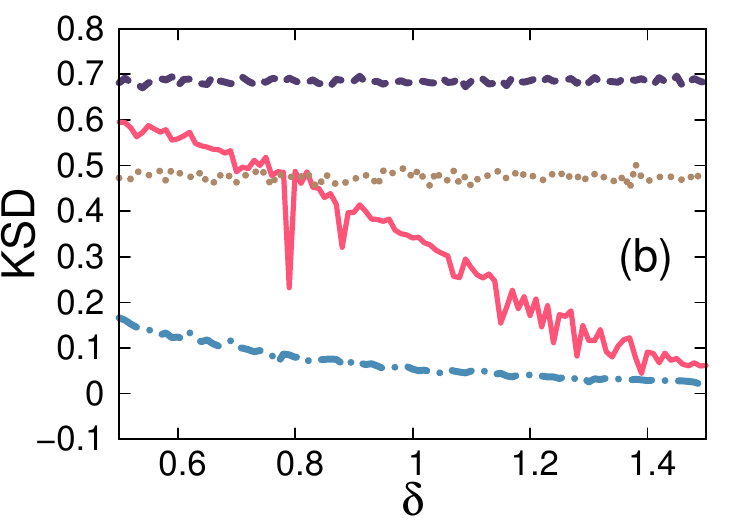}
  }
  \centerline{
      \includegraphics[width=0.5\columnwidth]{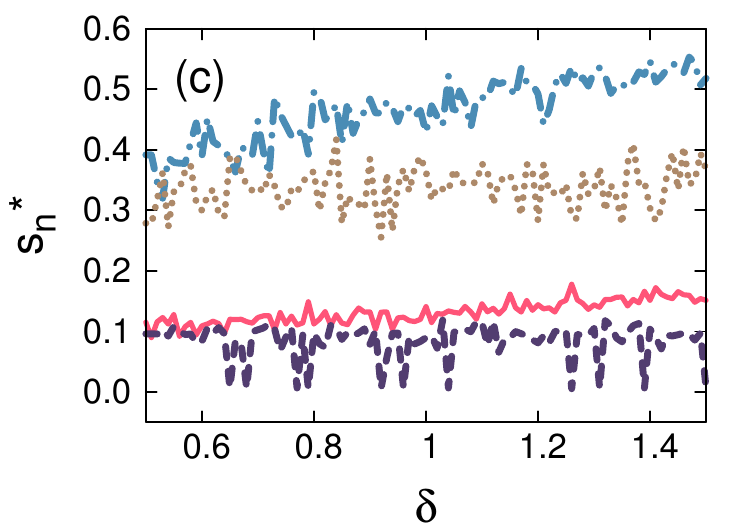}
      \includegraphics[width=0.5\columnwidth]{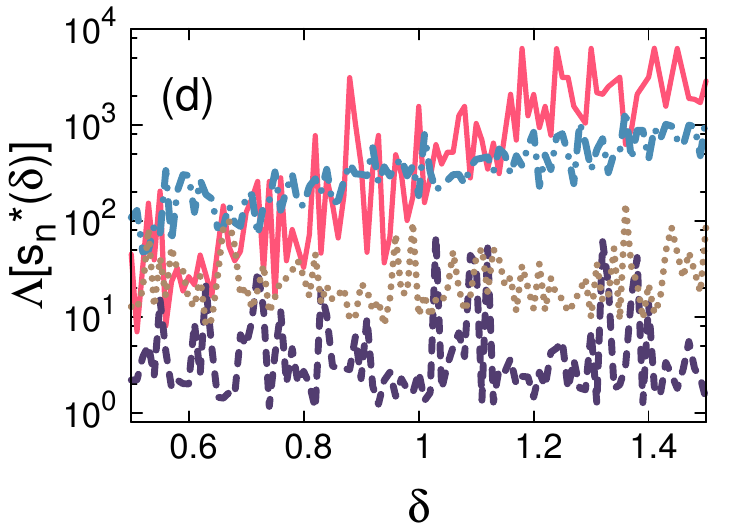}
  }
\caption{Magnitude dependence of the predictability indices of the QIF model. 
ROC-indices, AUC and KSD, are plotted in (a) and (b), and the best indicator $s_n^{*}$ and the likelihood ratio evaluated at the best indicator in (c) and (d). 
The indices for $k=1, w=10$ are shown with solid lines ($\color{magenta}{\boldsymbol{-}\hspace{-0.1cm}\boldsymbol{-}}$), for $k=1,w=500$ with dot-dashed lines ($\color{blue}{\boldsymbol{-}}$$\color{blue}{\boldsymbol{\cdot}}$$\color{blue}{\boldsymbol{-}}$), for $k=500, w=10$ with dashed lines ($\color{purple}{\boldsymbol{-\,-}}$) and for $k=500,w=500$ with dotted lines ($\color{brown}{\boldsymbol{\cdots}}$).
For small lead times we observe an increase in prediction quality with an increasing CT magnitude $\delta$.}
\label{fig:index_mag_QIF}
\end{figure}

Starting from time series of CTs and the indicator variable $s_n$ generated by the QIF model, we predict CTs and evaluate the quality of the predictions using ROC curves in Fig.~\ref{fig:roc_mag_QIF} and the related indices (AUC, KSD, and likelihood ratio) in Fig.~\ref{fig:index_mag_QIF}. 
If the prediction is made close to the transition (small lead times), we observe a clear positive magnitude dependence, {i.e.}, larger events can be better predicted as shown in Figs.~\ref{fig:roc_mag_QIF}(a) and \ref{fig:roc_mag_QIF}(c). 
For larger lead times the prediction quality decreases and the dependence on the magnitude vanishes completely.
The ROC indices in Fig.~\ref{fig:index_mag_QIF} nicely summarize these results by showing an increase in AUC and likelihood ratio and a decrease in KSD with increasing $\delta$ for small lead times. 
For larger lead times the indices do not show any clear trend and they are, up to small fluctuations, constant.
 
One can understand the improvement of the prediction quality close to the transitions ({i.e.}, with small $k$) as a consequence of the accelerating drifting of $x$ towards negative infinity (see Fig.~\ref{fig:phase_0502}) before the CT happens. 
As can be seen from the model's dynamics [see Eq.~\eqref{eq:QIFmodel1}] and Fig.~\ref{fig:phase_0502}, the drifting speed of the fast variable $x$ rises as the system has passed the fold point $(0,0)$. 
The larger the reset-threshold $\delta$ the longer this faster drifting continues, before the system is reset to $(1,1)$. 
As a consequence, the sliding $s_n$ at $k$ time steps before the reset ({i.e.}, the occurrence of a CT) becomes larger and larger, due to the accelerating drifting of $x$. 
In fact, the values of the maximum CPDF indicators $s_n^{*}$ increase with increasing threshold $\delta$ as can be seen in Fig.~\ref{fig:index_mag_QIF}(c).

Since these large indicatory $s_n$ are easier to distinguish from random fluctuations in the $s_n$, also the likelihood ratio estimated for $s_n$s, which are close to the transition (small lead times), increase with $\delta$ and are up to three orders of magnitude larger close to the event [as shown in Fig.~\ref{fig:index_mag_QIF}(d)] than likelihood ratios for large lead times.
In contrast to this, $s_n$s further away from the CT are not affected by the increase of the reset-threshold $\delta$ but follow the ($\delta$-independent) dynamics, given by 
Eqs.~\eqref{eq:QIFmodel1} and \eqref{eq:QIFmodel2}. 
Consequently, we observe a sensitive dependence on the event magnitude only for small lead times. 

\subsection{Results for the vdP Model}
\begin{figure}[t!]
\centerline{
  \includegraphics[width=0.5\columnwidth]{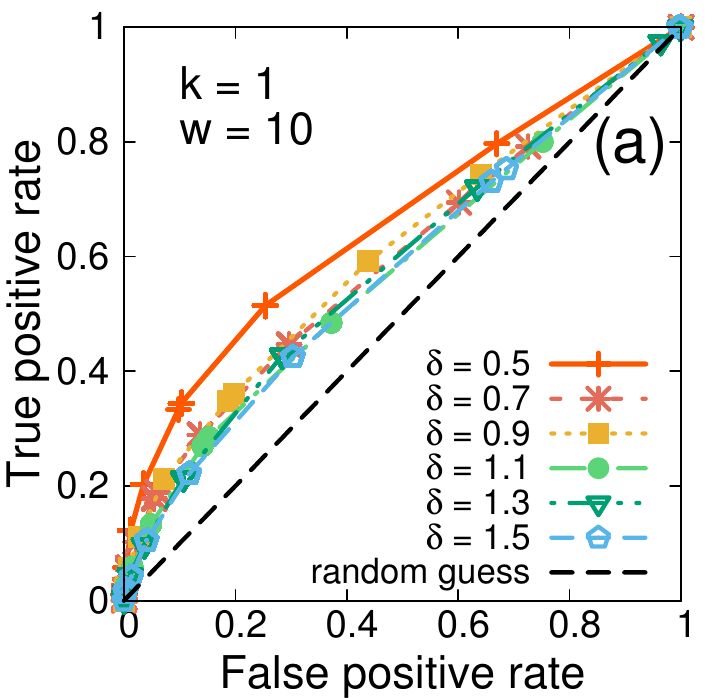}
  \includegraphics[width=0.5\columnwidth]{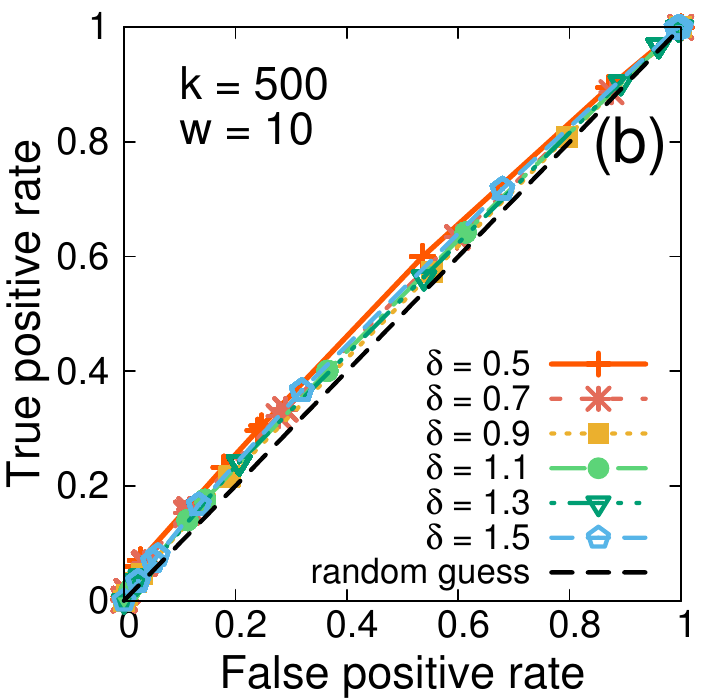}
}
\vspace{0.2cm}
  \centerline{
    \includegraphics[width=0.5\columnwidth]{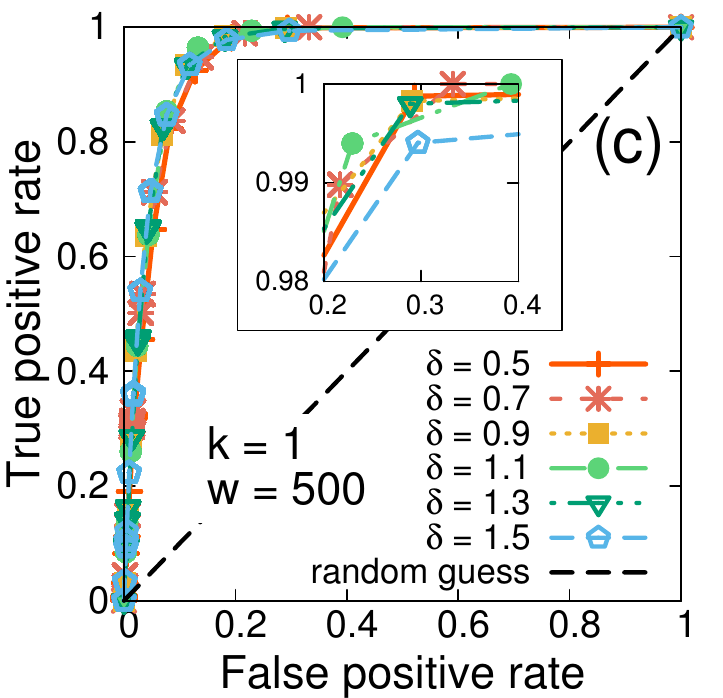}
    \includegraphics[width=0.5\columnwidth]{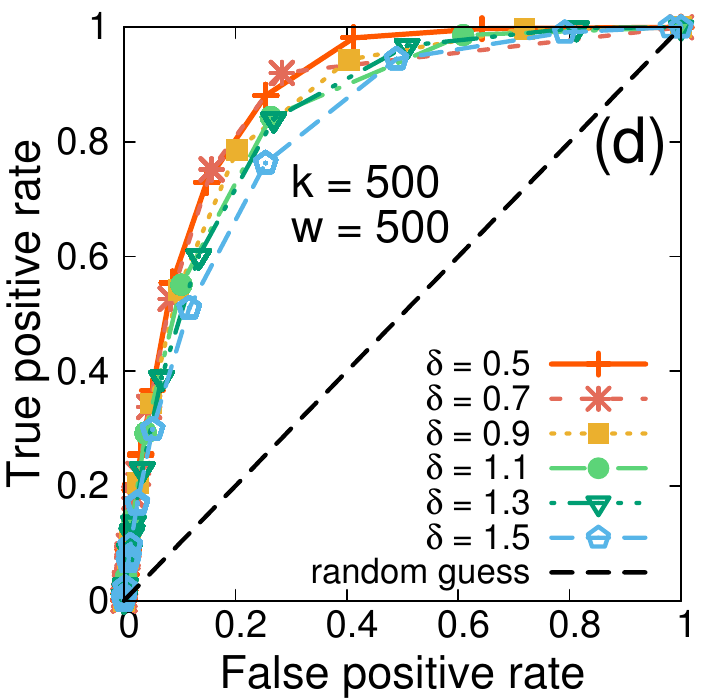}
  }
\vspace{-0.2cm}
\caption{Magnitude dependence of the ROC curves (a)-(d) of $s_n$ in the vdP model, for different combinations of lead times $k$ and window widths $w$. 
For large lead times and large window widths (d) and small lead times and small window widths (a) we observe a mild decrease in prediction quality with an increasing CT magnitude $\delta$.}
\label{fig:roc_mag_vdP}
\end{figure}
\begin{figure}[t!]
\centerline{
 \includegraphics[width=0.5\columnwidth]{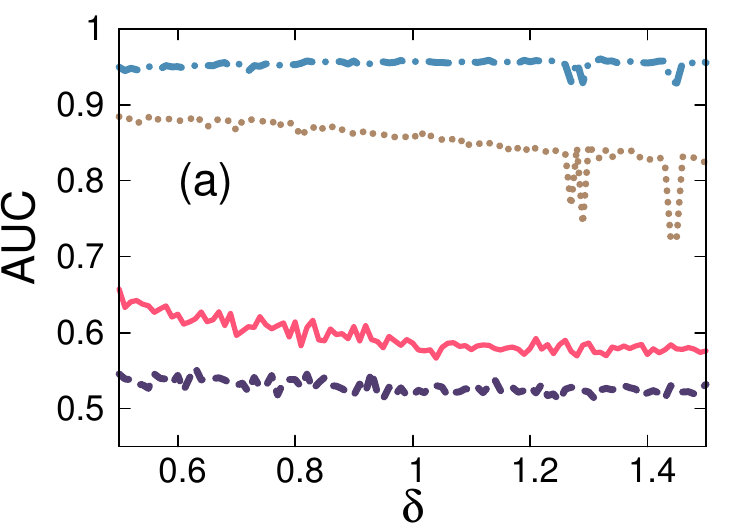}
 \includegraphics[width=0.5\columnwidth]{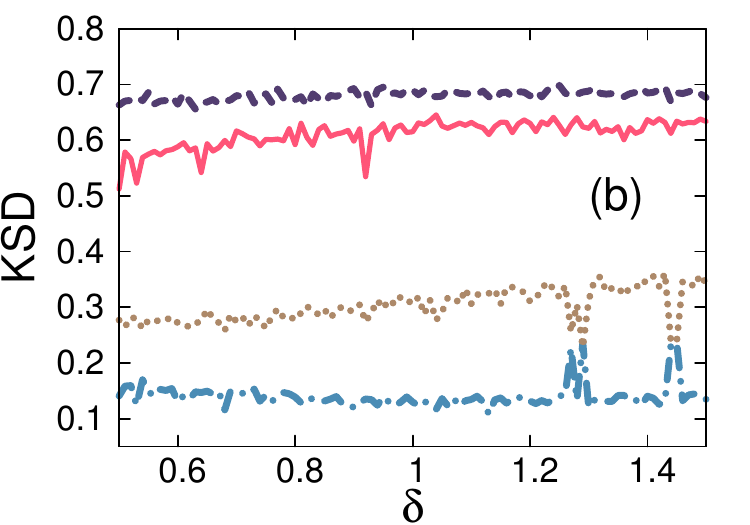}
}
\centerline{
 \includegraphics[width=0.5\columnwidth]{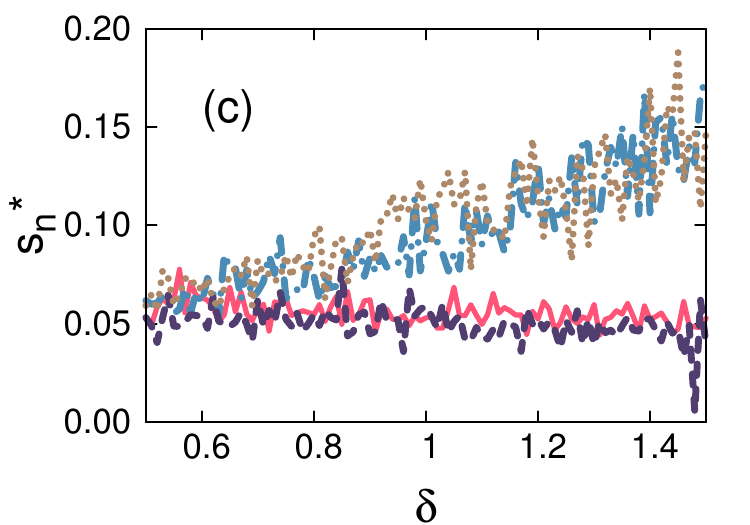}
 \includegraphics[width=0.5\columnwidth]{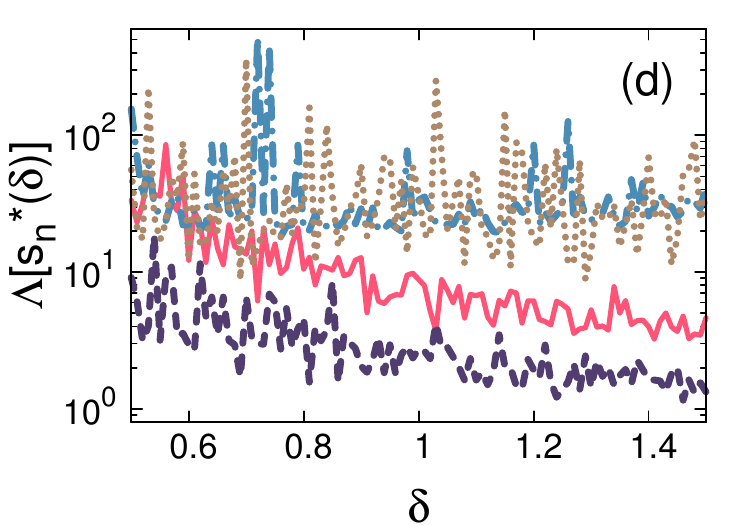}
}
\vspace{-0.2cm}
\caption{Magnitude dependence of the predictability indices of the vdP model. 
ROC-indices, AUC and KSD, are shown in (a) and (b), and the best indicators $s_n^{*}$ and the likelihood ratio evaluated at the best indicator in (c) and (d).
The indices for $k=1, w=10$ are shown with solid lines ($\color{magenta}{\boldsymbol{-}\hspace{-0.1cm}\boldsymbol{-}}$), for $k=1,w=500$ with dot-dashed lines ($\color{blue}{\boldsymbol{-}}$$\color{blue}{\boldsymbol{\cdot}}$$\color{blue}{\boldsymbol{-}}$), for $k=500, w=10$ with dashed lines ($\color{purple}{\boldsymbol{-\,-}}$), and for $k=500,w=500$ with dotted lines ($\color{brown}{\boldsymbol{\cdots}}$).
For large lead times and large window widths and small lead times and small window widths we observe a mild decrease in prediction quality with an increasing CT magnitude $\delta$ in (a) and (b).}
\label{fig:index_mag_vdP}
\end{figure}

Testing for the dependence of the prediction quality on the event magnitude in the vdP model, we find qualitatively different results. 
Instead of improving with increasing transition magnitude $\delta$, the quality of predictions remains the same or slightly decreases (\textit{negative magnitude dependence}) as illustrated in Fig.~\ref{fig:roc_mag_vdP}. 
The dependence on $\delta$ is especially pronounced for the combinations of small $k$ and small $w$ and large $k$ and large $w$. 
For the combinations of small $k$ and large $w$ or large $k$ and small $w$, there is no clearly visible dependence on the transition magnitude. 
Two of the ROC indices in Fig.~\ref{fig:index_mag_vdP} (AUC and KSD) support these findings. 
However, the likelihood ratio shows a decrease for small window widths and no clear trend for large window widths.
This deviation can be understood as a consequence of the definition of the likelihood ratio.
Since it is evaluated at the value of the maximum CPDF indicator $s_n^{*}$ it reflects the slope of the ROC curve in the region of small false positive rates, {i.e.}, close to the origin $(0,0)$.
In total we can conclude that the predictability of the vdP model is either independent or 
decreases with increasing magnitude of the transition.

\section{Towards a more detailed Understanding of the Magnitude Dependence}
\label{sec:detailedmag}

While the likelihood ratio $\Lambda(\psi_n^{\ast})$ is a useful summary index for analytically determined ROC curves, it is less reliable for numerically determined ROC curves, which may not be smooth. 
Nevertheless, $\Lambda[\psi_n^{\ast},\chi_n(\delta)]$ can still be used to investigate the magnitude dependences of the best indicators. 
Reformulating the partial derivative of the likelihood ratio with respect to $\delta$, one obtains the condition \cite{hallerberg2008predictability} 
\begin{equation}
    c(\psi_n,\delta):= \partial_{\delta}\,
		\ln\left[L(\psi_n,\delta)\right]-\frac{1-L(\psi_n,\delta)}{1-P(\delta)}\partial_{\delta} 
		\ln [P(\delta)].\;\label{eq:c}
\end{equation}
Here, $L(\psi_n,\delta) = \mathbb{P}(\chi_{n+k}(\delta)=1|\psi_n)$ and $P(\delta)$ denotes the total probability of finding transitions which are larger or equal than $\delta$. 
A positive $c$ at $\psi_n=\psi_n^{\ast}$ indicates that the best precursors' prediction quality improves with larger event magnitudes, and a negative one indicates the opposite.
%
\begin{figure}[!]
\centerline{
\includegraphics[width=0.5\columnwidth]{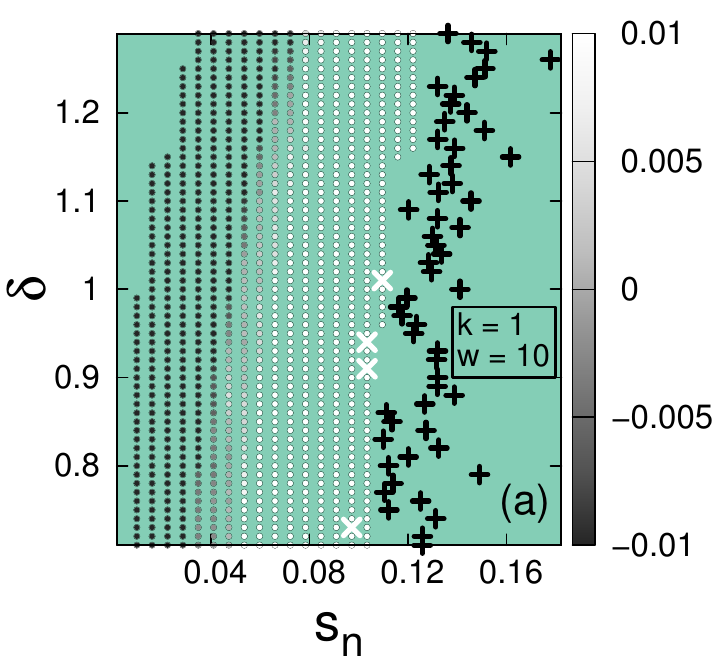}
\includegraphics[width=0.5\columnwidth]{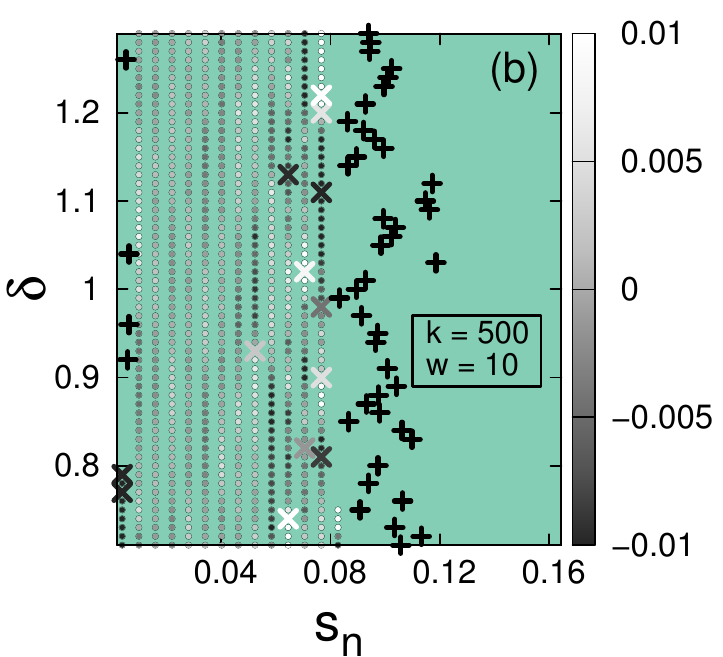}
}
\centerline{  
\includegraphics[width=0.5\columnwidth]{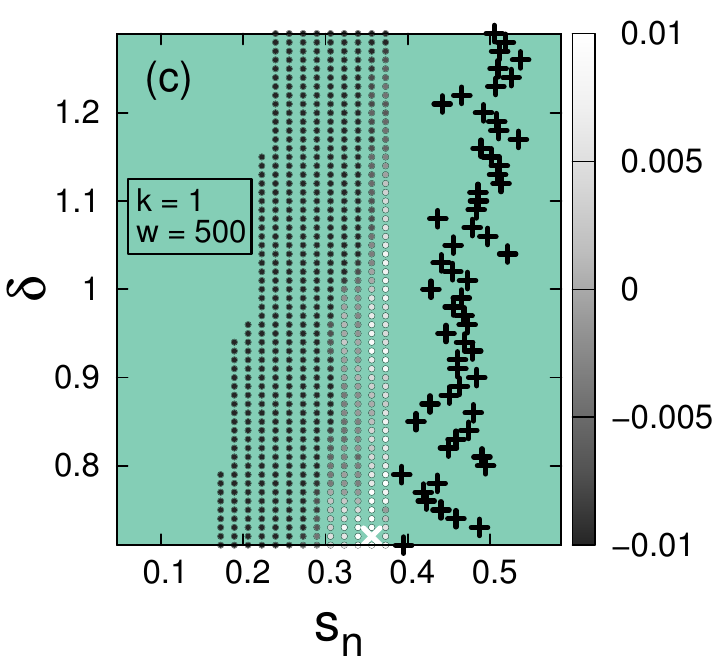}
\includegraphics[width=0.5\columnwidth]{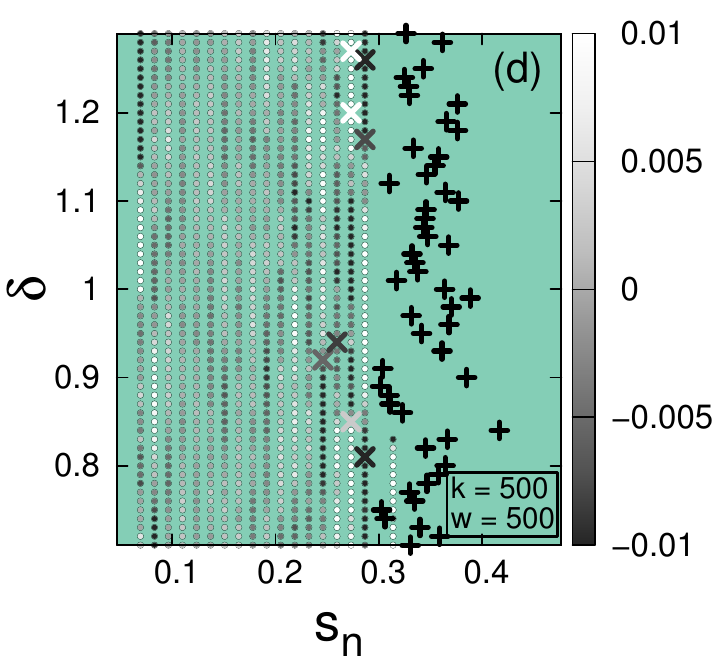}
}
\vspace{-0.4cm}
\caption{\label{fig:cQIF}Magnitude dependence of the QIF model according to the condition $c(\psi_n,\delta)$ [see Eq.~(\ref{eq:c})] for four representative combinations of $k$ and $w$.
The values of $c(\psi_n,\delta)$ are represented as grayscale-coded dots with a black border as long as they can be evaluated.
The location of the optimal indicator are marked as a \textquotedblleft$\times$\textquotedblright\, if the value of $c(\psi_n^{\ast}(\delta),\delta)$ is available, otherwise and it is marked as a black \textquotedblleft$+$\textquotedblright.
Note that $c(\psi_n,\delta_i)$ can only be evaluated, if PDFs, CPDFs and their derivatives could be estimated with sufficient accuracy for all values of $\psi_n$, $\delta_{i}$, and $\delta_{i+1}$, where the index $i$ enumerates discrete steps in $\delta$.
}
\end{figure}
Numerical estimates of $c=c(\psi_n,\delta)$ are presented for discrete values of $\psi_n$ and $\delta$ in the $\psi_n-\delta$ plane (Figs.~\ref{fig:cQIF} and \ref{fig:cvdP}).
The derivatives $\partial_{\delta}\ln[L(\psi_n,\delta)]$ and $\partial_{\delta}\ln[P(\delta)]$ are numerically estimated using PDFs and CPDFs for neighboring values of $\delta_i$ and $\delta_{i+1}$.
Additionally, all numerical estimates of distribution functions and their derivatives are smoothed using Savitzky-Golay filtering \cite{Savgol} of fourth order with $20$ supporting points to the left and to the right.
Note that the condition in Eq.~(\ref{eq:c}) can be only evaluated, if all numerical estimates of distribution functions can be obtained with sufficient accuracy for all values of $\psi_n$ and each neighboring pair of $\delta_{i}$ and $\delta_{i+1}$.
Blank (green/medium gray) areas in Figs.~\ref{fig:cQIF} and \ref{fig:cvdP} indicate that this is not always possible, given the number of events for specific values of $\psi_n$ and $\delta$.
Values of $c(\psi_n,\delta)$ are represented by grayscale-coded dots with a black border, locations of optimal indicators \cite{naivebayes}
\begin{equation}
  \psi_n^{\ast}(\delta) :=\arg\max_{\psi_n}\mathbb{P}(\chi_n=1|\psi_n, \delta),\label{eq:snopt}
\end{equation}
%
are indicated by crosses: If the value of $c[\psi_n^{\ast}(\delta),\delta]$ can be evaluated, then it is shown as a grayscale-coded \textquotedblleft$\times$\textquotedblright; if not, the location of $\psi_n^{\ast}(\delta)$ is shown as a black \textquotedblleft$+$\textquotedblright.
The results for the QIF model (\ref{fig:cQIF}) are mostly consistent with the previous findings presented in ROC plots (see Fig.~\ref{fig:roc_mag_QIF}).
If the lead time is small [$k=1$, see Figs.~\ref{fig:cQIF} (a) and \ref{fig:cQIF} (c)], then $c(\psi_n,\delta)$ is positive for relevant indicators.
For larger lead times ($k=500$) there are almost equally many positive and negative values of $c(\psi_n^{\ast},\delta)$ [see Fig.~\ref{fig:cQIF} (b) and \ref{fig:cQIF} (d)].
\begin{figure}
\centerline{
\includegraphics[width=0.5\columnwidth]{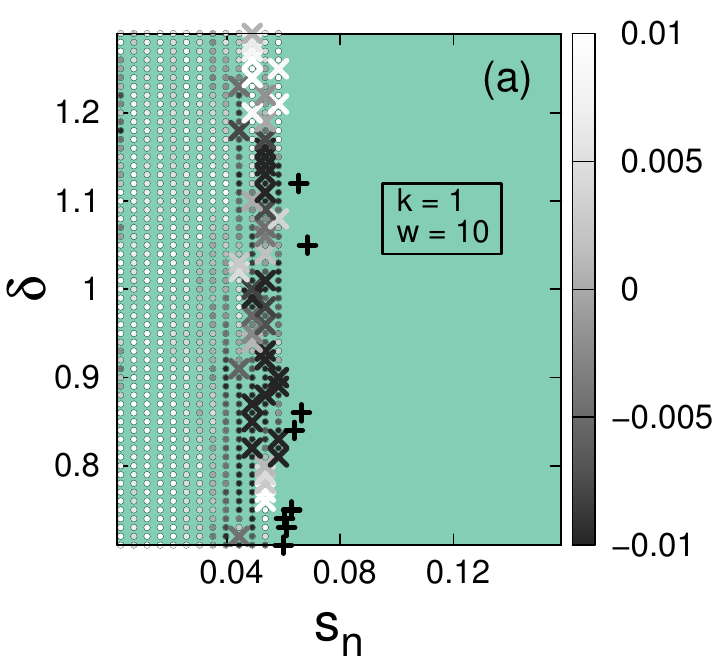}
\includegraphics[width=0.5\columnwidth]{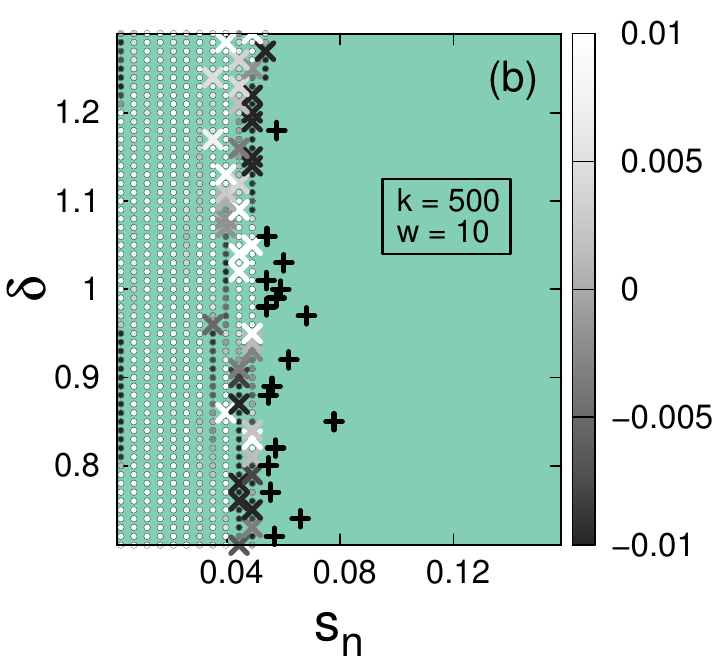}
}
\centerline{
\includegraphics[width=0.5\columnwidth]{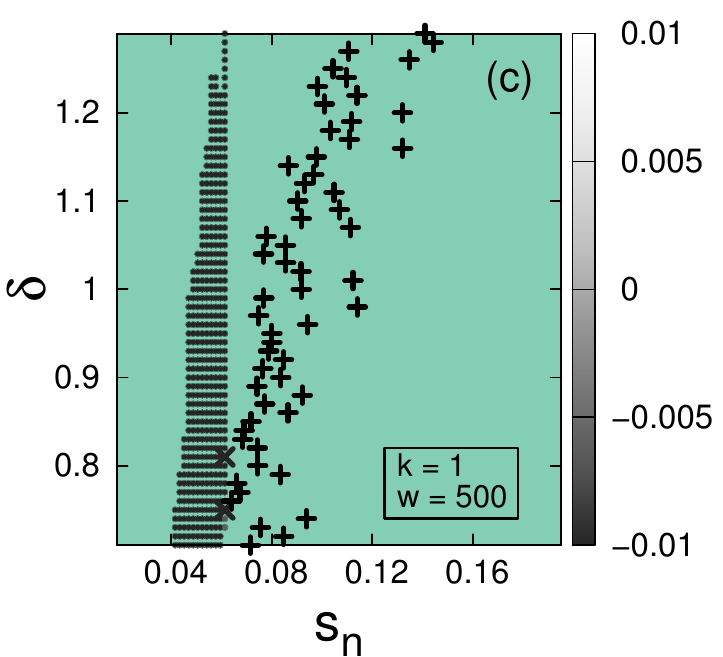}
\includegraphics[width=0.5\columnwidth]{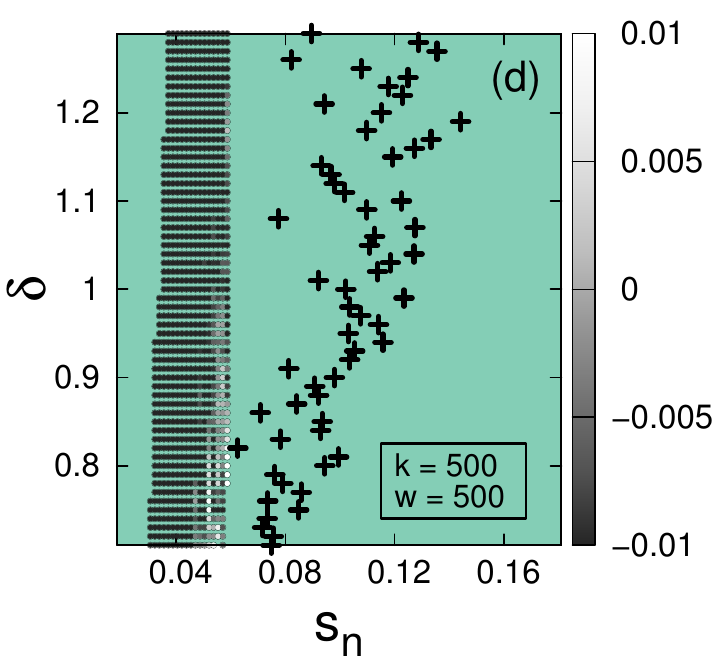}
}
\vspace{-0.4cm}
\caption{\label{fig:cvdP}Magnitude dependence of the vdP model according to the condition $c(\psi_n,\delta)$ [see Eq.~(\ref{eq:c})] for four representative combinations of $k$ and $w$.
Color coding as in Fig.~\ref{fig:cQIF}.
}
 \end{figure}

In contrast to this, the results for the vdP model (Fig.~\ref{fig:cvdP}) indicate different magnitude dependences than the ROC curves in Fig.~\ref{fig:roc_mag_vdP}.
For small window widths [$w=10$, see Fig.~\ref{fig:cvdP} (a) and \ref{fig:cvdP} (b)] no clear preference of the sign of $c(\psi_n^{\ast},\delta)$ can be seen while the ROC curve for $k=1,w=10$ shows a negative magnitude dependence.
If the window width is large [$w=500$, see Fig.~\ref{fig:cvdP} (c) and \ref{fig:cvdP} (d)], then it is not possible to evaluate Eq.~(\ref{eq:c}) for optimal indicators $\psi_n^{\ast}$, indicated as black crosses.
However, in the regions where it can be evaluated the $c(\psi_n,\delta)$ is negative.

In general, it is a drawback of the condition in Eq.~\eqref{eq:c} that it is evaluated only for {\sl one} specific indicator value, since in applications alarm {\sl volumes} containing {\sl ranges} of useful indicator values are relevant [see, e.g., Eq.~(\ref{eq:decision variable_roc})].
Additionally, we saw that the maximum CPDF indicator can also vary with the event magnitude $\delta$.
Therefore, we propose a quantity that summarizes Eq.~\eqref{eq:c} for a couple of suitable indicators.
This \textit{combined alarm volume} $V^{\ast}$ contains a range of indicator values comprising all best indicators $\psi_n^{\ast}(\delta)$ [see Eq.~(\ref{eq:snopt})] for all given range of $\delta$, i.e.,
\begin{equation}
V^{\ast}:=[\min_{\delta}\psi_n^{\ast}(\delta),\max_{\delta}\psi_n^{\ast}(\delta)].
\end{equation}
Analogously to Eq.~(\ref{eq:c}), a condition involving the combined alarm volume can be obtained using the derivative of the combined likelihood ratio, i.e.,
\begin{eqnarray}
    \partial_{\delta}\Lambda^{\ast}\left(V^{\ast},\chi_n(\delta)\right)&=
		\partial_{\delta}\left(\dfrac{\int_{V^{\ast}}\mathrm{d} \psi_n 
		\mathbb{P}\left(\psi_n|\chi_n(\delta)=1\right)}{\int_{V^{\ast}}\mathrm{d} \psi_n 
		\mathbb{P}\left(\psi_n|\chi_n(\delta)=0\right)}\right).\quad\quad\label{eq:towards_cv}
\end{eqnarray}
Since we are interested only in the sign of this expression, we can formulate the condition
\begin{eqnarray}
    c_{V}(V^{\ast}, \chi(\delta))&:=&I_{0}\partial_{\delta}I_{1}-I_{1}\partial_{\delta}I_{0}, \label{eq:cv} \\
\mbox{with}\; I_{1} &=& \int_{V^{\ast}}\mathrm{d} \psi_n p\left(\psi_n|\chi_n(\delta)=1\right), \nonumber\\
\mbox{and}\; I_{0} &=& \int_{V^{\ast}}\mathrm{d} \psi_n p\left(\psi_n|\chi_n(\delta)=0\right). \nonumber
\end{eqnarray}
%
\begin{figure}[t!]
\centerline{
\includegraphics[width=1.0\columnwidth]{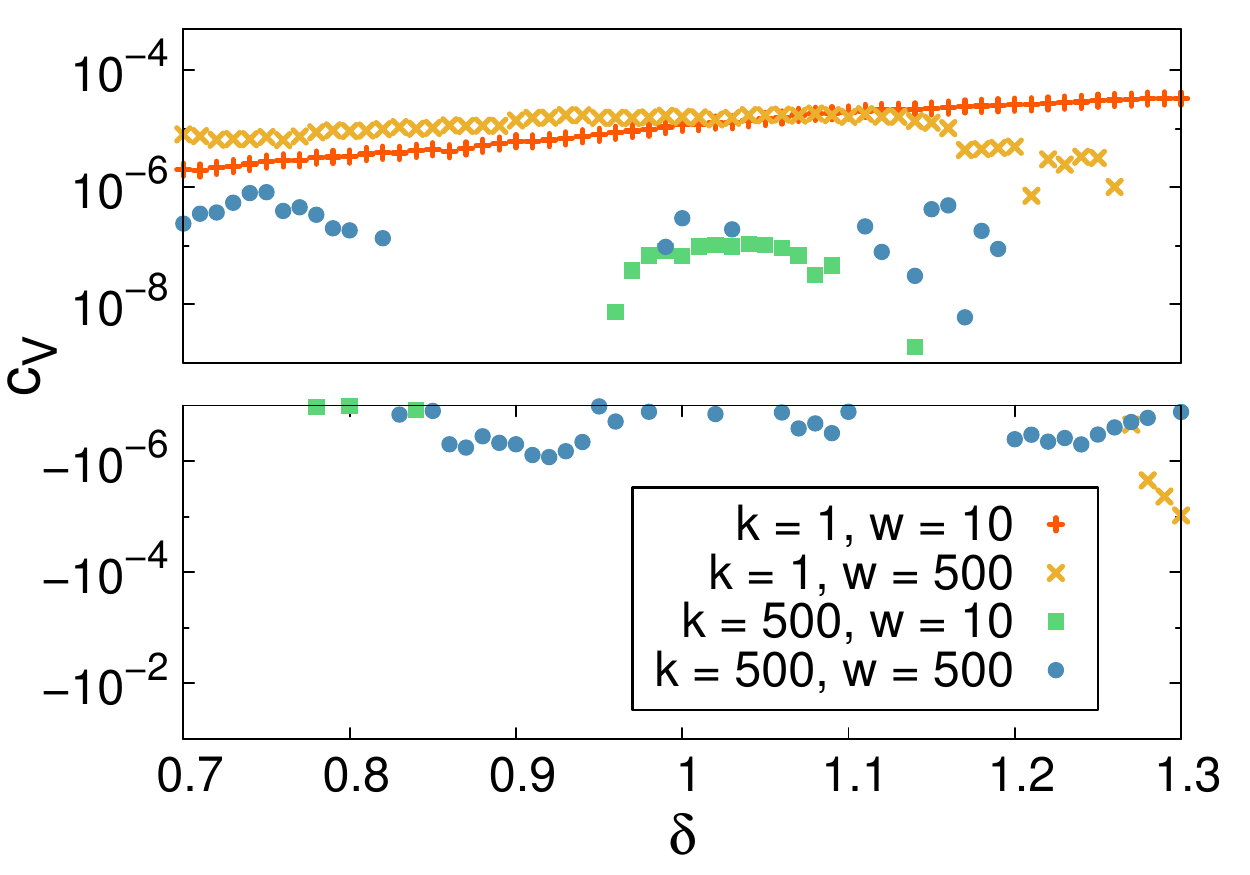}
}
\vspace{-0.4cm}
\caption{\label{fig:cvQIF}Magnitude dependence of the QIF model using combined alarm volumes $c^*:=c_{V}(V^{\ast}, \chi(\delta))$ [Eq.~(\ref{eq:cv})] for four representative values of $k$ and $w$.
}
\end{figure}
%
We evaluate the integrals $I_{1}(\delta)$, $I_{0}(\delta)$ numerically and estimate their 
derivatives by using the Savitzky-Golay filtering as described above.
It is worth mentioning that the combined likelihood ratio $\Lambda^{\ast}\left(V^{\ast},\chi(\delta)\right)$ is not the slope of the ROC curve, but an approximation of it, since the alarm volume for an ROC curve is defined through the discrimination threshold $d$ [see Eq.~(\ref{eq:decision variable_roc})], not the combined alarm volume.
\begin{figure}[t!!!]
\centerline{
\includegraphics[width=1.0\columnwidth]{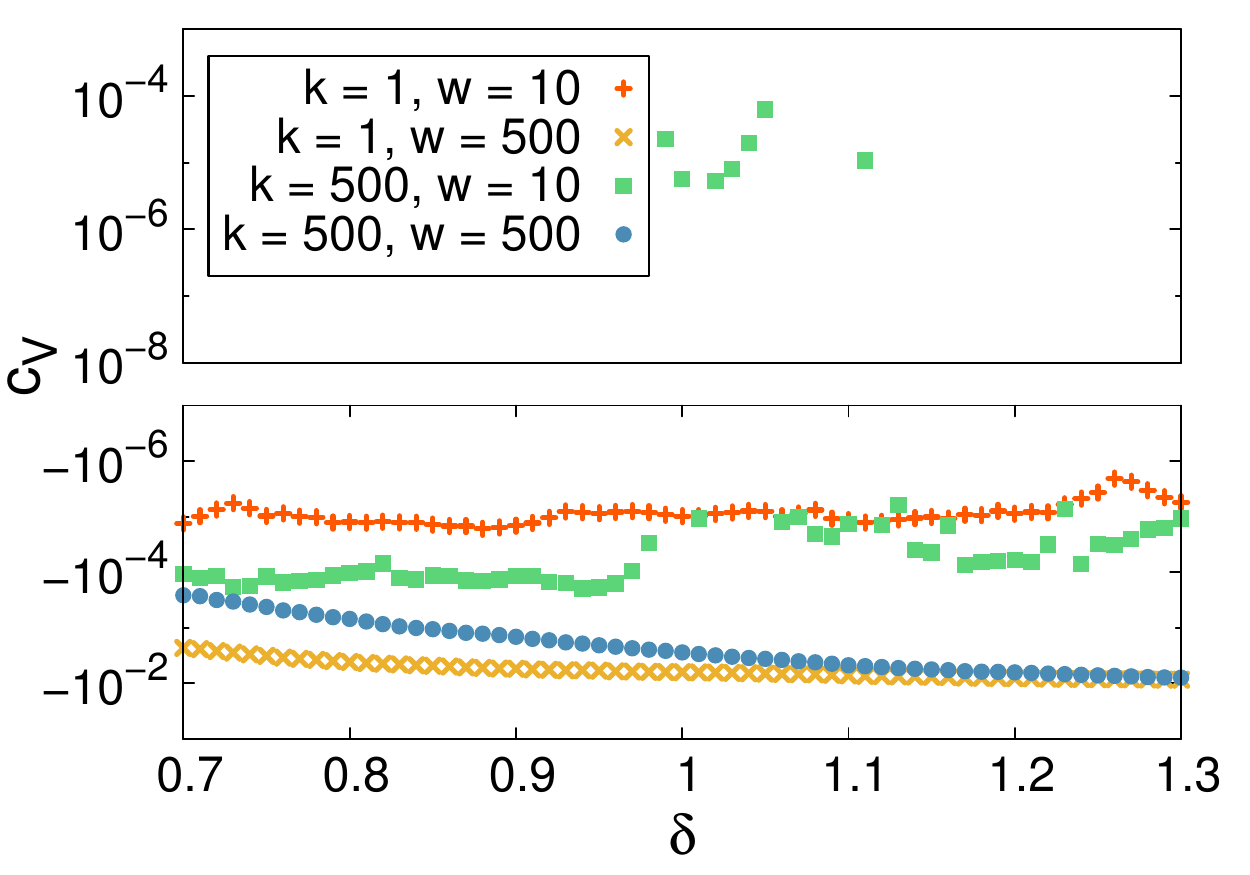}
}
\vspace{-0.4cm}
\caption{\label{fig:cvvdP}Magnitude dependence of the vdP model using combined alarm volumes $c^*:=c_{V}(V^{\ast}, \chi(\delta))$ [Eq.~(\ref{eq:cv})] for four representative values of $k$ and $w$.
}
\end{figure}
%
Plotting $c_{V}$ as a function of $\delta$ (see Figs.~\ref{fig:cvQIF} and \ref{fig:cvvdP}), we find that the sign of $c_{V}$ corresponds well to the magnitude dependence of the ROC curves in Figs.~\ref{fig:roc_mag_QIF} and \ref{fig:roc_mag_vdP}: mostly positive for $k=1$ in the QIF model and mostly negative for the vdP model.
%
%
\section{Conclusions}
\label{sec:conclusions}
CTs are abundant in nature and society and can have drastic consequences.
In contrast to forecasting future values of systems, predicting CTs on the basis of early-warning signals can be understood as a classification task.
From theoretical considerations one can expect that many classes of CTs are preceded by a vulnerability to external perturbations, i.e., by CSD \cite{Wiesenfeld1,wissel1984universal}.
As a consequence of CSD, several indicator variables such as lag-1 autocorrelation and variance were proposed \cite{Wiesenfeld1,wissel1984universal}.
In applications, these indicator variables have been observed to show a specific behavior previous to the observation of one single CT  \cite{Schefferetal1,SchefferCarpenter,Venegasetal,McSharrySmithTarassenko, MayLevinSugihara,Lentonetal,Alleyetal}.
However, the general quality of these indicators, their performance in the presence of noise and in the limit of short observation time series has so far not been accessed in a statistically significant way.
In this contribution we have evaluated the quality of indicators derived from CT by using long time series of many CTs, generated by models of two different fast-slow systems.
We have quantified the prediction success by computing ROC curves and related summary indices.

We found mechanisms influencing the prediction success to be either sensitive to the specific model or model-independent.
Effects that were observed in both models are the dependence of the length of the data record and the forecast horizon:
As common in forecasts, the prediction performance of all tested indicators and in both models decreases with increased forecast horizon.
Increasing the window width, and thus smoothing the indicator time series, all indicators show an increasingly strong ability to predict CTs.
Consequently, increasing the amount of available training data can be an effective way to improve predictions.
Surprisingly, we find that the performance of indicator variables is dependent on the specific model under study and the conditions of accessing it.
%
%
%
%
For the QIF model, the lag-1 autocorrelation produces the best results in our numerical study in the limit of small lead times and large window sizes for estimation.
However, for the vdP model, the sliding average of the standard deviation performs significantly better than other indicators in the limit of low data availability and long lead times, with a remarkable AUC score of 0.8 or higher.

Also the role of the transition magnitude depends sensitively on the model under study.
For the QIF model, we observed a positive magnitude dependence for short forecast horizons, while larger transitions cannot be predicted more easily if the forecast horizon is large.
For CTs generated by the vdP model, the quality of predictions can even decrease with an increasing transition magnitude. 
Consequently, high-impact events can not always be expected to be easier to predict, although results for short forecasts might indicate this.
%

\begin{acknowledgments}
C.K. thanks the Austrian Academy of Sciences (\"OAW) for support via an 
APART fellowship and the European Commission (EC/REA) for support by a Marie-Curie 
International Re-Integration Grant. C.K. also acknowledges support by Max-Planck Institute for Dynamics and Self-organization to facilitate his visit to G\"ottingen.
\end{acknowledgments}

\bibliography{zhang_kuehn_hallerberg_2015}

\begin{thebibliography}{64}%
\makeatletter
\providecommand \@ifxundefined [1]{%
 \@ifx{#1\undefined}
}%
\providecommand \@ifnum [1]{%
 \ifnum #1\expandafter \@firstoftwo
 \else \expandafter \@secondoftwo
 \fi
}%
\providecommand \@ifx [1]{%
 \ifx #1\expandafter \@firstoftwo
 \else \expandafter \@secondoftwo
 \fi
}%
\providecommand \natexlab [1]{#1}%
\providecommand \enquote  [1]{``#1''}%
\providecommand \bibnamefont  [1]{#1}%
\providecommand \bibfnamefont [1]{#1}%
\providecommand \citenamefont [1]{#1}%
\providecommand \href@noop [0]{\@secondoftwo}%
\providecommand \href [0]{\begingroup \@sanitize@url \@href}%
\providecommand \@href[1]{\@@startlink{#1}\@@href}%
\providecommand \@@href[1]{\endgroup#1\@@endlink}%
\providecommand \@sanitize@url [0]{\catcode `\\12\catcode `\$12\catcode
  `\&12\catcode `\#12\catcode `\^12\catcode `\_12\catcode `\%12\relax}%
\providecommand \@@startlink[1]{}%
\providecommand \@@endlink[0]{}%
\providecommand \url  [0]{\begingroup\@sanitize@url \@url }%
\providecommand \@url [1]{\endgroup\@href {#1}{\urlprefix }}%
\providecommand \urlprefix  [0]{URL }%
\providecommand \Eprint [0]{\href }%
\providecommand \doibase [0]{http://dx.doi.org/}%
\providecommand \selectlanguage [0]{\@gobble}%
\providecommand \bibinfo  [0]{\@secondoftwo}%
\providecommand \bibfield  [0]{\@secondoftwo}%
\providecommand \translation [1]{[#1]}%
\providecommand \BibitemOpen [0]{}%
\providecommand \bibitemStop [0]{}%
\providecommand \bibitemNoStop [0]{.\EOS\space}%
\providecommand \EOS [0]{\spacefactor3000\relax}%
\providecommand \BibitemShut  [1]{\csname bibitem#1\endcsname}%
\let\auto@bib@innerbib\@empty
\bibitem [{\citenamefont {Scheffer}\ \emph {et~al.}(2001)\citenamefont
  {Scheffer}, \citenamefont {Carpenter}, \citenamefont {Foley}, \citenamefont
  {Folke},\ and\ \citenamefont {Walker}}]{Schefferetal1}%
  \BibitemOpen
  \bibfield  {author} {\bibinfo {author} {\bibfnamefont {M.}~\bibnamefont
  {Scheffer}}, \bibinfo {author} {\bibfnamefont {S.}~\bibnamefont {Carpenter}},
  \bibinfo {author} {\bibfnamefont {J.}~\bibnamefont {Foley}}, \bibinfo
  {author} {\bibfnamefont {C.}~\bibnamefont {Folke}}, \ and\ \bibinfo {author}
  {\bibfnamefont {B.}~\bibnamefont {Walker}},\ }\href@noop {} {\bibfield
  {journal} {\bibinfo  {journal} {Nature}\ }\textbf {\bibinfo {volume} {413}},\
  \bibinfo {pages} {591} (\bibinfo {year} {2001})}\BibitemShut {NoStop}%
\bibitem [{\citenamefont {Scheffer}\ and\ \citenamefont
  {Carpenter}(2003)}]{SchefferCarpenter}%
  \BibitemOpen
  \bibfield  {author} {\bibinfo {author} {\bibfnamefont {M.}~\bibnamefont
  {Scheffer}}\ and\ \bibinfo {author} {\bibfnamefont {S.}~\bibnamefont
  {Carpenter}},\ }\href@noop {} {\bibfield  {journal} {\bibinfo  {journal}
  {TRENDS in Ecol. and Evol.}\ }\textbf {\bibinfo {volume} {18}},\ \bibinfo
  {pages} {648} (\bibinfo {year} {2003})}\BibitemShut {NoStop}%
\bibitem [{\citenamefont {Venegas}\ \emph {et~al.}(2005)\citenamefont
  {Venegas}, \citenamefont {Winkler}, \citenamefont {Musch}, \citenamefont
  {Melo}, \citenamefont {Layfield}, \citenamefont {Tgavalekos}, \citenamefont
  {Fischman}, \citenamefont {Callahan}, \citenamefont {Bellani},\ and\
  \citenamefont {Harris}}]{Venegasetal}%
  \BibitemOpen
  \bibfield  {author} {\bibinfo {author} {\bibfnamefont {J.}~\bibnamefont
  {Venegas}}, \bibinfo {author} {\bibfnamefont {T.}~\bibnamefont {Winkler}},
  \bibinfo {author} {\bibfnamefont {G.}~\bibnamefont {Musch}}, \bibinfo
  {author} {\bibfnamefont {M.~V.}\ \bibnamefont {Melo}}, \bibinfo {author}
  {\bibfnamefont {D.}~\bibnamefont {Layfield}}, \bibinfo {author}
  {\bibfnamefont {N.}~\bibnamefont {Tgavalekos}}, \bibinfo {author}
  {\bibfnamefont {A.}~\bibnamefont {Fischman}}, \bibinfo {author}
  {\bibfnamefont {R.}~\bibnamefont {Callahan}}, \bibinfo {author}
  {\bibfnamefont {G.}~\bibnamefont {Bellani}}, \ and\ \bibinfo {author}
  {\bibfnamefont {R.}~\bibnamefont {Harris}},\ }\href@noop {} {\bibfield
  {journal} {\bibinfo  {journal} {Nature}\ }\textbf {\bibinfo {volume} {434}},\
  \bibinfo {pages} {777} (\bibinfo {year} {2005})}\BibitemShut {NoStop}%
\bibitem [{\citenamefont {McSharry}\ \emph {et~al.}(2003)\citenamefont
  {McSharry}, \citenamefont {Smith},\ and\ \citenamefont
  {Tarassenko}}]{McSharrySmithTarassenko}%
  \BibitemOpen
  \bibfield  {author} {\bibinfo {author} {\bibfnamefont {P.}~\bibnamefont
  {McSharry}}, \bibinfo {author} {\bibfnamefont {L.}~\bibnamefont {Smith}}, \
  and\ \bibinfo {author} {\bibfnamefont {L.}~\bibnamefont {Tarassenko}},\
  }\href@noop {} {\bibfield  {journal} {\bibinfo  {journal} {Nature Med.}\
  }\textbf {\bibinfo {volume} {9}},\ \bibinfo {pages} {241} (\bibinfo {year}
  {2003})}\BibitemShut {NoStop}%
\bibitem [{\citenamefont {May}\ \emph {et~al.}(2008)\citenamefont {May},
  \citenamefont {Levin},\ and\ \citenamefont {Sugihara}}]{MayLevinSugihara}%
  \BibitemOpen
  \bibfield  {author} {\bibinfo {author} {\bibfnamefont {R.}~\bibnamefont
  {May}}, \bibinfo {author} {\bibfnamefont {S.}~\bibnamefont {Levin}}, \ and\
  \bibinfo {author} {\bibfnamefont {G.}~\bibnamefont {Sugihara}},\ }\href@noop
  {} {\bibfield  {journal} {\bibinfo  {journal} {Nature}\ }\textbf {\bibinfo
  {volume} {451}},\ \bibinfo {pages} {893} (\bibinfo {year}
  {2008})}\BibitemShut {NoStop}%
\bibitem [{\citenamefont {Lenton}\ \emph {et~al.}(2008)\citenamefont {Lenton},
  \citenamefont {Held}, \citenamefont {Kriegler}, \citenamefont {Hall},
  \citenamefont {Lucht}, \citenamefont {Rahmstorf},\ and\ \citenamefont
  {Schellnhuber}}]{Lentonetal}%
  \BibitemOpen
  \bibfield  {author} {\bibinfo {author} {\bibfnamefont {T.}~\bibnamefont
  {Lenton}}, \bibinfo {author} {\bibfnamefont {H.}~\bibnamefont {Held}},
  \bibinfo {author} {\bibfnamefont {E.}~\bibnamefont {Kriegler}}, \bibinfo
  {author} {\bibfnamefont {J.}~\bibnamefont {Hall}}, \bibinfo {author}
  {\bibfnamefont {W.}~\bibnamefont {Lucht}}, \bibinfo {author} {\bibfnamefont
  {S.}~\bibnamefont {Rahmstorf}}, \ and\ \bibinfo {author} {\bibfnamefont
  {H.}~\bibnamefont {Schellnhuber}},\ }\href@noop {} {\bibfield  {journal}
  {\bibinfo  {journal} {Proc. Natl. Acad. Sci. USA}\ }\textbf {\bibinfo
  {volume} {105}},\ \bibinfo {pages} {1786} (\bibinfo {year}
  {2008})}\BibitemShut {NoStop}%
\bibitem [{\citenamefont {Alley}\ \emph {et~al.}(2003)\citenamefont {Alley},
  \citenamefont {Marotzke}, \citenamefont {Nordhaus}, \citenamefont {Overpeck},
  \citenamefont {Peteet}, \citenamefont {Jr.}, \citenamefont {Pierrehumbert},
  \citenamefont {Rhines}, \citenamefont {Stocker}, \citenamefont {Talley},\
  and\ \citenamefont {Wallace}}]{Alleyetal}%
  \BibitemOpen
  \bibfield  {author} {\bibinfo {author} {\bibfnamefont {R.}~\bibnamefont
  {Alley}}, \bibinfo {author} {\bibfnamefont {J.}~\bibnamefont {Marotzke}},
  \bibinfo {author} {\bibfnamefont {W.}~\bibnamefont {Nordhaus}}, \bibinfo
  {author} {\bibfnamefont {J.}~\bibnamefont {Overpeck}}, \bibinfo {author}
  {\bibfnamefont {D.}~\bibnamefont {Peteet}}, \bibinfo {author} {\bibfnamefont
  {R.~P.}\ \bibnamefont {Jr.}}, \bibinfo {author} {\bibfnamefont
  {R.}~\bibnamefont {Pierrehumbert}}, \bibinfo {author} {\bibfnamefont
  {P.}~\bibnamefont {Rhines}}, \bibinfo {author} {\bibfnamefont
  {T.}~\bibnamefont {Stocker}}, \bibinfo {author} {\bibfnamefont
  {L.}~\bibnamefont {Talley}}, \ and\ \bibinfo {author} {\bibfnamefont
  {J.}~\bibnamefont {Wallace}},\ }\href@noop {} {\bibfield  {journal} {\bibinfo
   {journal} {Science}\ }\textbf {\bibinfo {volume} {299}},\ \bibinfo {pages}
  {2005} (\bibinfo {year} {2003})}\BibitemShut {NoStop}%
\bibitem [{\citenamefont {Wiesenfeld}(1985)}]{Wiesenfeld1}%
  \BibitemOpen
  \bibfield  {author} {\bibinfo {author} {\bibfnamefont {K.}~\bibnamefont
  {Wiesenfeld}},\ }\href@noop {} {\bibfield  {journal} {\bibinfo  {journal} {J.
  Stat. Phys.}\ }\textbf {\bibinfo {volume} {38}},\ \bibinfo {pages} {1071}
  (\bibinfo {year} {1985})}\BibitemShut {NoStop}%
\bibitem [{\citenamefont {Wissel}(1984)}]{wissel1984universal}%
  \BibitemOpen
  \bibfield  {author} {\bibinfo {author} {\bibfnamefont {C.}~\bibnamefont
  {Wissel}},\ }\href@noop {} {\bibfield  {journal} {\bibinfo  {journal}
  {Oecologia}\ }\textbf {\bibinfo {volume} {65}},\ \bibinfo {pages} {101}
  (\bibinfo {year} {1984})}\BibitemShut {NoStop}%
\bibitem [{\citenamefont {Scheffer}(2009)}]{scheffer2009critical}%
  \BibitemOpen
  \bibfield  {author} {\bibinfo {author} {\bibfnamefont {M.}~\bibnamefont
  {Scheffer}},\ }\href@noop {} {\emph {\bibinfo {title} {Critical transitions
  in nature and society}}}\ (\bibinfo  {publisher} {Princeton University
  Press},\ \bibinfo {year} {2009})\BibitemShut {NoStop}%
\bibitem [{\citenamefont {Biggs}\ \emph {et~al.}(2009)\citenamefont {Biggs},
  \citenamefont {Carpenter},\ and\ \citenamefont {Brock}}]{biggs2009turning}%
  \BibitemOpen
  \bibfield  {author} {\bibinfo {author} {\bibfnamefont {R.}~\bibnamefont
  {Biggs}}, \bibinfo {author} {\bibfnamefont {S.~R.}\ \bibnamefont
  {Carpenter}}, \ and\ \bibinfo {author} {\bibfnamefont {W.~A.}\ \bibnamefont
  {Brock}},\ }\href@noop {} {\bibfield  {journal} {\bibinfo  {journal}
  {Proceedings of the National academy of Sciences}\ }\textbf {\bibinfo
  {volume} {106}},\ \bibinfo {pages} {826} (\bibinfo {year}
  {2009})}\BibitemShut {NoStop}%
\bibitem [{\citenamefont {Carpenter}\ and\ \citenamefont
  {Brock}(2006)}]{carpenter2006rising}%
  \BibitemOpen
  \bibfield  {author} {\bibinfo {author} {\bibfnamefont {S.~R.}\ \bibnamefont
  {Carpenter}}\ and\ \bibinfo {author} {\bibfnamefont {W.~A.}\ \bibnamefont
  {Brock}},\ }\href@noop {} {\bibfield  {journal} {\bibinfo  {journal} {Ecology
  letters}\ }\textbf {\bibinfo {volume} {9}},\ \bibinfo {pages} {311} (\bibinfo
  {year} {2006})}\BibitemShut {NoStop}%
\bibitem [{\citenamefont {Scheffer}\ \emph {et~al.}(2009)\citenamefont
  {Scheffer}, \citenamefont {Bascompte}, \citenamefont {Brock}, \citenamefont
  {Brovkin}, \citenamefont {Carpenter}, \citenamefont {Dakos}, \citenamefont
  {Held}, \citenamefont {Van~Nes}, \citenamefont {Rietkerk},\ and\
  \citenamefont {Sugihara}}]{scheffer2009early}%
  \BibitemOpen
  \bibfield  {author} {\bibinfo {author} {\bibfnamefont {M.}~\bibnamefont
  {Scheffer}}, \bibinfo {author} {\bibfnamefont {J.}~\bibnamefont {Bascompte}},
  \bibinfo {author} {\bibfnamefont {W.~A.}\ \bibnamefont {Brock}}, \bibinfo
  {author} {\bibfnamefont {V.}~\bibnamefont {Brovkin}}, \bibinfo {author}
  {\bibfnamefont {S.~R.}\ \bibnamefont {Carpenter}}, \bibinfo {author}
  {\bibfnamefont {V.}~\bibnamefont {Dakos}}, \bibinfo {author} {\bibfnamefont
  {H.}~\bibnamefont {Held}}, \bibinfo {author} {\bibfnamefont {E.~H.}\
  \bibnamefont {Van~Nes}}, \bibinfo {author} {\bibfnamefont {M.}~\bibnamefont
  {Rietkerk}}, \ and\ \bibinfo {author} {\bibfnamefont {G.}~\bibnamefont
  {Sugihara}},\ }\href@noop {} {\bibfield  {journal} {\bibinfo  {journal}
  {Nature}\ }\textbf {\bibinfo {volume} {461}},\ \bibinfo {pages} {53}
  (\bibinfo {year} {2009})}\BibitemShut {NoStop}%
\bibitem [{\citenamefont {Ives}(1995)}]{ives1995measuring}%
  \BibitemOpen
  \bibfield  {author} {\bibinfo {author} {\bibfnamefont {A.~R.}\ \bibnamefont
  {Ives}},\ }\href@noop {} {\bibfield  {journal} {\bibinfo  {journal}
  {Ecological Monographs}\ ,\ \bibinfo {pages} {217}} (\bibinfo {year}
  {1995})}\BibitemShut {NoStop}%
\bibitem [{\citenamefont {Kuehn}(2013)}]{kuehn2013mathematical}%
  \BibitemOpen
  \bibfield  {author} {\bibinfo {author} {\bibfnamefont {C.}~\bibnamefont
  {Kuehn}},\ }\href@noop {} {\bibfield  {journal} {\bibinfo  {journal} {Journal
  of Nonlinear Science}\ }\textbf {\bibinfo {volume} {23}},\ \bibinfo {pages}
  {457} (\bibinfo {year} {2013})}\BibitemShut {NoStop}%
\bibitem [{\citenamefont {Donangelo}\ \emph {et~al.}(2010)\citenamefont
  {Donangelo}, \citenamefont {Fort}, \citenamefont {Dakos}, \citenamefont
  {Scheffer},\ and\ \citenamefont {Nes}}]{DonangeloFortDakosSchefferNes}%
  \BibitemOpen
  \bibfield  {author} {\bibinfo {author} {\bibfnamefont {R.}~\bibnamefont
  {Donangelo}}, \bibinfo {author} {\bibfnamefont {H.}~\bibnamefont {Fort}},
  \bibinfo {author} {\bibfnamefont {V.}~\bibnamefont {Dakos}}, \bibinfo
  {author} {\bibfnamefont {M.}~\bibnamefont {Scheffer}}, \ and\ \bibinfo
  {author} {\bibfnamefont {E.~V.}\ \bibnamefont {Nes}},\ }\href@noop {}
  {\bibfield  {journal} {\bibinfo  {journal} {Int. J. Bif. Chaos}\ }\textbf
  {\bibinfo {volume} {20}},\ \bibinfo {pages} {315} (\bibinfo {year}
  {2010})}\BibitemShut {NoStop}%
\bibitem [{\citenamefont {Guttal}\ and\ \citenamefont
  {Jayaprakash}(2009)}]{GuttalJayaprakash1}%
  \BibitemOpen
  \bibfield  {author} {\bibinfo {author} {\bibfnamefont {V.}~\bibnamefont
  {Guttal}}\ and\ \bibinfo {author} {\bibfnamefont {C.}~\bibnamefont
  {Jayaprakash}},\ }\href@noop {} {\bibfield  {journal} {\bibinfo  {journal}
  {Theor. Ecol.}\ }\textbf {\bibinfo {volume} {2}},\ \bibinfo {pages} {3}
  (\bibinfo {year} {2009})}\BibitemShut {NoStop}%
\bibitem [{\citenamefont {Drake}\ and\ \citenamefont
  {Griffen}(2010)}]{DrakeGriffen}%
  \BibitemOpen
  \bibfield  {author} {\bibinfo {author} {\bibfnamefont {J.}~\bibnamefont
  {Drake}}\ and\ \bibinfo {author} {\bibfnamefont {B.}~\bibnamefont
  {Griffen}},\ }\href@noop {} {\bibfield  {journal} {\bibinfo  {journal}
  {Nature}\ }\textbf {\bibinfo {volume} {467}},\ \bibinfo {pages} {456}
  (\bibinfo {year} {2010})}\BibitemShut {NoStop}%
\bibitem [{\citenamefont {Meisel}\ \emph {et~al.}(2015)\citenamefont {Meisel},
  \citenamefont {Klaus}, \citenamefont {Kuehn},\ and\ \citenamefont
  {Plenz}}]{MeiselKlausKuehnPlenz}%
  \BibitemOpen
  \bibfield  {author} {\bibinfo {author} {\bibfnamefont {C.}~\bibnamefont
  {Meisel}}, \bibinfo {author} {\bibfnamefont {A.}~\bibnamefont {Klaus}},
  \bibinfo {author} {\bibfnamefont {C.}~\bibnamefont {Kuehn}}, \ and\ \bibinfo
  {author} {\bibfnamefont {D.}~\bibnamefont {Plenz}},\ }\href@noop {}
  {\bibfield  {journal} {\bibinfo  {journal} {PLoS Comp. Biol.}\ }\textbf
  {\bibinfo {volume} {11}},\ \bibinfo {pages} {e1004097} (\bibinfo {year}
  {2015})}\BibitemShut {NoStop}%
\bibitem [{\citenamefont {Wang}\ \emph {et~al.}(2012)\citenamefont {Wang},
  \citenamefont {Dearing}, \citenamefont {Langdon}, \citenamefont {Zhang},
  \citenamefont {Yang}, \citenamefont {Dakos},\ and\ \citenamefont
  {Scheffer}}]{Wangetal}%
  \BibitemOpen
  \bibfield  {author} {\bibinfo {author} {\bibfnamefont {R.}~\bibnamefont
  {Wang}}, \bibinfo {author} {\bibfnamefont {J.}~\bibnamefont {Dearing}},
  \bibinfo {author} {\bibfnamefont {P.}~\bibnamefont {Langdon}}, \bibinfo
  {author} {\bibfnamefont {E.}~\bibnamefont {Zhang}}, \bibinfo {author}
  {\bibfnamefont {X.}~\bibnamefont {Yang}}, \bibinfo {author} {\bibfnamefont
  {V.}~\bibnamefont {Dakos}}, \ and\ \bibinfo {author} {\bibfnamefont
  {M.}~\bibnamefont {Scheffer}},\ }\href@noop {} {\bibfield  {journal}
  {\bibinfo  {journal} {Nature}\ }\textbf {\bibinfo {volume} {492}},\ \bibinfo
  {pages} {419} (\bibinfo {year} {2012})}\BibitemShut {NoStop}%
\bibitem [{\citenamefont {Kantz}\ \emph {et~al.}(2004)\citenamefont {Kantz},
  \citenamefont {Holstein}, \citenamefont {Ragwitz},\ and\ \citenamefont
  {Vitanov}}]{Physa}%
  \BibitemOpen
  \bibfield  {author} {\bibinfo {author} {\bibfnamefont {H.}~\bibnamefont
  {Kantz}}, \bibinfo {author} {\bibfnamefont {D.}~\bibnamefont {Holstein}},
  \bibinfo {author} {\bibfnamefont {M.}~\bibnamefont {Ragwitz}}, \ and\
  \bibinfo {author} {\bibfnamefont {N.~K.}\ \bibnamefont {Vitanov}},\
  }\href@noop {} {\bibfield  {journal} {\bibinfo  {journal} {Physica}\ }\textbf
  {\bibinfo {volume} {A 342}},\ \bibinfo {pages} {315} (\bibinfo {year}
  {2004})}\BibitemShut {NoStop}%
\bibitem [{\citenamefont {Hallerberg}\ \emph {et~al.}(2007)\citenamefont
  {Hallerberg}, \citenamefont {Altmann}, \citenamefont {Holstein},\ and\
  \citenamefont {Kantz}}]{hallerberg2007precursors}%
  \BibitemOpen
  \bibfield  {author} {\bibinfo {author} {\bibfnamefont {S.}~\bibnamefont
  {Hallerberg}}, \bibinfo {author} {\bibfnamefont {E.~G.}\ \bibnamefont
  {Altmann}}, \bibinfo {author} {\bibfnamefont {D.}~\bibnamefont {Holstein}}, \
  and\ \bibinfo {author} {\bibfnamefont {H.}~\bibnamefont {Kantz}},\
  }\href@noop {} {\bibfield  {journal} {\bibinfo  {journal} {Physical Review
  E}\ }\textbf {\bibinfo {volume} {75}},\ \bibinfo {pages} {016706} (\bibinfo
  {year} {2007})}\BibitemShut {NoStop}%
\bibitem [{\citenamefont {Hallerberg}\ and\ \citenamefont
  {Kantz}(2008{\natexlab{a}})}]{Hallerberg2008a}%
  \BibitemOpen
  \bibfield  {author} {\bibinfo {author} {\bibfnamefont {S.}~\bibnamefont
  {Hallerberg}}\ and\ \bibinfo {author} {\bibfnamefont {H.}~\bibnamefont
  {Kantz}},\ }\href {\doibase 10.5194/npg-15-321-2008} {\bibfield  {journal}
  {\bibinfo  {journal} {Nonlinear Processes in Geophysics}\ }\textbf {\bibinfo
  {volume} {15}},\ \bibinfo {pages} {321} (\bibinfo {year}
  {2008}{\natexlab{a}})}\BibitemShut {NoStop}%
\bibitem [{\citenamefont {Bogachev}\ and\ \citenamefont
  {Bunde}(2009{\natexlab{a}})}]{BogachevBundeEPL2009}%
  \BibitemOpen
  \bibfield  {author} {\bibinfo {author} {\bibfnamefont {M.~I.}\ \bibnamefont
  {Bogachev}}\ and\ \bibinfo {author} {\bibfnamefont {A.}~\bibnamefont
  {Bunde}},\ }\href@noop {} {\bibfield  {journal} {\bibinfo  {journal} {EPL}\
  }\textbf {\bibinfo {volume} {86}},\ \bibinfo {pages} {66002} (\bibinfo {year}
  {2009}{\natexlab{a}})}\BibitemShut {NoStop}%
\bibitem [{\citenamefont {Bogachev}\ and\ \citenamefont
  {Bunde}(2009{\natexlab{b}})}]{BogachevBundePRE2009}%
  \BibitemOpen
  \bibfield  {author} {\bibinfo {author} {\bibfnamefont {M.~I.}\ \bibnamefont
  {Bogachev}}\ and\ \bibinfo {author} {\bibfnamefont {A.}~\bibnamefont
  {Bunde}},\ }\href@noop {} {\bibfield  {journal} {\bibinfo  {journal} {Phys.
  Rev. E}\ }\textbf {\bibinfo {volume} {80}},\ \bibinfo {pages} {026131}
  (\bibinfo {year} {2009}{\natexlab{b}})}\BibitemShut {NoStop}%
\bibitem [{\citenamefont {Hallerberg}\ and\ \citenamefont
  {de~Wijn}(2014)}]{HallerbergDeWijn2014}%
  \BibitemOpen
  \bibfield  {author} {\bibinfo {author} {\bibfnamefont {S.}~\bibnamefont
  {Hallerberg}}\ and\ \bibinfo {author} {\bibfnamefont {A.~S.}\ \bibnamefont
  {de~Wijn}},\ }\href@noop {} {\bibfield  {journal} {\bibinfo  {journal}
  {Physical Review E}\ }\textbf {\bibinfo {volume} {90}},\ \bibinfo {pages}
  {062901} (\bibinfo {year} {2014})}\BibitemShut {NoStop}%
\bibitem [{\citenamefont {Miotto}\ and\ \citenamefont
  {Altmann}(2014)}]{Miotto2014}%
  \BibitemOpen
  \bibfield  {author} {\bibinfo {author} {\bibfnamefont {J.~M.}\ \bibnamefont
  {Miotto}}\ and\ \bibinfo {author} {\bibfnamefont {E.~G.}\ \bibnamefont
  {Altmann}},\ }\href {\doibase 10.1371/journal.pone.0111506} {\bibfield
  {journal} {\bibinfo  {journal} {PLoS ONE}\ }\textbf {\bibinfo {volume} {9}},\
  \bibinfo {pages} {e111506} (\bibinfo {year} {2014})}\BibitemShut {NoStop}%
\bibitem [{\citenamefont {Brier}(1950)}]{Brier}%
  \BibitemOpen
  \bibfield  {author} {\bibinfo {author} {\bibfnamefont {G.~W.}\ \bibnamefont
  {Brier}},\ }\href@noop {} {\bibfield  {journal} {\bibinfo  {journal} {Mon.
  Weather Rev.}\ }\textbf {\bibinfo {volume} {78}},\ \bibinfo {pages} {1}
  (\bibinfo {year} {1950})}\BibitemShut {NoStop}%
\bibitem [{\citenamefont {Roulston}(2002)}]{roulston}%
  \BibitemOpen
  \bibfield  {author} {\bibinfo {author} {\bibfnamefont {M.}~\bibnamefont
  {Roulston}},\ }\href@noop {} {\bibfield  {journal} {\bibinfo  {journal}
  {Monthly Weather Review}\ }\textbf {\bibinfo {volume} {130}},\ \bibinfo
  {pages} {1653} (\bibinfo {year} {2002})}\BibitemShut {NoStop}%
\bibitem [{\citenamefont {Egan}(1975)}]{egan1975signal}%
  \BibitemOpen
  \bibfield  {author} {\bibinfo {author} {\bibfnamefont {J.}~\bibnamefont
  {Egan}},\ }\href@noop {} {\emph {\bibinfo {title} {Signal Detection Theory
  and ROC-analysis}}},\ Academic Press series in cognition and perception\
  (\bibinfo  {publisher} {Academic Press},\ \bibinfo {year} {1975})\BibitemShut
  {NoStop}%
\bibitem [{\citenamefont {Boettinger}\ and\ \citenamefont
  {Hastings}(2012)}]{BoettingerHastings}%
  \BibitemOpen
  \bibfield  {author} {\bibinfo {author} {\bibfnamefont {C.}~\bibnamefont
  {Boettinger}}\ and\ \bibinfo {author} {\bibfnamefont {A.}~\bibnamefont
  {Hastings}},\ }\href@noop {} {\bibfield  {journal} {\bibinfo  {journal} {J.
  R. Soc. Interface}\ }\textbf {\bibinfo {volume} {9}},\ \bibinfo {pages}
  {2527} (\bibinfo {year} {2012})}\BibitemShut {NoStop}%
\bibitem [{\citenamefont {Kuehn}(2011)}]{kuehn2011mathematical}%
  \BibitemOpen
  \bibfield  {author} {\bibinfo {author} {\bibfnamefont {C.}~\bibnamefont
  {Kuehn}},\ }\href@noop {} {\bibfield  {journal} {\bibinfo  {journal} {Physica
  D: Nonlinear Phenomena}\ }\textbf {\bibinfo {volume} {240}},\ \bibinfo
  {pages} {1020} (\bibinfo {year} {2011})}\BibitemShut {NoStop}%
\bibitem [{\citenamefont {Kuehn}\ \emph {et~al.}(2014)\citenamefont {Kuehn},
  \citenamefont {Martens},\ and\ \citenamefont {Romero}}]{KuehnMartensRomero}%
  \BibitemOpen
  \bibfield  {author} {\bibinfo {author} {\bibfnamefont {C.}~\bibnamefont
  {Kuehn}}, \bibinfo {author} {\bibfnamefont {E.}~\bibnamefont {Martens}}, \
  and\ \bibinfo {author} {\bibfnamefont {D.}~\bibnamefont {Romero}},\
  }\href@noop {} {\bibfield  {journal} {\bibinfo  {journal} {J. Complex
  Networks}\ }\textbf {\bibinfo {volume} {2}},\ \bibinfo {pages} {141}
  (\bibinfo {year} {2014})}\BibitemShut {NoStop}%
\bibitem [{\citenamefont {Meisel}\ and\ \citenamefont
  {Kuehn}(2012)}]{MeiselKuehn}%
  \BibitemOpen
  \bibfield  {author} {\bibinfo {author} {\bibfnamefont {C.}~\bibnamefont
  {Meisel}}\ and\ \bibinfo {author} {\bibfnamefont {C.}~\bibnamefont {Kuehn}},\
  }\href@noop {} {\bibfield  {journal} {\bibinfo  {journal} {PLoS ONE}\
  }\textbf {\bibinfo {volume} {7}},\ \bibinfo {pages} {e30371} (\bibinfo {year}
  {2012})}\BibitemShut {NoStop}%
\bibitem [{\citenamefont {Ditlevsen}\ and\ \citenamefont
  {Johnsen}(2010)}]{DitlevsenJohnsen}%
  \BibitemOpen
  \bibfield  {author} {\bibinfo {author} {\bibfnamefont {P.}~\bibnamefont
  {Ditlevsen}}\ and\ \bibinfo {author} {\bibfnamefont {S.}~\bibnamefont
  {Johnsen}},\ }\href@noop {} {\bibfield  {journal} {\bibinfo  {journal}
  {Geophys. Res. Lett.}\ }\textbf {\bibinfo {volume} {37}},\ \bibinfo {pages}
  {19703} (\bibinfo {year} {2010})}\BibitemShut {NoStop}%
\bibitem [{\citenamefont {Ashwin}\ \emph {et~al.}(2012)\citenamefont {Ashwin},
  \citenamefont {Wieczorek}, \citenamefont {Vitolo},\ and\ \citenamefont
  {Cox}}]{AshwinWieczorekVitoloCox}%
  \BibitemOpen
  \bibfield  {author} {\bibinfo {author} {\bibfnamefont {P.}~\bibnamefont
  {Ashwin}}, \bibinfo {author} {\bibfnamefont {S.}~\bibnamefont {Wieczorek}},
  \bibinfo {author} {\bibfnamefont {R.}~\bibnamefont {Vitolo}}, \ and\ \bibinfo
  {author} {\bibfnamefont {P.}~\bibnamefont {Cox}},\ }\href@noop {} {\bibfield
  {journal} {\bibinfo  {journal} {Phil. Trans. R. Soc. A}\ }\textbf {\bibinfo
  {volume} {370}},\ \bibinfo {pages} {1166} (\bibinfo {year}
  {2012})}\BibitemShut {NoStop}%
\bibitem [{\citenamefont {Berglund}\ and\ \citenamefont
  {Gentz}(2006)}]{berglund2006noise}%
  \BibitemOpen
  \bibfield  {author} {\bibinfo {author} {\bibfnamefont {N.}~\bibnamefont
  {Berglund}}\ and\ \bibinfo {author} {\bibfnamefont {B.}~\bibnamefont
  {Gentz}},\ }\href@noop {} {\emph {\bibinfo {title} {Noise-induced phenomena
  in slow-fast dynamical systems}}}\ (\bibinfo  {publisher} {Springer},\
  \bibinfo {year} {2006})\BibitemShut {NoStop}%
\bibitem [{\citenamefont {Boettinger}\ and\ \citenamefont
  {Hastings}(2013)}]{BoettingerHastings1}%
  \BibitemOpen
  \bibfield  {author} {\bibinfo {author} {\bibfnamefont {C.}~\bibnamefont
  {Boettinger}}\ and\ \bibinfo {author} {\bibfnamefont {A.}~\bibnamefont
  {Hastings}},\ }\href@noop {} {\bibfield  {journal} {\bibinfo  {journal}
  {Proc. R. Soc. B}\ }\textbf {\bibinfo {volume} {280}},\ \bibinfo {pages}
  {20131372} (\bibinfo {year} {2013})}\BibitemShut {NoStop}%
\bibitem [{\citenamefont {Freidlin}\ and\ \citenamefont
  {Wentzell}(1998)}]{FreidlinWentzell}%
  \BibitemOpen
  \bibfield  {author} {\bibinfo {author} {\bibfnamefont {M.}~\bibnamefont
  {Freidlin}}\ and\ \bibinfo {author} {\bibfnamefont {A.}~\bibnamefont
  {Wentzell}},\ }\href@noop {} {\emph {\bibinfo {title} {Random Perturbations
  of Dynamical Systems}}}\ (\bibinfo  {publisher} {Springer},\ \bibinfo {year}
  {1998})\BibitemShut {NoStop}%
\bibitem [{\citenamefont {Kuehn}(2015)}]{KuehnBook}%
  \BibitemOpen
  \bibfield  {author} {\bibinfo {author} {\bibfnamefont {C.}~\bibnamefont
  {Kuehn}},\ }\href@noop {} {\emph {\bibinfo {title} {Multiple Time Scale
  Dynamics}}}\ (\bibinfo  {publisher} {Springer},\ \bibinfo {year}
  {2015})\BibitemShut {NoStop}%
\bibitem [{\citenamefont {Berglund}\ and\ \citenamefont
  {Gentz}(2002)}]{berglund2002metastability}%
  \BibitemOpen
  \bibfield  {author} {\bibinfo {author} {\bibfnamefont {N.}~\bibnamefont
  {Berglund}}\ and\ \bibinfo {author} {\bibfnamefont {B.}~\bibnamefont
  {Gentz}},\ }\href@noop {} {\bibfield  {journal} {\bibinfo  {journal}
  {Stochastics and Dynamics}\ }\textbf {\bibinfo {volume} {2}},\ \bibinfo
  {pages} {327} (\bibinfo {year} {2002})}\BibitemShut {NoStop}%
\bibitem [{\citenamefont {Sacerdote}\ and\ \citenamefont
  {Giraudo}(2013)}]{SacerdoteGiraudo}%
  \BibitemOpen
  \bibfield  {author} {\bibinfo {author} {\bibfnamefont {L.}~\bibnamefont
  {Sacerdote}}\ and\ \bibinfo {author} {\bibfnamefont {M.}~\bibnamefont
  {Giraudo}},\ }in\ \href@noop {} {\emph {\bibinfo {booktitle} {Stochastic
  Biomathematical Models}}}\ (\bibinfo  {publisher} {Springer},\ \bibinfo
  {year} {2013})\ pp.\ \bibinfo {pages} {99--148}\BibitemShut {NoStop}%
\bibitem [{\citenamefont {Latham}\ \emph {et~al.}(2000)\citenamefont {Latham},
  \citenamefont {Richmond}, \citenamefont {Nelson},\ and\ \citenamefont
  {Nirenberg}}]{latham2000intrinsic}%
  \BibitemOpen
  \bibfield  {author} {\bibinfo {author} {\bibfnamefont {P.~E.}\ \bibnamefont
  {Latham}}, \bibinfo {author} {\bibfnamefont {B.~J.}\ \bibnamefont
  {Richmond}}, \bibinfo {author} {\bibfnamefont {P.~G.}\ \bibnamefont
  {Nelson}}, \ and\ \bibinfo {author} {\bibfnamefont {S.}~\bibnamefont
  {Nirenberg}},\ }\href@noop {} {\bibfield  {journal} {\bibinfo  {journal}
  {Journal of Neurophysiology}\ }\textbf {\bibinfo {volume} {83}},\ \bibinfo
  {pages} {808} (\bibinfo {year} {2000})}\BibitemShut {NoStop}%
\bibitem [{\citenamefont {Ermentrout}(1996)}]{ermentrout1996type}%
  \BibitemOpen
  \bibfield  {author} {\bibinfo {author} {\bibfnamefont {B.}~\bibnamefont
  {Ermentrout}},\ }\href@noop {} {\bibfield  {journal} {\bibinfo  {journal}
  {Neural computation}\ }\textbf {\bibinfo {volume} {8}},\ \bibinfo {pages}
  {979} (\bibinfo {year} {1996})}\BibitemShut {NoStop}%
\bibitem [{\citenamefont {Higham}(2001)}]{Higham}%
  \BibitemOpen
  \bibfield  {author} {\bibinfo {author} {\bibfnamefont {D.}~\bibnamefont
  {Higham}},\ }\href@noop {} {\bibfield  {journal} {\bibinfo  {journal} {SIAM
  Review}\ }\textbf {\bibinfo {volume} {43}},\ \bibinfo {pages} {525} (\bibinfo
  {year} {2001})}\BibitemShut {NoStop}%
\bibitem [{\citenamefont {Ermentrout}\ and\ \citenamefont
  {Terman}(2010)}]{ErmentroutTerman}%
  \BibitemOpen
  \bibfield  {author} {\bibinfo {author} {\bibfnamefont {G.}~\bibnamefont
  {Ermentrout}}\ and\ \bibinfo {author} {\bibfnamefont {D.}~\bibnamefont
  {Terman}},\ }\href@noop {} {\emph {\bibinfo {title} {Mathematical Foundations
  of Neuroscience}}}\ (\bibinfo  {publisher} {Springer},\ \bibinfo {year}
  {2010})\BibitemShut {NoStop}%
\bibitem [{\citenamefont {van~der Pol}(1926)}]{vanderPol1}%
  \BibitemOpen
  \bibfield  {author} {\bibinfo {author} {\bibfnamefont {B.}~\bibnamefont
  {van~der Pol}},\ }\href@noop {} {\bibfield  {journal} {\bibinfo  {journal}
  {Philosophical Magazine}\ }\textbf {\bibinfo {volume} {7}},\ \bibinfo {pages}
  {978} (\bibinfo {year} {1926})}\BibitemShut {NoStop}%
\bibitem [{\citenamefont {FitzHugh}(1955)}]{FitzHugh}%
  \BibitemOpen
  \bibfield  {author} {\bibinfo {author} {\bibfnamefont {R.}~\bibnamefont
  {FitzHugh}},\ }\href@noop {} {\bibfield  {journal} {\bibinfo  {journal}
  {Bull. Math. Biophysics}\ }\textbf {\bibinfo {volume} {17}},\ \bibinfo
  {pages} {257} (\bibinfo {year} {1955})}\BibitemShut {NoStop}%
\bibitem [{\citenamefont {Grasman}(1987)}]{Grasman}%
  \BibitemOpen
  \bibfield  {author} {\bibinfo {author} {\bibfnamefont {J.}~\bibnamefont
  {Grasman}},\ }\href@noop {} {\emph {\bibinfo {title} {Asymptotic Methods for
  Relaxation Oscillations and Applications}}}\ (\bibinfo  {publisher}
  {Springer},\ \bibinfo {year} {1987})\BibitemShut {NoStop}%
\bibitem [{\citenamefont {Laney}(2001)}]{laney01controlling3v}%
  \BibitemOpen
  \bibfield  {author} {\bibinfo {author} {\bibfnamefont {D.}~\bibnamefont
  {Laney}},\ }\href
  {http://blogs.gartner.com/doug-laney/files/2012/01/ad949-3D-Data-Management-Controlling-Data-Volume-Velocity-and-Variety.pdf}
  {\emph {\bibinfo {title} {{3D} Data Management: Controlling Data Volume,
  Velocity, and Variety}}},\ \bibinfo {type} {Tech. Rep.}\ (\bibinfo
  {institution} {META Group},\ \bibinfo {year} {2001})\BibitemShut {NoStop}%
\bibitem [{\citenamefont {Hirata}(2014)}]{Hirata2014}%
  \BibitemOpen
  \bibfield  {author} {\bibinfo {author} {\bibfnamefont {Y.}~\bibnamefont
  {Hirata}},\ }\href@noop {} {\bibfield  {journal} {\bibinfo  {journal}
  {Physical Review E}\ }\textbf {\bibinfo {volume} {89}},\ \bibinfo {pages}
  {052916} (\bibinfo {year} {2014})}\BibitemShut {NoStop}%
\bibitem [{\citenamefont {Runge}\ \emph {et~al.}(2015)\citenamefont {Runge},
  \citenamefont {Donner},\ and\ \citenamefont {Kurths}}]{Runge2015}%
  \BibitemOpen
  \bibfield  {author} {\bibinfo {author} {\bibfnamefont {J.}~\bibnamefont
  {Runge}}, \bibinfo {author} {\bibfnamefont {V.~R.}\ \bibnamefont {Donner}}, \
  and\ \bibinfo {author} {\bibfnamefont {J.}~\bibnamefont {Kurths}},\
  }\href@noop {} {\bibfield  {journal} {\bibinfo  {journal} {Physical Review
  E}\ }\textbf {\bibinfo {volume} {89}},\ \bibinfo {pages} {052916} (\bibinfo
  {year} {2015})}\BibitemShut {NoStop}%
\bibitem [{\citenamefont {Rish}(2001)}]{Rish2001}%
  \BibitemOpen
  \bibfield  {author} {\bibinfo {author} {\bibfnamefont {I.}~\bibnamefont
  {Rish}},\ }\href@noop {} {\emph {\bibinfo {title} {An empirical study of the
  naive Bayes classifier}}},\ \bibinfo {type} {Tech. Rep.}\ (\bibinfo
  {institution} {IBM Research Division Thomas J. Watson Research Center P.O.
  Box 218 Yorktown Heights, NY 10598},\ \bibinfo {year} {2001})\BibitemShut
  {NoStop}%
\bibitem [{\citenamefont {LeCun}\ \emph {et~al.}(1989)\citenamefont {LeCun},
  \citenamefont {Boser}, \citenamefont {Denker}, \citenamefont {Henderson},
  \citenamefont {Howard}, \citenamefont {Hubbard},\ and\ \citenamefont
  {Jackel}}]{LeCun1989}%
  \BibitemOpen
  \bibfield  {author} {\bibinfo {author} {\bibfnamefont {Y.}~\bibnamefont
  {LeCun}}, \bibinfo {author} {\bibfnamefont {B.}~\bibnamefont {Boser}},
  \bibinfo {author} {\bibfnamefont {J.~S.}\ \bibnamefont {Denker}}, \bibinfo
  {author} {\bibfnamefont {D.}~\bibnamefont {Henderson}}, \bibinfo {author}
  {\bibfnamefont {R.~E.}\ \bibnamefont {Howard}}, \bibinfo {author}
  {\bibfnamefont {W.}~\bibnamefont {Hubbard}}, \ and\ \bibinfo {author}
  {\bibfnamefont {L.~D.}\ \bibnamefont {Jackel}},\ }\href {\doibase
  10.1162/neco.1989.1.4.541} {\bibfield  {journal} {\bibinfo  {journal} {Neural
  Comput.}\ }\textbf {\bibinfo {volume} {1}},\ \bibinfo {pages} {541} (\bibinfo
  {year} {1989})}\BibitemShut {NoStop}%
\bibitem [{\citenamefont {Hinton}\ and\ \citenamefont
  {Salakhutdinov}(2006)}]{HinSal06}%
  \BibitemOpen
  \bibfield  {author} {\bibinfo {author} {\bibfnamefont {G.}~\bibnamefont
  {Hinton}}\ and\ \bibinfo {author} {\bibfnamefont {R.}~\bibnamefont
  {Salakhutdinov}},\ }\href@noop {} {\bibfield  {journal} {\bibinfo  {journal}
  {Science}\ }\textbf {\bibinfo {volume} {313}},\ \bibinfo {pages} {504 }
  (\bibinfo {year} {2006})}\BibitemShut {NoStop}%
\bibitem [{\citenamefont {Cortes}\ and\ \citenamefont
  {Vapnik}(1995)}]{Cortes1995}%
  \BibitemOpen
  \bibfield  {author} {\bibinfo {author} {\bibfnamefont {C.}~\bibnamefont
  {Cortes}}\ and\ \bibinfo {author} {\bibfnamefont {V.}~\bibnamefont
  {Vapnik}},\ }\href {\doibase 10.1023/A:1022627411411} {\bibfield  {journal}
  {\bibinfo  {journal} {Mach. Learn.}\ }\textbf {\bibinfo {volume} {20}},\
  \bibinfo {pages} {273} (\bibinfo {year} {1995})}\BibitemShut {NoStop}%
\bibitem [{\citenamefont {Zhang}(2004)}]{naivebayes}%
  \BibitemOpen
  \bibfield  {author} {\bibinfo {author} {\bibfnamefont {H.}~\bibnamefont
  {Zhang}},\ }\href@noop {} {\bibfield  {journal} {\bibinfo  {journal}
  {American Association of Artificial Intelligence}\ } (\bibinfo {year}
  {2004})}\BibitemShut {NoStop}%
\bibitem [{\citenamefont {Priestley}(2001)}]{priestley2001spectral}%
  \BibitemOpen
  \bibfield  {author} {\bibinfo {author} {\bibfnamefont {B.}~\bibnamefont
  {Priestley}},\ }\href {http://books.google.de/books?id=z9h4nQEACAAJ} {\emph
  {\bibinfo {title} {Spectral Analysis and Time Series}}}\ (\bibinfo
  {publisher} {Elsevier},\ \bibinfo {year} {2001})\BibitemShut {NoStop}%
\bibitem [{\citenamefont {Bayes}(1763)}]{bayes}%
  \BibitemOpen
  \bibfield  {author} {\bibinfo {author} {\bibfnamefont {T.}~\bibnamefont
  {Bayes}},\ }\href@noop {} {\bibfield  {journal} {\bibinfo  {journal} {Phil.
  Trans. Roy. Soc. London}\ }\textbf {\bibinfo {volume} {53}},\ \bibinfo
  {pages} {370} (\bibinfo {year} {1763})}\BibitemShut {NoStop}%
\bibitem [{\citenamefont {Fawcett}(2006)}]{fawcett2006introduction}%
  \BibitemOpen
  \bibfield  {author} {\bibinfo {author} {\bibfnamefont {T.}~\bibnamefont
  {Fawcett}},\ }\href@noop {} {\bibfield  {journal} {\bibinfo  {journal}
  {Pattern recognition letters}\ }\textbf {\bibinfo {volume} {27}},\ \bibinfo
  {pages} {861} (\bibinfo {year} {2006})}\BibitemShut {NoStop}%
\bibitem [{\citenamefont {Metz}(1978)}]{metz1978basic}%
  \BibitemOpen
  \bibfield  {author} {\bibinfo {author} {\bibfnamefont {C.~E.}\ \bibnamefont
  {Metz}},\ }in\ \href@noop {} {\emph {\bibinfo {booktitle} {Seminars in
  nuclear medicine}}},\ Vol.~\bibinfo {volume} {8}\ (\bibinfo {organization}
  {Elsevier},\ \bibinfo {year} {1978})\ pp.\ \bibinfo {pages}
  {283--298}\BibitemShut {NoStop}%
\bibitem [{\citenamefont {Hallerberg}\ and\ \citenamefont
  {Kantz}(2008{\natexlab{b}})}]{hallerberg2008influence}%
  \BibitemOpen
  \bibfield  {author} {\bibinfo {author} {\bibfnamefont {S.}~\bibnamefont
  {Hallerberg}}\ and\ \bibinfo {author} {\bibfnamefont {H.}~\bibnamefont
  {Kantz}},\ }\href@noop {} {\bibfield  {journal} {\bibinfo  {journal}
  {Physical Review E}\ }\textbf {\bibinfo {volume} {77}},\ \bibinfo {pages}
  {011108} (\bibinfo {year} {2008}{\natexlab{b}})}\BibitemShut {NoStop}%
\bibitem [{\citenamefont {Hallerberg}(2008)}]{hallerberg2008predictability}%
  \BibitemOpen
  \bibfield  {author} {\bibinfo {author} {\bibfnamefont {S.}~\bibnamefont
  {Hallerberg}},\ }\emph {\bibinfo {title} {Predictability of Extreme Events in
  Time Series}},\ \href@noop {} {Ph.D. thesis},\ \bibinfo  {school} {Wuppertal,
  Univ., Diss., 2008} (\bibinfo {year} {2008})\BibitemShut {NoStop}%
\bibitem [{\citenamefont {Savitzky}\ and\ \citenamefont
  {Golay}(1964)}]{Savgol}%
  \BibitemOpen
  \bibfield  {author} {\bibinfo {author} {\bibfnamefont {A.}~\bibnamefont
  {Savitzky}}\ and\ \bibinfo {author} {\bibfnamefont {M.~J.~E.}\ \bibnamefont
  {Golay}},\ }\href@noop {} {\bibfield  {journal} {\bibinfo  {journal}
  {Analytical Chemistry}\ }\textbf {\bibinfo {volume} {36}},\ \bibinfo {pages}
  {1627} (\bibinfo {year} {1964})}\BibitemShut {NoStop}%
\end{thebibliography}%

\end{document}